\documentclass[12pt]{article}

\usepackage{latexsym}
\usepackage{amssymb,amsfonts,amsmath}
\usepackage{graphicx} 
\usepackage{indentfirst}
\usepackage{bbm}
\usepackage{amssymb}
\usepackage{verbatim}
\usepackage{amsmath, amsthm,amssymb}
\usepackage{mathrsfs}
\usepackage{hyperref}
\usepackage{amsfonts}
\usepackage{dsfont}
\usepackage{cite}
\usepackage{xcolor}
\usepackage[multiple]{footmisc}

\usepackage{arydshln}

\parskip=\medskipamount

\parskip=\medskipamount

\usepackage[mathscr]{euscript}

\numberwithin{equation}{section}

\newcommand {\cD}{{\cal D}}
\newcommand {\cE}{{\cal E}}

\newcommand {\cG}{{\cal G}}
\newcommand {\cH}{{\cal H}}

\newcommand {\cJ}{{\cal J}}

\newcommand {\cM}{{\cal M}}
\newcommand {\cN}{{\cal N}}

\newcommand {\cP}{{\cal P}}

\newcommand {\cR}{{\cal R}}

\newcommand {\cT}{{\cal T}}

\newcommand {\cW}{{\cal W}}

\def\a{\alpha}
\def\b{\beta}

\def\d{\delta}
\def\e{\epsilon}

\def\g{\gamma}

\def\l{\lambda}
\def\m{\mu}
\def\n{\nu}
\def\o{\omega}
\def\p{\pi}
\def\q{\theta}

\def\s{\sigma}

\def\z{\zeta}

\def\L{\Lambda}
\def\O{\Omega}

\def\S{\Sigma}

\def\ri{{\rm i}}
\def\re{{\rm e}}

\newcommand{\sSL}{\mathsf{SL}}
\newcommand{\sGL}{\mathsf{GL}}
\newcommand{\sO}{\mathsf{O}}
\newcommand{\sSO}{\mathsf{SO}}
\newcommand{\sISO}{\mathsf{ISO}}

\newcommand{\sOSp}{\mathsf{OSp}}

\newcommand{\Iu}{\underline{I}}
\newcommand{\Au}{\underline{A}}
\newcommand{\Bu}{\underline{B}}

\newcommand{\Ju}{\underline{J}}
\newcommand{\alu}{{\underline{\a}}}
\newcommand{\beu}{{\underline{\b}}}

\newcommand{\Io}{{\overline{I}}}
\newcommand{\Ao}{\overline{A}}
\newcommand{\Bo}{\overline{B}}

\newcommand{\Jo}{\overline{J}}
\newcommand{\alo}{{\overline{\a}}}
\newcommand{\beo}{{\overline{\b}}}

\newcommand{\ve}{\varepsilon}

\newcommand{\pa}{\partial}
\newcommand{\hf}{\frac12}

\newcommand{\vf}{\varphi}

\newcommand{\be}{\begin{equation}}
\newcommand{\ee}{\end{equation}}
\newcommand{\bea}{\begin{eqnarray}}
\newcommand{\eea}{\end{eqnarray}}
\newcommand{\non}{\nonumber}
\newcommand{\ba}{\begin{array}}
\newcommand{\ea}{\end{array}}

\newcommand{\bm}[1]{\mbox{\boldmath$#1$}}

\def\double #1{#1{\hbox{\kern-2pt $#1$}}}

\newcommand{\te}{{\theta}}

\newcommand{\bsubeq}{\begin{subequations}}
\newcommand{\esubeq}{\end{subequations}}

\newcommand{\rd}{\mathrm d}
\newcommand{\Exp}[1]{\langle #1 \rangle}

\newcommand{\fff}{\vrule width0.5pt height5pt depth1pt} 

\newcommand{\pp}{{{ =\hskip-3.75pt{\fff}}\hskip3.75pt }}

\def\tr{{\rm tr}}

\begin{document}
\begin{titlepage}
\begin{flushright}
March, 2023\\
\end{flushright}
\vspace{5mm}

\begin{center}
{\Large \bf 
Embedding formalism for $\bm{(p,q)}$ AdS superspaces \\
in three dimensions}
\end{center}

\begin{center}

{\bf
Sergei M. Kuzenko and Kai Turner
} \\
\vspace{5mm}

\footnotesize{
{\it Department of Physics M013, The University of Western Australia,\\
35 Stirling Highway, Perth W.A. 6009, Australia}}  
~\\
\vspace{2mm}
\texttt{sergei.kuzenko@uwa.edu.au, kai.turner@research.uwa.edu.au
}\\
\vspace{2mm}

\end{center}

\begin{abstract}
\baselineskip=14pt

We develop an embedding formalism for $(p,q)$ anti-de Sitter (AdS) superspaces in three dimensions by using a modified version of their supertwistor description given in the literature. A coset construction for these superspaces is worked out. We put forward a program of constructing a supersymmetric analogue of the Ba\~nados metric, which  
is expected to be a deformation of the $(p,q)$ AdS superspace geometry by a two-dimensional conformal $(p,q)$ supercurrent multiplet. 
\end{abstract} 
\vspace{5mm}

\vfill
\end{titlepage}

\newpage
\renewcommand{\thefootnote}{\arabic{footnote}}
\setcounter{footnote}{0}

\tableofcontents{}
\vspace{1cm}
\bigskip\hrule


\allowdisplaybreaks

\section{Introduction} \label{section1}

To study field theories in a $d$-dimensional de Sitter space ${\rm dS}_d = \sO(d,1)/\sO(d-1,1)$ or anti-de Sitter space
${\rm AdS}_d = \sO(d-1,2)/\sO(d-1,1)$,  it is often useful to deal with their embeddings as hypersurfaces in pseudo-Euclidean spaces  ${\mathbb R}^{d,1}$  and
${\mathbb R}^{d-1,2}$, respectively.
These embeddings are defined by 
\begin{subequations}\label{1.1}
\bea
 {\rm dS}_d : \qquad   -(X^0)^2 + (X^1)^2 + \dots +(X^{d-1})^2 + (X^d)^2 
 &=& \ell^2 
  ~,\\
{\rm AdS}_d :\qquad -(X^0)^2 + (X^1)^2 + \dots +(X^{d-1})^2 - (X^d)^2 
 &=& -\ell^2 ~, 
 \label{1.1b}
 \eea
 \end{subequations}
 with $\ell >0$ a constant parameter. Supersymmetric analogues of ${\rm AdS}_d $, known as AdS superspaces, exist in the special spacetime dimensions $d$
 that are related to those dimensions $\tilde{d}= d-1 \leq 6$ 
 which support finite-dimensional superconformal groups. 
 To the best of our knowledge, their superembeddings  have not been studied much in the literature.  For possible AdS$_d$ superembeddings, one should not expect to have a universal functional result  like \eqref{1.1b}. In other words, their structure should be $d$-dependent in a non-trivial way. In addition, it is worth expecting that AdS$_d$ superembeddings be realised in terms of supertwistors in $(d-1)$ dimensions. 
 It should be pointed out that the (super)twistor descriptions of (super)particles in AdS had been given in the literature much earlier \cite{CGKRZ,CRZ,CKR,BLPS,Zunger,Cederwall1,Cederwall2,AB-GT1,AB-GT2,Uvarov}.

 In a recent paper \cite{KT-M21}, supertwistor realisations were proposed for 
  $(p,q)$ anti-de Sitter (AdS) superspaces in three dimensions, ${\rm AdS}^{(3|p,q)}$, and $\cal N$-extended AdS superspaces in four dimensions, $\rm{AdS}^{4|4\cN}$. 
  Making use of the latter construction,  the bi-supertwistor formulation of $\rm{AdS}^{4|4\cN}$
 was derived. It yielded the supersymmetric analogue of \eqref{1.1b} in the $d=4 $ case. 
In this paper we present a bi-supertwistor formulation of ${\rm AdS}^{(3|p,q)}$, which 
 provides the supersymmetric analogue of \eqref{1.1b} in the $d=3 $ case. 
We also work out a coset construction for ${\rm AdS}^{(3|p,q)}$ and demonstrate that its geometry agrees with that described in \cite{KLT-M12}.

The superspaces ${\rm AdS}^{(3|p,q)}$ were introduced  in \cite{KLT-M12}
as  
backgrounds of the $\cN$-extended off-shell conformal supergravity in three dimensions 
\cite{HIPT,KLT-M11} with  covariantly constant and Lorentz invariant torsion and curvature tensors, with $\cN= p+q$.
These superspaces were demonstrated to be conformally flat \cite{KLT-M12}, see \cite{BILS} for the earlier alternative analysis.\footnote{In the  $(p,q) = (\cN,0)$ case, there also exist non-conformally flat AdS superspaces for $\cN\geq 4$ \cite{KLT-M12}.} 
The infinitesimal isometries of ${\rm AdS}^{(3|p,q)} $ were 
 shown \cite{KLT-M12} to span the superalgebra\footnote{There are more general AdS$_3$ superlagebras \cite{GST}.}  
 \bea
\mathfrak{osp} (p|2; {\mathbb R} ) \oplus  \mathfrak{osp} (q|2; {\mathbb R} )~.
\eea
However, a direct study of ${\rm AdS}^{(3|p,q)} $ as the homogeneous space
\bea
 \frac{ {\sOSp} (p|2; {\mathbb R} ) \times  {\sOSp} (q|2; {\mathbb R} ) } 
{ {\sSL}( 2, {\mathbb R}) \times {\sSO}(p) \times {\sSO}(q)}~,
\eea
was not given in \cite{KLT-M12}. This will be done in the present paper using the supertwistor realisation of ${\rm AdS}^{(3|p,q)} $.

One of the motivations to study AdS superspaces in three dimensions is to obtain supersymmetric analogues of the Ba\~nados metric \cite{Banados} 
\bea
\rd s^2 =\ell^2 \bigg\{ \Big(\frac{ \rd z}{z}\Big)^2
- \Big[ \frac{\rd x^{\pp}}{z} + z{\cT}_{==}(x^{=}) \rd x^{=}\Big]
\Big[ \frac{\rd x^{=}}{z} + z \cT_{\pp \pp}(x^{\pp}) \rd x^{\pp}\Big] \bigg\}~,
\label{1.4}
\eea
where ${\cT}_{==}(x)$ and ${\cT}_{\pp \pp}(x)$ are arbitrary functions of a real variable.
For any choice of ${\cT}_{==}(x)$ and ${\cT}_{\pp\pp}(x)$, 
this metric  is a solution of the Einstein equations with a negative cosmological term, which can
be written as  the algebra of AdS$_3$ covariant derivatives 
\bea
\big[ \nabla_a , \nabla_b \big]  = -\frac{1}{\ell^2} \cM_{ab}~.
\label{1.5}
\eea
The choice ${\cT}_{==}=0$ and ${\cT}_{\pp\pp}=0$ in \eqref{1.4} corresponds to an AdS background.

A supersymmetric analogue of \eqref{1.5} is 
the algebra of the covariant derivatives $\cD_A = (\cD_a , \cD_\a^I)$
of  ${\rm AdS}^{(3|p,q)} $, 
which was derived in \cite{KLT-M12}:
\bsubeq
\label{alg-AdS}
\bea
\{\cD_\a^I,\cD_\b^J\}&=&
2\ri\d^{IJ} (\g^c)_{\a\b} \cD_{c}
-4\ri S^{IJ}\cM_{\a\b}
-4 \ri\ve_{\a\b}
S^{K}{}^{[I}\d^{J]L}\cN_{KL}
~,
\label{alg-AdS-1}
\\
{[}\cD_{a},\cD_\b^J{]}
&=&
S^{J}{}_{K}(\g_a)_\b{}^\g\cD_{\g}^K
~,
\label{alg-AdS-3/2}
\\
{[}\cD_{a},\cD_b{]}
&=&
- \frac{1}{\ell^2}
\cM_{ab}
~,
\label{alg-AdS-2}
\eea
where 
$\cM_{ab}=-\cM_{ba} $ and $\cN_{IJ}=-\cN_{JI}$ are the Lorentz\footnote{There are two alternative ways to write the Lorentz generator: (i) as a three-vector $\cM_a=\hf\ve_{abc}\cM^{bc}$; 
and (ii) as a symmetric second rank spinor
$\cM_{\a\b}:=
\hf(\g^a)_{\a\b}\ve_{abc}\cM^{bc}$. For more details see the appendix.}
and $\sSO(p+q)$ generators,
respectively. The algebra \eqref{alg-AdS} is determined by a symmetric  tensor field $S^{IJ}=S^{JI}$, which is covariantly constant, $\cD_A S^{IJ}=0$, and has the following algebraic properties: 
\bea
{\hat S}^2 =S^2 {\mathbbm 1}~, \qquad
S^2 :=\frac{1}{\cN}\, \tr ({\hat S}^2) =\frac{1}{4\ell^2} 
>0
~,
\label{alg-constr-SS}
\eea
\esubeq
where ${\hat S}:= (S^{IJ} ) ={\hat S}^{\rm T}$.
Applying a local $\sSO(\cN)$ transformation allows one to bring $S^{IJ}$ to the diagonal form
\bea \label{diagonal}
S^{IJ}=S\,{\rm{diag}}(\,  \overbrace{+1,\cdots,+1}^{p} \, , \overbrace{-1,\cdots,-1}^{q} \,)
~.
\label{diag-S}
\eea
In such a frame, one is left with an unbroken local group 
$\sSO(p)\times \sSO(q)$.  It is an interesting problem to find a $(p,q)$ supersymmetric generalisation of the metric \eqref{1.4}. The starting point to address this problem should be to derive a Poincar\'e coordinate patch for  ${\rm AdS}^{(3|p,q)} $ in which the AdS superspace covariant derivatives $\cD_A = (\cD_a, \cD_\a^I)$ are conformally related to those of Minkowski superspace ${\mathbb M}^{3|p+q}$.

This paper is organised as follows. In section \ref{Section2} we describe 
a revised version of the supertwistor description of 
 ${\rm AdS}^{(3|p,q)}$ presented in \cite{KT-M21}. 
The bi-supertwistor formulation for ${\rm AdS}^{(3|p,q)}$ is introduced in section 
 \ref{Section3}. Sections \ref{Section4} and \ref{Section5} are devoted to the
  the coset construction of  ${\rm AdS}^{(3|p,q)}$ and its use to describe the superspace geometry.
The geometric aspects of ${\rm AdS}^{(3|p,q)}$ in a Poincar\'e coordinate chart are studied in 
section \ref{Section6}. Our notation and conventions are described in the appendix.


\section{A review of the supertwistor construction} \label{Section2}

In this section we present a slightly modified version of the supertwistor description of 
 ${\rm AdS}^{(3|p,q)}$ given in \cite{KT-M21}.
 

 \subsection{Two realisations of ${\sOSp} (n|2; {\mathbb R} ) $}

In order to introduce the supertwistor description of ${\rm AdS}^{(3|p,q)}$, it is useful to work with two different, but equivalent, realisations of ${\sOSp} (n|2; {\mathbb R} ) $, 
for which we will use the notation ${\sOSp}_+ (n|2; {\mathbb R} ) $ and ${\sOSp}_- (n|2; {\mathbb R} ) $.
Both supergroups naturally act on the space of even  supertwistors and on the space of odd supertwistors. An arbitrary  supertwistor looks like
\bea
T = (T_{A}) =\left(
\begin{array}{c}
T_{\a} \\ \hline
 T_{I}
\end{array}
\right)~, 
 \qquad \a  = 1,2 ~, \quad I = 1, \dots, n~.
\eea
For pure supertwistors (even or odd), the components $T_\a$ and $T_I$ have certain Grassmann parities. 
If $T$ is even, the components $ T_\a$ are bosonic
and $T_I$  fermionic. If $T$ is odd, the components $ T_\a$ are fermionic
and $T_I$  bosonic.  Equivalently, the components $T_{A}$ of a pure supertwistor
 have the following Grassmann parities:
\bea
\e ( T_A) = \e ( T ) + \e_{A} \qquad (\mbox{mod 2})~,\quad 
 \e_{A} := \left\{
\begin{array}{c}
 0 \qquad A=\a \\
 1 \qquad A=I
\end{array}
\right.{}~.
\eea
Here the parity function $\e ( T )$ is defined by the rule:
$\e ( T ) = 0$ if $ T$ is even, and $\e ( T ) =1$ if $T $ is odd.
A pure  supertwistor is said to be real if its components obey the reality condition
\bea
(T_{A})^* = (-1)^{\e(T) \e_{A} + \e_{A}} \,T_{A}~.
\label{realcon}
\eea 

Let us introduce two graded antisymmetric supermatrices ${\mathbb J}_+$
and ${\mathbb J}_-$ defined by 
\bea
{\mathbb J}_{\pm} = ({\mathbb J}^{AB}) = \left(
\begin{array}{c |c}
\ve  ~&~ 0 \\\hline 
0 ~& ~\pm {\rm i} {\mathbbm 1}_p
\end{array} \right) ~, \qquad
\ve
=\big(\ve^{\a  \b} \big) = \ri \s_2
=\left(
\begin{array}{cc}
0  & { 1}\\
 -{ 1}  &    0
\end{array}
\right) ~,
\label{symplectic}
\eea
and associate with them two inner products defined by 
\bea
\langle  T| S \rangle_\pm : = {T}^{\rm sT}{\mathbb J}_{\pm} {S}
~,
\qquad { T}^{\rm sT} := \big( T_{\a} , - (-1)^{\ve(T)}  T_{I} \big)~,
\eea
for arbitrary pure supertwistors $T$ and $S$. Here 
  ${ T}^{\rm sT} $ denotes the super-transpose of $T$.
These inner products are characterised by the symmetry property
\bea
\langle { T}_1 | { T}_2  \rangle_\pm
= -(-1)^{\ve(T_1)  \ve (T_2)} \langle { T}_2 | { T}_1  \rangle_\pm ~,
\eea
for arbitrary pure supertwistors $T_1$ and $T_2$. 
If $T_1$ and $T_2$ are real supertwistors, their inner products
obey the reality conditions
\bea
\big( \langle { T}_1 | { T}_2  \rangle_\pm \big)^* 
= - \langle { T}_2 | { T}_1  \rangle_\pm~.
\eea

By definition, the supergroup  $\sOSp_+(n|2; {\mathbb R})$
consists of those even $(2|n) \times (2|n)$ supermatrices
\bea
g = (g_{A}{}^{B}) ~, \qquad \e(g_A{}^B) = \e_{A} + \e_{B} ~,
\eea
which are characterised by the properties
\begin{subequations}\label{groupcond}
\bea
g^{\rm sT} {\mathbb J}_+ g &=& {\mathbb J}_+ ~, \qquad
(g^{\rm sT})^{A}{}_{B} := (-1)^{\e_{A} \e_{B} + \e_{B}} g_{B}{}^{A}~, \\
g^\dagger {\mathbb J}_+ g &=& {\mathbb J}_+ ~.
\eea
\end{subequations}
Every group element $g \in \sOSp_+(n|2; {\mathbb R})$ takes every real even (odd)
supertwistor to a real even (odd) supertwistor, 
\bea
T =(T_A) ~\to ~ g\cdot T = (g_{A}{}^{B} T_B)~,
\eea
such that the inner product $\langle  T| S \rangle_+$ is preserved. 
The  supergroup  $\sOSp_-(n|2; {\mathbb R})$ is defined similarly by replacing 
${\mathbb J}_+  \to {\mathbb J}_-$.

The realisations $\sOSp_+(n|2; {\mathbb R})$ and $\sOSp_-(n|2; {\mathbb R})$ are equivalent. They can be related to each other by performing a transformation of the supertwistor space
\bea
T_A ~\to ~\tilde{T}_A = \left(
\begin{array}{c}
\s_\a{}^\b T_{\b} \\ \hline
 T_{I}
\end{array}
\right)~, \qquad \s =(\s_\a{}^\b) := \cos \vf\, \s_1 + \sin \vf \,\s_3~,
\label{equiv2.11}
\eea
for some  angle $\vf \in {\mathbb R}$ and the Pauli matrices $\s_1$ and $\s_3$.
Then it is easy to see that 
\bea
\langle  T| S \rangle_- = - \langle  \tilde{T}| \tilde{S} \rangle_+~.
\label{equiv2.12}
\eea


\subsection{Supertwistor description of ${\rm AdS}^{(3|p,q)}$}

 In this paper we choose the isometry group of ${\rm AdS}^{(3|p,q)} $ to be
 one of the 
 following two
 supergroups:
\begin{subequations}\label{isometry1+2}
\bea
(a) \qquad G_\pm&=& {\sOSp}_+ (p|2; {\mathbb R} ) \times  {\sOSp}_- (q|2; {\mathbb R} )
\equiv G_{\rm L}^+ \times G_{\rm R}^-~,
\label{isometry} \\
(b) \qquad G_\mp&=& {\sOSp}_- (p|2; {\mathbb R} ) \times  {\sOSp}_+ (q|2; {\mathbb R} )
\equiv G_{\rm L}^- \times G_{\rm R}^+~,
\label{isometry2}
\eea
\end{subequations}
in agreement with \cite{AT1,AT2}. 
This differs from the supergroup, which was  used in \cite{KT-M21}:
$G_\pp= {\sOSp}_+ (p|2; {\mathbb R} ) \times  {\sOSp}_+ (q|2; {\mathbb R} )$. 
The main reason for the new choice, say,  \eqref{isometry} is that 
the vector covariant derivative appears  with the same sign in the anti-commutators of spinor covariant derivatives, eqs.  \eqref{4.30e} and \eqref{4.30f}.
More comments on this difference will be given in due course. 

In what follows, we will mostly work with the supergroup \eqref{isometry}, and then explain what changes occur when the supergroup \eqref{isometry2} is chosen instead.  

We will use the notation 
\bea
T_{\rm L} = (T_{\Ao}) =\left(
\begin{array}{c}
T_{\alo} \\
 T_{\Io}
\end{array}
\right)~, 
 \qquad \alo  = 1,2 ~, \quad \Io = 1, \dots, p ~,
\eea
for the supertwistors associated with the subgroup $G_{\rm L}$ in \eqref{isometry}, 
while the right supertwistors will be denoted as
\bea
T_{\rm R} = (T_{\Au}) =\left(
\begin{array}{c}
T_\alu \\
 T_{\Iu}
\end{array}
\right)~, 
 \qquad \alu  = 1,2 ~, \quad \Iu = 1, \dots, q ~.
\eea
In the case of the supergroups in \eqref{isometry}, the symplectic supermatrices \eqref{symplectic} will be denoted 
\bea
{\mathbb J}_{\rm L} = ({\mathbb J}^{\Ao\, \Bo}) = \left(
\begin{array}{c ||c}
\ve_{\rm L}  ~&~ 0 \\\hline \hline
0 ~& ~{\rm i} \,{\mathbbm 1}_p
\end{array} \right) ~, \qquad
\ve_{\rm L}
=\big(\ve^{\alo  \beo} \big)
=\left(
\begin{array}{cc}
0  & { 1}\\
 -{ 1}  &    0
\end{array}
\right) ~,
\eea
and similarly for ${\mathbb J}_{\rm R}$.

Following \cite{KT-M21}, we identify the $(p,q)$ AdS superspace with
the quotient space 
\bea
{\rm AdS}^{(3|p,q)} = {\mathfrak L}_{(p,q)}/ \sim~.
\label{quotient}
\eea
Here the space $ {\mathfrak L}_{(p,q)}$
  consists of all pairs 
$(\cP_{\rm L}, \cP_{\rm R})$, where 
\begin{subequations}
\bea
\cP_{\rm L} &=& (X_{\Ao}{}^\m ) ~, \qquad \m =1,2 ~,
\eea
is a left real even two-plane, and 
\bea
\cP_{\rm R} &=& (Y_{\Au}{}^\m ) ~, \qquad \m =1,2~,
\eea
\end{subequations}
is a  right real even two-plane, with the additional property 
\bea
\cP_{\rm L}^{\rm sT} {\mathbb J}_{\rm L} \cP_{\rm L} 
= \cP_{\rm R}^{\rm sT} {\mathbb J}_{\rm R} \cP_{\rm R}~.
\label{two-plane-equation}
\eea
When saying that  $\cP_{\rm L}$ is even real, we mean that the two supertwistors $X_{\rm L}^\m$ are even and real. The property of $\cP_{\rm L}$ being a two-plane means that
\bea
\det ( X_{\alo}{}^\m ) \neq 0~.
\eea 
Similar statements hold for the right planes.
In the space ${\mathfrak L}_{(p,q)}$ we introduce the following equivalence relation 
\bea
(\cP_{\rm L}, \cP_{\rm R}) \sim (\cP_{\rm L} M, \cP_{\rm R}M)~, \qquad 
M \in \sGL( 2, {\mathbb R})~.
\label{equivalence}
\eea

The supergroup \eqref{isometry}  acts on ${\mathfrak L}_{(p,q)}$ by the rule
\bea
(g_{\rm L}, g_{\rm R}) (\cP_{\rm L}, \cP_{\rm R}) := 
(g_{\rm L} \cP_{\rm L}, g_{\rm R} \cP_{\rm R}) ~,\quad (g_{\rm L}, g_{\rm R}) \in 
{\sOSp}_+ (p|2; {\mathbb R} ) \times  {\sOSp}_- (q|2; {\mathbb R} )~.
\eea
This action is naturally extended to the quotient space \eqref{quotient}.
The latter proves to be a homogeneous space for
$G_\pm ={\sOSp}_+ (p|2; {\mathbb R} ) \times  {\sOSp}_- (q|2; {\mathbb R} )$. 

As pointed out earlier, the two supergroup realisations $\sOSp_+(n|2; {\mathbb R})$ and $\sOSp_-(n|2; {\mathbb R})$ are equivalent. We remind the reader that the equivalence may be obtained by applying the transformation \eqref{equiv2.11}. 
However, if such a transformation is applied to $\sOSp_-(n|2; {\mathbb R})$ 
in order to convert $G_\pm ={\sOSp}_+ (p|2; {\mathbb R} ) \times  {\sOSp}_- (q|2; {\mathbb R} )$ into $G_\pp ={\sOSp}_+ (p|2; {\mathbb R} ) \times  {\sOSp}_+ (q|2; {\mathbb R} )$, then the AdS relation \eqref{two-plane-equation} will become
\bea
\cP_{\rm L}^{\rm sT} {\mathbb J}_{\rm L} \cP_{\rm L} 
=- \tilde \cP_{\rm R}^{\rm sT} {\mathbb J}_{\rm L} \tilde \cP_{\rm R}~,
\eea
due to \eqref{equiv2.12}. Thus, the equivalence is not extended to the AdS superspaces.
These conclusions are parallel to those given in the literature for $(p,q)$ AdS supergravities as Chern-Simons theories \cite{AT1,AT2}.
It is clear that the supergroups $G_\pm$ and $G_\mp$ lead to equivalent descriptions 
of  the $(p,q)$ AdS superspaces, while the supergroups $G_\pp$ and $G_=$ provide equivalent descriptions of ``exotic'' AdS superspaces (adopting the terminology of  \cite{Witten}).


\section{Bi-supertwistor construction} \label{Section3}

Let $\cP_{\rm L} = ( X_{\Ao}^{\phantom \m \m})$ and $\cP_{\rm R} = ( Y_{\Au}^{\phantom \m \m})$ be left and right even two-planes constrained by 
\eqref{two-plane-equation}.
We adopt the notation that
\bea
\Exp{\cP | \cP} &=& \Exp{ \cP_{\rm L} | \cP_{\rm L} } = \Exp{ X^{\m} | X^{\n} }_{\rm L} \ve_{\m\n}~, \\
&=&  \Exp{ \cP_{\rm R} | \cP_{\rm R} } = \Exp{ Y^{\m} | Y^{\n} }_{\rm R} \ve_{\m\n}~,
\eea
and then introduce the bi-supertwistors
\begin{subequations} \label{bisupertwistors}
\bea 
\mathbb{Z}_{\Ao \Bu} &=& -2 \frac{X_{\Ao}^{\phantom \m \m} Y_{\Bu}^{\phantom \n \n} \ve_{\m\n}}{ \Exp{\cP | \cP}}~, \\
\mathbb{Z}_{\Au \Bo} &=& -2 \frac{Y_{\Au}^{\phantom \m \m} X_{\Bo}^{\phantom \n \n} \ve_{\m\n}}{ \Exp{\cP | \cP}}~, \\
\mathbb{X}_{\Ao \Bo} &=& -2 \frac{X_{\Ao}^{\phantom \m \m} X_{\Bo}^{\phantom \n \n} \ve_{\m\n}}{ \Exp{\cP | \cP}}~, \\
\mathbb{Y}_{\Au \Bu} &=& -2 \frac{Y_{\Au}^{\phantom \m \m} Y_{\Bu}^{\phantom \n \n} \ve_{\m\n}}{ \Exp{\cP | \cP}}~.
\eea \end{subequations}
Note that these bi-supertwistors are invariant under equivalence transformations
\bea \label{supertwistorgauge}
( \cP_{\rm L}, \cP_{\rm R}) \rightarrow ( \cP_{\rm L}^{\prime}, \cP_{\rm R}^{\prime}) = ( \cP_{\rm L} M, \cP_{\rm R} M)~, \quad M \in \sGL(2, \mathbb{R})~,
\eea
and so can be used to parameterise ${\rm AdS}^{(3|p,q)}$. The bi-supertwistors \eqref{bisupertwistors} satisfy a plethora of algebraic properties. They are graded antisymmetric supermatrices
\begin{subequations}
\bea
\mathbb{Z}_{\Ao \Bu} &=& - (-1)^{\e_{\Ao} \ve_{\Bu} } \mathbb{Z}_{\Bu \Ao}~, \\
\mathbb{X}_{\Ao \Bo} &=& - (-1)^{\e_{\Ao} \ve_{\Bo} } \mathbb{X}_{\Bo \Ao}~, \\
\mathbb{Y}_{\Au \Bu} &=& - (-1)^{\e_{\Au} \ve_{\Bu} } \mathbb{Y}_{\Bu \Au}~,
\eea
\end{subequations}
and under the graded anti-symmetrisation of indices satisfy
\begin{subequations}
\bea
\mathbb{X}_{ [ \Ao \Bo} \mathbb{Z}_{\overline{C} \} \underline{D}} &=& 0~, \\
\mathbb{X}_{ [ \Ao \Bo} \mathbb{X}_{\overline{C} \} \overline{D}} &=& 0~, \\
\mathbb{Y}_{ [ \Au \Bu} \mathbb{Z}_{\underline{C} \} \overline{D}} &=& 0~, \\
\mathbb{Y}_{ [ \Au \Bu} \mathbb{Y}_{\underline{C} \} \underline{D}} &=& 0~.
\eea
\end{subequations}
Additionally, these bi-supertwistors satisfy
\begin{subequations}
\bea
\mathbb{J}^{\Ao \Bo} \mathbb{X}_{\Bo \Ao} &=& 2~, \\
\mathbb{J}^{\Au \Bu} \mathbb{Y}_{\Bu \Au} &=& 2~, \\
(-1)^{\e_{\Bu}} \mathbb{Z}_{\Ao \Bu} \mathbb{J}^{ \Bu \underline{C} } \mathbb{Z}_{\underline{C} \overline{D} } &=& \mathbb{X}_{\overline{A} \overline{D} }~, \\
(-1)^{\e_{\Bo}} \mathbb{Z}_{\Au \Bo} \mathbb{J}^{\Bo \overline{C}} \mathbb{Z}_{\overline{C} \underline{D} } &=& \mathbb{Y}_{\underline{A} \underline{D}}~.
\eea
\end{subequations}
These relations define the superembedding of ${\rm AdS}^{(3|p,q)}$.

In the non-supersymmetric case, $p=q=0$, $\mathbb{X}_{ \Ao \Bo} $ and
$\mathbb{Y}_{\Au \Bu} $ are constant matrices, 
\bea
p=q=0: \qquad {\mathbb X}_{\alo  \beo} = \ve_{\alo  \beo} ~, \qquad 
{\mathbb Y}_{\alu \beu} = \ve_{\alu \beu}~.
\eea 
Modulo these constant sectors, $\mathbb{X}_{ \Ao \Bo} $ and
$\mathbb{Y}_{\Au \Bu} $ are purely fermionic in the supersymmetric case, when at least one of $p$ and $q$ is non-zero.
The explicit expressions for $\mathbb{Z}_{\Ao \Bu} $, $\mathbb{X}_{ \Ao \Bo} $ and
$\mathbb{Y}_{\Au \Bu} $
are given in section \ref{section6.4} in the Poincar\'e coordinate patch. 

Let us now introduce bi-supertwistors with a raised index. Considering $\mathbb{X}_{\Ao \Bo}$ we define
\bea
\mathbb{X}_{\Ao}^{\phantom C \overline{C}} = (-1)^{\e_{\Bo}} \mathbb{X}_{\Ao \Bo} \mathbb{J}^{\Bo \overline{C}}~,
\eea
and analogously for the other bi-supertwistors. Under a group transformation we then have
\bea
\mathbb{X}_{\Ao}^{\phantom C \overline{B}} \rightarrow g_{\Ao}^{\phantom C \overline{C}} \mathbb{X}_{\overline{C}}^{\phantom D \overline{D}} (g^{-1})_{\overline{D}}^{\phantom D \overline{B}}~.
\eea
We may associate with any supermatrix $\mathbb{X} = (\mathbb{X}_{\Ao}^{\phantom B \Bo})$ its supertrace defined by
\bea
\mathrm{ str}\, \mathbb{X} &= (-1)^{\e_{\Ao}} \mathbb{X}_{\Ao}^{\phantom \Ao \Ao}~.
\eea
The supertrace of a supermatrix is invariant under group transformations
\bea
\mathrm{str}\, \mathbb{X}^{\prime} = \mathrm{str} \mathbb{X}~.
\label{supertrace}
\eea

Given two arbitrary points in  ${\rm AdS}^{(3|p,q)}$,  
we can construct the following two-point functions
\begin{subequations} \label{twopoint}
\bea
\mathrm{str} (\tilde{\mathbb{Z}} \mathbb{Z} ) &=& (-1)^{\e_{\Ao}} \tilde{\mathbb{Z}}_{\Ao}^{\phantom \Bu \Bu} \mathbb{Z}_{\Bu}^{\phantom \Au \Ao}  ~, \\
\mathrm{str} (\tilde{\mathbb{X}} \mathbb{X} ) &=& (-1)^{\e_{\Ao}} \tilde{\mathbb{X}}_{\Ao}^{\phantom \Bu \Bo} \mathbb{X}_{\Bo}^{\phantom \Au \Ao}  ~, \\
\mathrm{str} (\tilde{\mathbb{Y}} \mathbb{Y} ) &=& (-1)^{\e_{\Au}} \tilde{\mathbb{Y}}_{\Au}^{\phantom \Bu \Bu} \mathbb{Y}_{\Bu}^{\phantom \Au \Au} ~,
\eea
\end{subequations}
which are invariant under arbitrary ${\sOSp}_+ (p|2; {\mathbb R} ) \times  {\sOSp}_- (q|2; {\mathbb R} )$ transformations, in accordance with \eqref{supertrace}.


\section{Coset construction} \label{Section4}

Given a homogeneous space $\mathfrak X$ for a group $G$, it can always be realised as a coset space $G/H_o$, where $H_o$ is the stabiliser of some marked point $o \in \mathfrak X$. 
In this section we develop a coset construction for 
${\rm AdS}^{(3|p,q)}$, which is a homogeneous space for the supergroup \eqref{isometry}.

As a marked/preferred point $Z^{(0)} = (\cP_{\rm L}^{(0)}, \cP_{\rm R}^{(0)})$ 
of ${\rm AdS}^{(3|p,q)}$, we choose
\bea
\cP_{\rm L}^{(0)} =\left(
\begin{array}{c}
{\mathbbm 1}_2 \\ \hline
0
\end{array}
\right)~, \qquad 
\cP_{\rm R}^{(0)} =\left(
\begin{array}{c}
{\mathbbm 1}_2 \\ \hline
0
\end{array}
\right)~.
\eea
The stabiliser $H$ of $Z^{(0)} $ consists of those elements $h= (h_{\rm L} , h_{\rm R} ) $ of the AdS supergroup $ {\sOSp}_{+} (p|2; {\mathbb R} ) \times  {\sOSp}_{-} (q|2; {\mathbb R} )$,
\bea
h_{\rm L}=\left(
\begin{array}{c|c}
 A _{\rm L} & B_{\rm L}\\
 \hline 
C_{\rm L} &    D_{\rm L} 
\end{array}
\right) \in {\sOSp}_{+} (p|2; {\mathbb R} ) ~,\qquad 
h_{\rm R}=\left(
\begin{array}{c|c}
 A _{\rm R} & B_{\rm R}\\
 \hline 
C_{\rm R} &    D_{\rm R} 
\end{array}
\right) \in {\sOSp}_{-} (q|2; {\mathbb R} ) ~,
\eea
which satisfy the conditions 
\bea
h_{\rm L} \cP_{\rm L}^{(0)} =\left(
\begin{array}{c}
M \\ \hline
0
\end{array}
\right)~, \qquad 
h_{\rm R} \cP_{\rm R}^{(0)} =\left(
\begin{array}{c}
M \\ \hline
0
\end{array}
\right)~,
\eea
for some $M \in \sGL( 2, {\mathbb R})$. These conditions imply that 
\begin{subequations}\label{IsotropySubroup}
\bea
h_{\rm L}=\left(
\begin{array}{c|c}
 N & 0\\
 \hline 
 0 &    R_{\rm L} 
\end{array}
\right) 
~,\qquad 
h_{\rm R}=\left(
\begin{array}{c|c}
 N & 0\\
 \hline 
0 &    R_{\rm R} 
\end{array}
\right)~, 
\eea
where 
\bea
 N\in \sSL (2, {\mathbb R}) ~,\quad
R_{\rm L}  \in {\sSO} (p) ~, \quad R_{\rm R}  \in {\sSO} (q)~.
\eea
\end{subequations}
Thus the stability subgroup $H$ is isomorphic to
\bea
 \sSL (2, {\mathbb R}) \times
 {\sSO} (p) \times  {\sSO} (q)~.
 \eea

In what follows,  it is useful to work with normalised two-planes 
\bea
\cP_{\rm L}^{\rm sT} {\mathbb J}_{\rm L} \cP_{\rm L} 
= \cP_{\rm R}^{\rm sT} {\mathbb J}_{\rm R} \cP_{\rm R} = \ve~.
\eea
The normalisation condition is achieved by performing an equivalence 
transformation \eqref{equivalence}. This condition means the following:
\begin{subequations} \label{normalised}
\bea
\cP_{\rm L} &=&\left(
\begin{array}{c}
{\bm x} \\ \hline
\ri \q_{\rm L}
\end{array}
\right)
=\left(
\begin{array}{c}
{\bm x}_\alo{}^\m \\
\ri  \q_{\Io}{}^\m 
\end{array}
\right)
~, \qquad \det {\bm x} = 1 + \frac{\ri}{2} {\rm tr} \big(\q_{\rm L} \ve^{-1} \q_{\rm L}{}^{\rm T} \big) ~; 
\label{normalised.a} \\
\cP_{\rm R} &=&\left(
\begin{array}{c}
{\bm y} \\ \hline
\ri \q_{\rm R}
\end{array}
\right) 
=\left(
\begin{array}{c}
{\bm y}_\alu{}^\m \\
 \ri \theta_{\Iu}{}^\m 
\end{array}
\right)
~, \qquad \det {\bm y} = 1 - \frac{\ri}{2} {\rm tr} \big(\q_{\rm R} \ve^{-1} \q_{\rm R}{}^{\rm T} \big) ~.
\label{normalised.b}
\eea
\end{subequations} 
Then the equivalence relation becomes
\bea
(\cP_{\rm L}, \cP_{\rm R}) \sim (\cP_{\rm L} N, \cP_{\rm R}N)~, \qquad 
N \in \sSL( 2, {\mathbb R})~.
\label{ER2}
\eea
It is useful to represent the matrices $\bm x$ and $\bm y$ in the form 
\begin{subequations}\label{unim}
\bea
{\bm x} &=& x \sqrt{ 1 + \frac{\ri}{2} {\rm tr} \big(\q_{\rm L} \ve^{-1} \q_{\rm L}{}^{\rm T}}\big)\equiv x\, \l_{\rm L} (\q_{\rm L} )
~,  \quad x \in \sSL(2,{\mathbb R})~;
\label{unim.a}\\
{\bm y} &=& y \sqrt{ 1 - \frac{\ri}{2} {\rm tr} \big(\q_{\rm R} \ve^{-1} \q_{\rm R}{}^{\rm T} \big) } \equiv y \,\l_{\rm R}(\q_{\rm R})~, \quad y \in \sSL(2,{\mathbb R})
~. \label{unim.b}
\eea
\end{subequations}
Here $x$ and $y$ are purely bosonic unimodular matrices. 

We now turn to constructing a global cross section (or, equivalently, a coset representative). 
In general, given a homogeneous space ${\mathfrak X} = G/H_o$ for a group $G$,
a global cross section $\mathfrak S$ is a map  
\bea
{\mathfrak S}: G/H_o \to G \quad \text{such that} \quad \p \circ {\mathfrak S} = {\rm id}~,
\eea
where $\p : G \to G/H_o$ is the natural projection.\footnote{For many homogeneous spaces, a global cross section does not exist, only local cross sections can always be defined.
For example, no global cross section exists in the case of the homogeneous space 
$S^2 = \sSO(3) /\sSO(2) $ for  $\sSO(3)$.}
If a global coset representative exists, it encodes the differential geometry of the homogeneous space $\mathfrak X$.

Associated with the normalised two-planes \eqref{normalised} are the following group 
elements:
\begin{subequations}\label{CosRep}
\bea
S_{\rm L}(X) 
&=&\left(
\begin{array}{c|c}
{\bm  x} & -\ve_{\rm L}^{-1} ({\bm x}^{-1})^{\rm T}\q_{\rm L}{}^{\rm T} U_{\rm L}  \\
 \hline 
\ri \q_{\rm L} &    U_{\rm L} 
\end{array}
\right)~, \quad 
U_{\rm L} (\q_{\rm L}) := 
\Big( {\mathbbm 1}_p +\ri \frac{ \q_{\rm L} \ve^{-1} \q_{\rm L}{}^{\rm T} }{\det {\bm x}} \Big)^{-\hf} ~;~~~
\\
S_{\rm R}(Y) 
&=&\left(
\begin{array}{c|c}
{\bm  y} & \ve_{\rm R}^{-1} ({\bm y}^{-1})^{\rm T}\q_{\rm R}{}^{\rm T} U_{\rm R}  \\
 \hline 
\ri \q_{\rm R} &    U_{\rm R} 
\end{array}
\right)~, \quad 
U_{\rm R} (\q_{\rm R} ):= 
\Big( {\mathbbm 1}_q -\ri \frac{ \q_{\rm R} \ve^{-1} \q_{\rm R}{}^{\rm T} }{\det {\bm y}} \Big)^{-\hf} ~.~~~
\eea
\end{subequations}
We point out that the matrices $U_{\rm L} $ and $U_{\rm R} $ are symmetric, 
$U_{\rm L}{}^{\rm T} = U_{\rm L} $ and $U_{\rm R}{}^{\rm T} = U_{\rm R} $.
It is easy to  check the following identities:
\bea
 U_{\rm L} \q_{\rm L} = \l_{\rm L}  \q_{\rm L} ~, \quad
 U_{\rm R} \q_{\rm R} = \l_{\rm R} \q_{\rm R}~. 
 \label{312}
\eea
The important properties of $S_{\rm L}(X) $ and $S_{\rm R}(Y) $ are
\begin{subequations} \label{2.8}
\bea
S_{\rm L}(X N ) &=& S_{\rm L}(X) \cN_{\rm L} ~, \quad 
 \cN_{\rm L} =
 \left(
\begin{array}{c|c}
 N & 0\\
 \hline 
 0 &   {\mathbbm 1}_p  
\end{array}
\right) 
~,
 \\
S_{\rm R}(Y N ) &=& S_{\rm R}(Y) \cN_{\rm R} ~, \quad 
 \cN_{\rm R} =
 \left(
\begin{array}{c|c}
 N & 0\\
 \hline 
 0 &   {\mathbbm 1}_q  
\end{array}
\right) 
~, 
\eea  
\end{subequations}
with $N\in \sSL (2, {\mathbb R}) $.
In addition, we have the properties 
\bea
S_{\rm L}(X)\cP^{(0)}_{\rm L} = \cP_{\rm L} ~,\qquad 
S_{\rm R}(Y)\cP^{(0)}_{\rm R} = \cP_{\rm R}~. 
\eea

The freedom \eqref{ER2} may be fixed, e.g., by choosing
\bea
{\bm y} = \l_{\rm R} (\q_{\rm R} ) {\mathbbm 1}_2 ~, 
\label{gauge-fixedY}
\eea
where $\l_{\rm R} (\q_{\rm R} ) $ is given by \eqref{unim.b}.
Then the expressions \eqref{CosRep} define a global coset representative for ${\rm AdS}^{(3|p,q)}$.

Given a group element $g= (g_{\rm L} , g_{\rm R} ) \in {\sOSp}_{+} (p|2; {\mathbb R} ) \times  {\sOSp}_{-} (q|2; {\mathbb R} )$, it can be uniquely represented in the form 
\bea 
(g_{\rm L} , g_{\rm R} ) = (S_{\rm L} (X) h_{\rm L} , S_{\rm R}(Y) h_{\rm R} ) ~,
\label{group_decomposition}
\eea
where $h= (h_{\rm L} , h_{\rm R} )  $ belongs to the isotropy subgroup \eqref{IsotropySubroup}, and $Y$ is constrained to have the form \eqref{gauge-fixedY}.
However, if $X$ and $Y$ are only required to be normalised, as in eq. \eqref{normalised}, then the decomposition \eqref{group_decomposition} 
is not unique, and the available freedom is described by  
\bea 
(g_{\rm L} , g_{\rm R} ) = \big(S_{\rm L} (XN) \cN_{\rm L}{}^{-1} h_{\rm L} , S_{\rm R}(YN) \cN_{\rm R}{}^{-1} h_{\rm R} \big) ~,\qquad N\in \sSL (2, {\mathbb R}) ~,
\label{group_decomposition2}
\eea
where $ \cN_{\rm L}$ and $ \cN_{\rm R}$ are given in \eqref{2.8}.
 

\section{Torsion and curvature tensors} \label{Section5}

In this section we give explicit expressions for the vielbein, connection, torsion and curvature tensors. 

\subsection{Geometric objects of ${\rm AdS}^{(3|p,q)}$}

Let us denote by $\cG$ the superalgebra 
of the AdS supergroup \eqref{isometry},
and 
by $\cH$ the algebra 
 of the stability group \eqref{IsotropySubroup}.
Let $\cW$ be
a
complement of $\cH$ in $\cG$, $\cG = \cH \oplus \cW$. 
With the freedom \eqref{ER2}   fixed, 
we define $\cW$ to consist of 
elements $X= (X_{\rm L}, X_{\rm R}) $ 
of the form 
\bea \label{algcompform}
X_{\rm L}
=\left(
\begin{array}{c|c}
 A_{\rm L} & - \ve^{-1} B_{\rm L}^{\rm T} \\
 \hline 
\ri B_{\rm L} &    0
\end{array}
\right)~, \quad 
X_{\rm R}
=\left(
\begin{array}{c|c}
 0 &  \ve^{-1} B_{\rm R}^{\rm T} \\
 \hline 
\ri B_{\rm R} &    0
\end{array}
\right)~, \quad A_{\rm L} \in \mathfrak{sl} (2,\mathbb{R})~.
\eea
The elements $Y= (Y_{\rm L}, Y_{\rm R}) \in \cH$ take the form
\bea \label{algform}
Y_{\rm L}
=\left(
\begin{array}{c|c}
 n & 0 \\
 \hline 
0 &    r_{\rm L}
\end{array}
\right)~, \quad 
Y_{\rm R}
=\left(
\begin{array}{c|c}
 n & 0 \\
 \hline 
0 &   r_{\rm R}
\end{array}
\right)~, \quad
n  \in \mathfrak{sl} (2,\mathbb{R})~, \quad r_{\rm L} \in  \mathfrak{so} (p)~, \quad
 r_{\rm R} \in  \mathfrak{so} (q)
~.~~
\eea
It is straightforward to verify that $[ \cW, \cH] \subset \cW$. We may uniquely decompose the Maurer-Cartan one-form $\o = S^{-1}\rd S$ as a sum 
$\o = E+ \O$, where $E = S^{-1}\rd S|_{\cW}$ is the vielbein taking its values in $\cW$, 
and $\O = S^{-1}\rd S|_{\cH}$ is the connection taking its values in $\cH$.
The Maurer-Cartan one-form is
\begin{subequations}
\bea
\o_{\rm L} 
&=& \left(
\begin{array}{c|c}
 \l_{\rm L}^2
 {\bm x}^{-1} {\rm d} {\bm x} + \ri \ve^{-1} \te_{\rm L}^{\rm T} {\rm d} \te_{\rm L} & 
 ~ - \ve^{-1} [ {\rm d} \te_{\rm L}^{\rm T}  - {\rm d} {\bm x}^{\rm T} ({\bm x}^{-1})^{\rm T} \te_{\rm L}^{\rm T} ] U_{\rm L}   \\
 \hline 
\ri U_{\rm L} \big[ {\rm d} \te_{\rm L} -  \te_{\rm L} {\bm x}^{-1} {\rm d} {\bm x} \big] & \begin{aligned} &
~ \ri U_{\rm L} \te_{\rm L} {\bm x}^{-1} \ve^{-1} 
{\rm d}
\big[ ({\bm x}^{-1})^{\rm T} \te_{\rm L}^{\rm T} \big]
U_{\rm L} 
 + U_{\rm L}^{-1} {\rm d} U_{\rm L} \end{aligned}
\end{array}
\right)~, \\
\o_{\rm R} 
&=& \left(
\begin{array}{c|c}
  \hf {\rm d}\l_{\rm R}^2
  \mathbbm{1} - \ri \ve^{-1} \te_{\rm R}^{\rm T} {\rm d} \te_{\rm R} & 
  \ve^{-1} \big[ {\rm d} \te_{\rm R}^{\rm T}  - \l_{\rm R}^{-1}{\rm d} \l_{\rm R}  \te_{\rm R}^{\rm T} \big] U_{\rm R}  \\
 \hline 
\ri  U_{\rm R} \big[{\rm d} \te_{\rm R} - \l_{\rm R}^{-1} {\rm d} \l_{\rm R}  \te_{\rm R}  \big] & 
\begin{aligned} &~ - \ri \l_{\rm R}^{-1} U_{\rm R} \te_{\rm R} \ve^{-1} {\rm d}
\big[
\l_{\rm R}^{-1} \te_{\rm R}^{\rm T} \big] U_{\rm R}
+U_{\rm R}^{-1} {\rm d} U_{\rm R} \end{aligned}
\end{array}
\right)~,
\eea
\end{subequations}
which we decompose in to matrices with the forms \eqref{algcompform} and \eqref{algform} to obtain the vielbein
\begin{subequations}\label{vielbein}
\bea 
E_{\rm L} 
= \left(
\begin{array}{c|c}
\tilde{E} & - \ve^{-1} \cE_{\rm L}^{\rm T} \\
 \hline 
\ri \cE_{\rm L}  & 0
\end{array}
\right)~, \quad 
E_{\rm R} 
&=& \left(
\begin{array}{c|c}
0 &  \ve^{-1} \cE_{\rm R}^{\rm T} \\
 \hline 
\ri \cE_{\rm R}  & 0
\end{array}
\right)~,
\eea
where
\bea
\tilde{E} &=& 
\l^2_{\rm L}
{\bm x}^{-1} {\rm d} {\bm x} + \ri \ve^{-1} \te_{\rm L}^{\rm T} {\rm d} \te_{\rm L} 
- \hf  {\rm d}\l_{\rm R}^2 \mathbbm{1} 
+ \ri \ve^{-1} \te_{\rm R}^{\rm T} {\rm d} \te_{\rm R}~, \\
\cE_{\rm L} &=& 
U_{\rm L} \big[  {\rm d} \te_{\rm L} -\te_{\rm L} {\bm x}^{-1} {\rm d} {\bm x} \big]
~, \\
\cE_{\rm R} &=& 
U_{\rm R}\big[ {\rm d} \te_{\rm R} - \l_{\rm R}^{-1} {\rm d} \l_{\rm R}  \te_{\rm R} \big]
~,
\eea
\end{subequations}
and the connection
\begin{subequations}
\bea
\O_{\rm L} 
= \left(
\begin{array}{c|c}
\tilde{\O} & 0 \\
 \hline 
0  & \O_{\sSO(p)}
\end{array}
\right)~, \quad 
\O_{\rm R} 
&=& \left(
\begin{array}{c|c}
\tilde{\O} & 0 \\
 \hline 
0 & \O_{\sSO(q)}
\end{array}
\right)~,
\eea
where
\bea 
\tilde{\O} &=& \hf {\rm d}\l_{\rm R}^2 \mathbbm{1} - \ri \ve^{-1} \te_{\rm R}^{\rm T} {\rm d} \te_{\rm R}~, \\
\O_{\sSO(p)} &=& \ri U_{\rm L} \te_{\rm L} {\bm x}^{-1} \ve^{-1} 
\big[
{\rm d} ({\bm x}^{-1})^{\rm T} \te_{\rm L}^{\rm T} 
+ ({\bm x}^{-1})^{\rm T} {\rm d} \te_{\rm L}^{\rm T}  \big] U_{\rm L} 
+ U_{\rm L}^{-1} {\rm d} U_{\rm L}~, \\
\O_{\sSO(q)} &=& 
\ri \l_{\rm R}^{-2} U_{\rm R} \te_{\rm R} \ve^{-1} 
\big[
\l_{\rm R}^{-1}{\rm d} \l_{\rm R} \te_{\rm R}^{\rm T} 
-{\rm d} \te_{\rm R}^{\rm T} \big] U_{\rm R}
+ U_{\rm R}^{-1} {\rm d} U_{\rm R}~.
\eea
\end{subequations}
The expressions for the connections may be simplified using various  identities such as \eqref{312}.
However, the above expressions appear most convenient to prove the required properties
of the connections
\bea
\tr\, \tilde{\O} =0~, \quad \O^{\rm T}_{\sSO(p)} = -\O_{\sSO(p)} , \quad \O^{\rm T}_{\sSO(q)} = -\O_{\sSO(q)} ~.
\eea

We now turn to computing the torsion $\cT$ and curvature $\cR$ tensors. 
They 
are defined as follows
\bea \label{torandcur}
-\cT = {\rm d} E - E \wedge \O - \O \wedge E~, \quad \cR = {\rm d} \O - \O \wedge \O~,
\eea
and transform covariantly, 
\begin{subequations}
\bea
\cT' = h \cT h^{-1}~, \quad  \cR' = h \cR h^{-1}~,
\eea
under {\it local} $H$-transformations 
\bea
E' = h Eh^{-1}~, \quad  \O' = h \O h^{-1} - \rd h  h^{-1}~,
\eea
with $h \in H$. 
\end{subequations}

For the torsion tensor we obtain
\begin{subequations}
\bea
\cT_{\rm L} = \left(
\begin{array}{c|c}
\cT_{1} & - \ve^{-1} \cT_{2}^{\rm T} \\
 \hline 
\ri \cT_{2} & 0
\end{array}
\right)~, \quad 
\cT_{\rm R} 
&=& 0~,
\eea
where
\bea
\cT_{1} &=&  {\rm d} [ \l_{\rm L}^2
 {\bm x}^{-1}] {\rm d} {\bm x} + \ri \ve^{-1} {\rm d} \te_{\rm L}^{\rm T} {\rm d} \te_{\rm L} 
+ \ri \ve^{-1} {\rm d} \te_{\rm R}^{\rm T} {\rm d} \te_{R} 
- \ri 
\l_{\rm L}^2
\big\{ {\bm x}^{-1} {\rm d} {\bm x} \,, \,\ve^{-1} \te_{\rm R}^{\rm T} {\rm d} \te_{\rm R} \big\}
 \nonumber \\
&&
+ \big\{ \ve^{-1} \te_{\rm R}^{\rm T} {\rm d} \te_{\rm R} \,,\, \ve^{-1} \te_{\rm L}^{\rm T} {\rm d} \te_{\rm L} \big\}
+2 \ve^{-1} \te_{\rm R}^{\rm T} {\rm d} \te_{\rm R} \ve^{-1} \te_{\rm R}^{\rm T} {\rm d} \te_{\rm R}~, \\
\cT_{2} &=& 
-\hf  {\rm d} \l_{\rm R}^2 U_{\rm L} \big(
{\rm d} \te_{\rm L} - \te_{\rm L} {\bm x}^{-1} {\rm d} {\bm x} \big)
+\ri U_{\rm L} \te_{\rm L} {\bm x}^{-1} {\rm d} {\bm x} \ve^{-1} \big( \te_{\rm{L}}^{\rm T} {\rm d} \te_{\rm L} + \te_{\rm{R}}^{\rm T} {\rm d} \te_{\rm R} \big) \non \\
&-&  \ri U_{\rm L} {\rm d} \te_{\rm L} \ve^{-1} \big( \te_{\rm{L}}^{\rm T} {\rm d} \te_{\rm L} + \te_{\rm{R}}^{\rm T} {\rm d} \te_{\rm R} \big)
- \l_{\rm L}^{2} U_{\rm L} {\rm d} \te_{\rm L} {\bm x}^{-1} {\rm d} {\bm x} + \l_{\rm L}^{2} U_{\rm L} \te_{\rm L}{\bm x}^{-1} {\rm d} {\bm x} {\bm x}^{-1} {\rm d} {\bm x}~,
\eea
\end{subequations}
whilst the curvature is given by
\begin{subequations}
\begin{align}
\cR_{\rm L} &= \left(
\begin{array}{c|c}
\ri \ve^{-1} \big( {\rm d} \te_{\rm R}^{\rm T} U_{\rm R}^{2} {\rm d} \te_{\rm R}^{\phantom {\rm R}} - {\rm d} \l_{\rm R}^{2} \te_{\rm R}^{\rm T} {\rm d} \te_{\rm R} \big)
 & 0 \\
 \hline 
0 & \begin{aligned} &\ri U_{\rm L} [ \te_{\rm L} {\bm x}^{-1} {\rm d} {\bm x} \ve^{-1} {\rm d} \te_{\rm L}^{\rm T} 
- {\rm d} \te_{\rm L} \ve^{-1} {\rm d} \te_{\rm L}^{\rm T}   \\ 
&\qquad + {\rm d} \te_{\rm L} \ve^{-1} {\rm d} {\bm x}^{\rm T} ({\bm x}^{-1})^{\rm T} \te_{\rm L}^{\rm T}  \\ 
&-  \te_{\rm L} {\bm x}^{-1} {\rm d} {\bm x} \ve^{-1} {\rm d} {\bm x}^{\rm T} ({\bm x}^{-1})^{\rm T} \te_{\rm L}^{\rm T} ]U_{\rm L} \end{aligned}
\end{array}
\right)~, \\
\cR_{\rm R} 
&= \left(
\begin{array}{c|c}
\ri \ve^{-1} \big( {\rm d} \te_{\rm R}^{\rm T} U_{\rm R}^{2} {\rm d} \te_{\rm R}^{\phantom {\rm R}} - {\rm d} \l_{\rm R}^{2} \te_{\rm R}^{\rm T} {\rm d} \te_{\rm R} \big) & 0 \\
 \hline 
0 & \begin{aligned} &\ri U_{\rm R} [ {\rm d} \te_{\rm R} \ve^{-1} {\rm d} \te_{\rm R}^{\rm T} 
-  \l_{\rm R}^{-1} {\rm d} \l_{\rm R}  \te_{\rm R} \ve^{-1} {\rm d} \te_{\rm R}^{\rm T}  \\
&\qquad +  \l_{\rm R}^{-1} {\rm d} \l_{\rm R}  {\rm d} \te_{\rm R} \ve^{-1} \te_{\rm R}^{\rm T} ] U_{\rm R}  \end{aligned}
\end{array}
\right)~.
\end{align}
\end{subequations}
It is possible to express both the torsion and the curvature in terms of the vielbein \eqref{vielbein}. They read
\begin{align} \label{Torsionvielbein}
\cT_{\rm L} &= \left(
\begin{array}{c|c}
-\tilde{E} \wedge \tilde{E} + \ri \ve^{-1} \cE_{\rm L}^{\rm T} \wedge \cE_{\rm L} + \ri \ve^{-1} \cE_{\rm R}^{\rm T} \wedge \cE_{\rm R} & \tilde{E} \wedge \ve^{-1} \cE_{\rm L}^{\rm T} \\
 \hline 
-\ri \cE_{\rm L} \wedge \tilde{E} & 0
\end{array}
\right)~, \quad 
\cT_{\rm R} 
= 0~, \\ \label{Curvaturevielbein}
\cR_{\rm L} &= \left(
\begin{array}{c|c}
\ri \ve^{-1} \cE_{\rm R}^{\rm T} \wedge \cE_{\rm R} & 0 \\
 \hline 
0 & - \ri \cE_{\rm L} \wedge \ve^{-1} \cE_{\rm L}^{\rm T}
\end{array}
\right)~, \quad 
\cR_{\rm R} 
=  \left(
\begin{array}{c|c}
\ri \ve^{-1} \cE_{\rm R}^{\rm T} \wedge \cE_{\rm R} & 0 \\
 \hline 
0 &  \ri \cE_{\rm R} \wedge \ve^{-1} \cE_{\rm R}^{\rm T}
\end{array}
\right)~.
\end{align}

In order to reconcile the coset construction of ${\rm AdS}^{(3|p,q)}$ with the supergravity approach employed in \cite{KLT-M12} we would like to use the torsion and curvature tensors to construct the (anti-)commutation relations of the covariant derivatives. Accordingly, we must choose a basis $W_{A} = (W_{ab}, W_{\alpha \Io}, W_{\alpha \Iu})$ for the subspace $\cW$ and likewise a basis $H_{i}$ for the algebra $\cH$. Elements in $\cW$, such as the vielbein $E$ and torsion $\cT$, may then be decomposed according to
\begin{subequations}
\bea \label{Wdecomp}
E &=& - \frac 12 E^{ab} W_{ab} + E^{\a \Io} W_{\a \Io} + E^{\a \Iu} W_{\a \Iu}
 = 
E^{a} W_{a} + E^{\a \Io} W_{\a \Io} + E^{\a \Iu} W_{\a \Iu}~, \\
\cT &=& -\frac12 \cT^{ab} W_{ab} + \cT^{\a \Io} W_{\a \Io} + \cT^{\a \Iu} W_{\a \Iu}
= \cT^{a} W_{a} + \cT^{\a \Io} W_{\a \Io} + \cT^{\a \Iu} W_{\a \Iu}~,
\eea
\end{subequations}
in order to obtain the covariant one-forms $E^{A} = (E^{ab}, E^{\a \Io}, E^{\a \Iu})$ and torsion components $\cT^{A} = (\cT^{ab}, \cT^{\a \Io}, \cT^{\a \Iu})$. A similar decomposition is performed for the curvature. The components of the torsion and curvature may then be further decomposed as super two-forms according to 
\bea
\cT^{A} &=& \frac{1}{2} E^{B} \wedge E^{C} \cT_{C B}^{\phantom A \phantom A A}~,\\
\cR &=& \frac{1}{2}  E^{B} \wedge E^{C} \cR_{C B}^{\phantom C \phantom B i} H_{i}~,
\eea
and these components used to construct the (anti-)commutation relations of the covariant derivatives
\bea
\{ \cD_{A}, \cD_{B} ] &=& \cT_{{A} {B}}^{\phantom A \phantom B  C} \cD_{C} + \frac{1}{2} \cR_{{A} {B}}^{\phantom A \phantom B a b} \cM_{ab} + \frac{1}{2} \cR_{{A} {B}}^{\phantom A \phantom B \Io \Jo } \cN_{\Io \Jo} + \frac{1}{2} \cR_{{A} {B}}^{\phantom A \phantom B \Iu \Ju } \cN_{\Iu \Ju}~,
\eea
where $\cM_{a b}$, $\cN_{\Io \Jo}$ and $\cN_{\Iu \Ju}$ are the generators of the structure group ($\sSL(2, \mathbb{R})$, $\sSO(p)$ and $\sSO(q)$ respectively).

At this stage we introduce the generators of the ${\sOSp}_+ (p|2; {\mathbb R} )$ and ${\sOSp}_- (q|2; {\mathbb R} )$ algebras. Let $\mathfrak{m}^{\rm L}_{ab}$, $Q_{\a \Io}$ and $\cN_{\Io \Jo}$ be the generators of the $\mathfrak{sl} (2,\mathbb{R})$, fermionic, and $\mathfrak{so}(p)$ parts of the ${\sOSp}_+ (p|2; {\mathbb R} )$ algebra respectively, whilst $\mathfrak{m}^{\rm R}_{ab}$, $Q_{\a \Iu}$ and $\cN_{\Iu \Ju}$ are the corresponding generators for the ${\sOSp}_- (q|2; {\mathbb R} )$ algebra. Given the forms of the vielbein and connection, it is useful to then define the objects $\mathfrak{m}_{ab} = \mathfrak{m}^{\rm L}_{ab}$ and $\mathcal{M}_{ab} = \mathfrak{m}^{\rm L}_{ab} \oplus \mathfrak{m}^{\rm R}_{ab}$. We take $\mathfrak{m}_{ab}$, $Q_{\a \Io}$ and $Q_{\a \Iu}$ as basis elements for $\cW$ whilst $\mathcal{M}_{ab}$, $\cN_{\Io \Jo}$ and $\cN_{\Iu \Ju}$ form a basis of $\cH$.

These basis elements satisfy the following graded commutation relations:
\begin{subequations} \label{algcommutators}
\begin{align}
\left[ \mathfrak{m}_{ab}, \mathfrak{m}_{cd} \right] &= \eta_{ad} \mathfrak{m}_{bc} - \eta_{ac} \mathfrak{m}_{bd} + \eta_{bc} \mathfrak{m}_{ad} - \eta_{bd} \mathfrak{m}_{ac}~, \\
\left[ \mathcal{M}_{ab}, \mathfrak{m}_{cd} \right] &= \eta_{ad} \mathfrak{m}_{bc} - \eta_{ac} \mathfrak{m}_{bd} + \eta_{bc} \mathfrak{m}_{ad} - \eta_{bd} \mathfrak{m}_{ac}~, \\
\left[ \mathcal{M}_{ab}, \mathcal{M}_{cd} \right] &= \eta_{ad} \mathcal{M}_{bc} - \eta_{ac} \mathcal{M}_{bd} + \eta_{bc} \mathcal{M}_{ad} - \eta_{bd} \mathcal{M}_{ac}~, \\
\left[ \mathfrak{m}_{ab}, Q_{\a \Io} \right] &= - (\S_{ab})_{\a}^{\phantom \a \b} Q_{\b \Io}~, \quad \left[ \mathfrak{m}_{ab}, Q_{\a \Iu} \right] = 0~, \\
\{ Q_{\a \Io}, Q_{\b \Jo} \} &= 2 \ri \d_{\Io \Jo} (\S^{ab})_{\a\b} \mathfrak{m}_{ab} + \ri \ve_{\a\b} \cN_{\Io \Jo}~, \\
\{ Q_{\a \Iu}, Q_{\b \Ju} \} &= -2 \ri \d_{\Iu \Ju} (\S^{ab})_{\a\b} ( \mathcal{M}_{ab} - \mathfrak{m}_{ab} )- \ri \ve_{\a\b} \cN_{\Iu \Ju}~, \\
\left[ \mathcal{M}_{ab}, Q_{\a \Io} \right] &= - (\S_{ab})_{\a}^{\phantom \a \b} Q_{\b \Io}~, \quad \left[ \mathcal{M}_{ab}, Q_{\a \Iu} \right] = - (\S_{ab})_{\a}^{\phantom \a \b} Q_{\b \Iu}~, \\
\left[ \cN_{\Io \Jo}, Q_{\a \overline{K}} \right] &= - 2 \d_{\overline{K} [\Io} Q_{\a \Jo]}~, \quad \left[ \cN_{\Iu \Ju}, Q_{\a \underline{K}} \right] = - 2 \d_{\underline{K} [\Iu} Q_{\a \Ju]}~, \\
\left[ \cN_{\Io \Jo}, \cN_{\overline{M} \overline{N}} \right] &=  
\d_{\Io \overline{N}} \cN_{\Jo \overline{M}}
- \d_{\Io \overline{M}} \cN_{\Jo \overline{N}}
+\d_{\Jo \overline{M}} \cN_{\Io \overline{N}} 
- \d_{\Jo \overline{N}} \cN_{\Io \overline{M}} 
 ~, \\
\left[ \cN_{\Iu \Ju}, \cN_{\underline{M} \underline{N}} \right] &= 
 \d_{\Iu \underline{N}} \cN_{\Ju \underline{M}}
- \d_{\Iu \underline{M}} \cN_{\Ju \underline{N}} 
+\d_{\Ju \underline{M}} \cN_{\Iu \underline{N}} - \d_{\Ju \underline{N}} \cN_{\Iu \underline{M}} ~,
\end{align}
\end{subequations}
with all other (anti-)commutators vanishing.

Using the Maurer-Cartan structure equation
\bea
\rm{d} \o - \o \wedge \o = 0~,
\eea
the decomposition $\o = E + \O$, and the definitions of the torsion and curvature \eqref{torandcur} it is straightforward to show that
\bea
-\cT = (E \wedge E)|_{\cW}~, \\
\cR = (E \wedge E)|_{\cH}~.
\eea
Expanding the vielbein as 
\begin{subequations}
\bea
E &=& -\frac{1}{2} E^{ab} \mathfrak{m}_{ab} + E^{\a \Io} Q_{\a \Io} + E^{\a \Iu} Q_{\a \Iu}~, \\
&=& E^{a} \mathfrak{m}_{a} + E^{\a \Io} Q_{\a \Io} + E^{\a \Iu} Q_{\a \Iu}~,
\eea
\end{subequations}
computing $E \wedge E$ and making use of the (anti-)commutation relations \eqref{algcommutators} we then obtain the non-vanishing (dualised) components of the torsion and curvature
\begin{subequations}
\bea
\cT_{ab}^{\phantom a \phantom b c} &=&  -\ve_{ab}^{\phantom a \phantom b c}~, \\
\cT_{\a \Io \b \Jo}^{\phantom \a \phantom \Io \phantom \b \phantom \Jo a} = 2\ri \d_{\Io \Jo} (\g^{a})_{\a\b}~, &\quad& \cT_{\a \Iu \b \Ju}^{\phantom \a \phantom \Iu \phantom \b \phantom \Ju a} = 2 \ri \d_{\Iu \Ju} (\g^{a})_{\a\b}~, \\
\cT_{a \a \Io}^{\phantom a \phantom \a \phantom \Io \b \Jo} &=& -\frac{1}{2} \d^{\Jo}_{\Io} (\g_{a})_{\a}^{\phantom \a \b}~, \\
\cR_{\a \Iu \b \Ju}^{\phantom \a \phantom \Iu \phantom \b \phantom \Ju a} &=& -2\ri \d_{\Iu \Ju} (\g^{a})_{\a\b}~, \\
\cR_{\a \overline{I} \b \overline{J}}^{\phantom \a \phantom {\underline{I}} \phantom \b \phantom {\underline{J}} \overline{M} \overline{N}} = -2 \ri \ve_{\a\b} \d_{[\overline{I}}^{\overline{M}} \d_{\overline{J}]}^{\overline{N}}, &\quad& \cR_{\a \underline{I} \b \underline{J}}^{\phantom \a \phantom {\underline{I}} \phantom \b \phantom {\underline{J}} \underline{M} \underline{N}} = 2 \ri \ve_{\a\b}\d_{[\underline{I}}^{\underline{M}} \d_{\underline{J}]}^{\underline{N}} ~.
\eea
\end{subequations}
The graded commutation relations of the covariant derivatives are thus
\begin{subequations} \label{derivativecommutators}
\bea
\left[ \cD_{a}, \cD_{b} \right] &=& \ve_{a b}^{\phantom b \phantom a c} \cD_{c}~, \\
\left[ \cD_{a}, \cD_{\a \Io} \right] &=& - \frac{1}{2} (\g_{a})_{\a}^{\phantom \a \b} \cD_{\b \Io}~, \\
\left[ \cD_{a}, \cD_{\a \Iu} \right] &=& 0~, \\
\{ \cD_{\a \Io}, \cD_{\b \Jo} \}  &=& - 2 \ri \d_{\Io \Jo} \cD_{\a\b} - \ri \ve_{\a\b} \cN_{\Io\Jo}~, \\
\{ \cD_{\a \Iu}, \cD_{\b \Ju} \}  &=& - 2 \ri \d_{\Iu \Ju} \cD_{\a\b} + 2 \ri \d_{\Iu \Ju} \cM_{\a\b} + \ri \ve_{\a\b} \cN_{\Iu\Ju}~.
\eea
\end{subequations}

To make contact with the results of \cite{KLT-M12}, we
redefine the vector covariant derivative such that the vector commutator is torsion-free. With the choice
\bea
\tilde{\cD}_{a} &=& \cD_{a} - \hf \cM_{a}~, 
\eea
the graded commutation relations become
\begin{subequations} \label{4.30}
\bea
\big[ \tilde{\cD}_{a}, \tilde{\cD}_{b} \big] &=& \frac 14 \ve_{a b c} \cM^{c}~, \\
\big[ \tilde{\cD}_{a}, {\cD}_{\a \Io} \big] &=& - \frac 14 (\g_{a})_{\a}^{\phantom \a \b} {\cD}_{\b \Io}~, \\
\big[ \tilde{\cD}_{a}, {\cD}_{\a \Iu} \big] &=& \frac 14 (\g_{a})_{\a}^{\phantom \a \b} {\cD}_{\b \Iu}~, \\
\big\{ {\cD}_{\a \Io}, {\cD}_{\b \Jo} \big\}  &=& - 2 \ri \d_{\Io \Jo} \tilde{\cD}_{\a\b} 
- \ri \d_{\Io \Jo} \cM_{\a\b} -  \ri \ve_{\a\b} \cN_{\Io\Jo}~, \label{4.30e}\\
\big\{ {\cD}_{\a \Iu}, {\cD}_{\b \Ju} \big\}  &=& - 2 \ri \d_{\Iu \Ju} \tilde{\cD}_{\a\b} 
+ \ri \d_{\Iu \Ju} \cM_{\a\b} +  \ri \ve_{\a\b} \cN_{\Iu\Ju}~.\label{4.30f}
\eea
\end{subequations}

On the other hand, the algebra of the covariant derivatives of ${\rm AdS}^{(3|p,q)}$ 
 given in \cite{KLT-M12} has the form: 
 \begin{subequations} \label{algebra-KLT-M}
\bea
\big[ {\cD}_{a}, {\cD}_{b} \big] &=& 4 S^2 \ve_{a b c} \cM^{c}~, \\
\big[ {\cD}_{a}, {\cD}_{\a \Io} \big] &=&S (\g_{a})_{\a}^{\phantom \a \b} {\cD}_{\b \Io}~, \\
\left[ {\cD}_{a}, {\cD}_{\a \Iu} \right] &=& - S (\g_{a})_{\a}^{\phantom \a \b} {\cD}_{\b \Iu}~, \\
\big\{ {\cD}_{\a \Io}, {\cD}_{\b \Jo} \big\}  &=& 2 \ri \d_{\Io \Jo} {\cD}_{\a\b} 
- 4\ri S\d_{\Io \Jo} \cM_{\a\b} -  4\ri S\ve_{\a\b} \cN_{\Io\Jo}~, \\
\big\{ {\cD}_{\a \Iu}, {\cD}_{\b \Ju} \big\}  &=& 2 \ri \d_{\Iu \Ju} {\cD}_{\a\b} 
+ 4 \ri S\d_{\Iu \Ju} \cM_{\a\b} + 4 \ri S\ve_{\a\b} \cN_{\Iu\Ju}~,
\eea
\end{subequations}
with $S \neq 0$ a constant curvature parameter.
These graded commutation relations are equivalent to \eqref{alg-AdS}.
We thus observe that the graded commutation relations \eqref{4.30} are obtained from \eqref{algebra-KLT-M} by setting $S=-1/4$, with an overall negative sign occurring in the anti-commutation relations of spinor covariant derivatives. In \cite{KLT-M12} $S$ was chosen to be positive, however a negative value of $S$ is just as valid. 


\subsection{Alternate choices}

Had we instead chosen as our isometry group
\bea
G_\mp = {\sOSp}_- (p|2; {\mathbb R} ) \times  {\sOSp}_+ (q|2; {\mathbb R} )~,
\eea
then the algebra \eqref{algcommutators} would differ in the $QQ$ anti-commutators
\begin{subequations}
\bea
\{ Q_{\a \Io}, Q_{\b \Jo} \} &=& -2 \ri \d_{\Io \Jo} (\S^{ab})_{\a\b} \mathfrak{m}_{ab} - \ri \ve_{\a\b} \cN_{\Io \Jo}~, \\
\{ Q_{\a \Iu}, Q_{\b \Ju} \} &=& 2 \ri \d_{\Iu \Ju} (\S^{ab})_{\a\b} ( \mathcal{M}_{ab} - \mathfrak{m}_{ab} )+ \ri \ve_{\a\b} \cN_{\Iu \Ju}~.
\eea
\end{subequations}
With this choice of isometry group we instead obtain the following graded commutation relations
\begin{subequations}
\bea
\left[ \cD_{a}, \cD_{b} \right] &=& \ve_{a b}^{\phantom b \phantom a c} \cD_{c}~, \\
\left[ \cD_{a}, \cD_{\a \Io} \right] &=& - \frac{1}{2} (\g_{a})_{\a}^{\phantom \a \b} \cD_{\b \Io}~, \\
\{ \cD_{\a \Io}, \cD_{\b \Jo} \}  &=& 2 \ri \d_{\Io \Jo} \cD_{\a\b} + \ri \ve_{\a\b} \cN_{\Io\Jo}~, \\
\{ \cD_{\a \Iu}, \cD_{\b \Ju} \}  &=& 2 \ri \d_{\Iu \Ju} \cD_{\a\b} - 2 \ri \d_{\Iu \Ju} \cM_{\a\b} - \ri \ve_{\a\b} \cN_{\Iu\Ju}~.
\eea
\end{subequations}
Again redefining the vector covariant derivative as
\bea
\tilde{\cD}_{a} &=& \cD_{a} - \hf \cM_{a}~, 
\eea
the graded commutation relations become 
\begin{subequations}
\bea
\big[ \tilde{\cD}_{a}, \tilde{\cD}_{b} \big] &=& \frac 14 \ve_{a b c} \cM^{c}~, \\
\big[ \tilde{\cD}_{a}, {\cD}_{\a \Io} \big] &=& - \frac 14 (\g_{a})_{\a}^{\phantom \a \b} {\cD}_{\b \Io}~, \\
\big[ \tilde{\cD}_{a}, {\cD}_{\a \Iu} \big] &=& \frac 14 (\g_{a})_{\a}^{\phantom \a \b} {\cD}_{\b \Iu}~, \\
\big\{ {\cD}_{\a \Io}, {\cD}_{\b \Jo} \big\}  &=& 2 \ri \d_{\Io \Jo} \tilde{\cD}_{\a\b} 
 + \ri \d_{\Io \Jo} \cM_{\a\b} + \ri \ve_{\a\b} \cN_{\Io\Jo}~, \\
\big\{ {\cD}_{\a \Iu}, {\cD}_{\b \Ju} \big\}  &=& 2 \ri \d_{\Iu \Ju} \tilde{\cD}_{\a\b} 
- \ri \d_{\Iu \Ju} \cM_{\a\b} - \ri \ve_{\a\b} \cN_{\Iu\Ju}~,
\eea
\end{subequations}
which coincide with \eqref{algebra-KLT-M} for $S = - \frac 14$. 

We observe that neither choice of isometry group $G_{\pm}$ or $G_{\mp}$ result in an algebra of covariant derivatives with a positive $S$ parameter. A non-negative $S$ value may be obtained by instead defining decomposition in the subspace $\cW$ by
\begin{align} 
E &= \frac 12 E^{ab} W_{ab} + E^{\a \Io} W_{\a \Io} + E^{\a \Iu} W_{\a \Iu}
 = 
-E^{a} W_{a} + E^{\a \Io} W_{\a \Io} + E^{\a \Iu} W_{\a \Iu}~,
\end{align}
in contrast with \eqref{Wdecomp}. This results in all torsion components picking up an additional negative sign, and hence the (anti-)commutation relations of the covariant derivatives become 
\begin{subequations}
\bea
\left[ \cD_{a}, \cD_{b} \right] &=& -\ve_{a b}^{\phantom b \phantom a c} \cD_{c}~, \\
\left[ \cD_{a}, \cD_{\a \Io} \right] &=& \frac{1}{2} (\g_{a})_{\a}^{\phantom \a \b} \cD_{\b \Io}~, \\
\left[ \cD_{a}, \cD_{\a \Iu} \right] &=& 0~, \\
\{ \cD_{\a \Io}, \cD_{\b \Jo} \}  &=& 2 \ri \d_{\Io \Jo} \cD_{\a\b} - \ri \ve_{\a\b} \cN_{\Io\Jo}~, \\
\{ \cD_{\a \Iu}, \cD_{\b \Ju} \}  &=& 2 \ri \d_{\Iu \Ju} \cD_{\a\b} + 2 \ri \d_{\Iu \Ju} \cM_{\a\b} + \ri \ve_{\a\b} \cN_{\Iu\Ju}~.
\eea
\end{subequations}
In this case we must redefine the vector covariant derivative as
\bea
\tilde{\cD}_{a} &=& \cD_{a} + \hf \cM_{a}~, 
\eea
and the graded commutation relations then read
\begin{subequations}
\bea
\big[ \tilde{\cD}_{a}, \tilde{\cD}_{b} \big] &=& \frac 14 \ve_{a b c} \cM^{c}~, \\
\big[ \tilde{\cD}_{a}, {\cD}_{\a \Io} \big] &=& \frac 14 (\g_{a})_{\a}^{\phantom \a \b} {\cD}_{\b \Io}~, \\
\big[ \tilde{\cD}_{a}, {\cD}_{\a \Iu} \big] &=& -\frac 14 (\g_{a})_{\a}^{\phantom \a \b} {\cD}_{\b \Iu}~, \\
\big\{ {\cD}_{\a \Io}, {\cD}_{\b \Jo} \big\}  &=& 2 \ri \d_{\Io \Jo} \tilde{\cD}_{\a\b} 
- \ri \d_{\Io \Jo} \cM_{\a\b} -  \ri \ve_{\a\b} \cN_{\Io\Jo}~, \\
\big\{ {\cD}_{\a \Iu}, {\cD}_{\b \Ju} \big\}  &=& 2 \ri \d_{\Iu \Ju} \tilde{\cD}_{\a\b} 
+ \ri \d_{\Iu \Ju} \cM_{\a\b} +  \ri \ve_{\a\b} \cN_{\Iu\Ju}~,
\eea
\end{subequations}
which agree with \eqref{algebra-KLT-M} for $S = \frac 14 >0$. Ultimately, the sign of $S$ is a matter of convention, and is chosen to be negative in this paper for convenience in later calculations.

We could also consider having fixed the freedom \eqref{ER2} in the left sector
\bea
\bm{x} = \l_{\rm L}(\q_{\rm L}) {\mathbbm 1}_2 ~.
\eea
With this choice we would have instead used the following definition of the basis element $\mathfrak{m}_{ab} = \mathfrak{m}_{ab}^{\rm R}$ and as a result obtained the (anti-)commutation relations
\begin{subequations}
\bea
\left[ \cD_{a}, \cD_{b} \right] &=& \ve_{a b}^{\phantom b \phantom a c} \cD_{c}~, \\
\left[ \cD_{a}, \cD_{\a \Iu} \right] &=&  - \frac{1}{2} (\g_{a})_{\a}^{\phantom \a \b} \cD_{\b \Iu}~, \\
\{ \cD_{\a \Io}, \cD_{\b \Jo} \}  &=& 2 \ri \d_{\Io \Jo} \cD_{\a\b} - 2 \ri \d_{\Iu \Ju} \cM_{\a\b} - \ri \ve_{\a\b} \cN_{\Io\Jo}~, \\
\{ \cD_{\a \Iu}, \cD_{\b \Ju} \}  &=& 2 \ri \d_{\Iu \Ju} \cD_{\a\b}  + \ri \ve_{\a\b} \cN_{\Iu\Ju}~.
\eea
\end{subequations}
With the redefinition
\bea
\tilde{\cD}_{a} &=& \cD_{a} - \hf \cM_{a}~, 
\eea
we obtain
\begin{subequations}
\bea
\big[ \tilde{\cD}_{a}, \tilde{\cD}_{b} \big] &=& \frac 14 \ve_{a b c} \cM^{c}~, \\
\big[ \tilde{\cD}_{a}, {\cD}_{\a \Io} \big] &=& \frac 14 (\g_{a})_{\a}^{\phantom \a \b} {\cD}_{\b \Io}~, \\
\big[ \tilde{\cD}_{a}, {\cD}_{\a \Iu} \big] &=& - \frac 14 (\g_{a})_{\a}^{\phantom \a \b} {\cD}_{\b \Iu}~, \\
\big\{ {\cD}_{\a \Io}, {\cD}_{\b \Jo} \big\}  &=& 2 \ri \d_{\Io \Jo} \tilde{\cD}_{\a\b} 
 - \ri \d_{\Io \Jo} \cM_{\a\b} - \ri \ve_{\a\b} \cN_{\Io\Jo}~, \\
\big\{ {\cD}_{\a \Iu}, {\cD}_{\b \Ju} \big\}  &=& 2 \ri \d_{\Iu \Ju} \tilde{\cD}_{\a\b} 
+ \ri \d_{\Iu \Ju} \cM_{\a\b} + \ri \ve_{\a\b} \cN_{\Iu\Ju}~.
\eea
\end{subequations}
This agrees with \eqref{algebra-KLT-M} for $S = \frac 14$.

Let us briefly explore what differences arise if the orthosymplectic groups are not chosen to be in different realisations. Suppose that we had chosen \begin{align}
G_\pp = \sOSp_{+} (p| 2; \mathbb{R}) \times \sOSp_{+} (q| 2; \mathbb{R})~,
\end{align}
as the isometry group of $\mathrm{AdS}^{(3|p,q)}$.
The $QQ$ anti-commutators would then have the same form in both sectors
\begin{subequations}
\bea
\{ Q_{\a \Io}, Q_{\b \Jo} \} &=& 2 \ri \d_{\Io \Jo} (\S^{ab})_{\a\b} \mathfrak{m}_{ab} + \ri \ve_{\a\b} \cN_{\Io \Jo}~, \\
\{ Q_{\a \Iu}, Q_{\b \Ju} \} &=& 2 \ri \d_{\Iu \Ju} (\S^{ab})_{\a\b} ( \mathcal{M}_{ab} - \mathfrak{m}_{ab} ) + \ri \ve_{\a\b} \cN_{\Iu \Ju}~,
\eea
\end{subequations}
and would in turn give rise to the following (anti-)commutation relations
\begin{subequations} 
\bea
\left[ \cD_{a}, \cD_{b} \right] &=& \ve_{a b}^{\phantom b \phantom a c} \cD_{c}~, \\
\left[ \cD_{a}, \cD_{\a \Io} \right] &=& - \frac{1}{2} (\g_{a})_{\a}^{\phantom \a \b} \cD_{\b \Io}~, \\
\{ \cD_{\a \Io}, \cD_{\b \Jo} \}  &=& - 2 \ri \d_{\Io \Jo} \cD_{\a\b} - \ri \ve_{\a\b} \cN_{\Io\Jo}~, \\
\{ \cD_{\a \Iu}, \cD_{\b \Ju} \}  &=& 2 \ri \d_{\Iu \Ju} \cD_{\a\b} - 2 \ri \d_{\Iu \Ju} \cM_{\a\b} - \ri \ve_{\a\b} \cN_{\Iu\Ju}~.
\eea
\end{subequations}
After once again redefining the vector covariant derivatives
\bea
\tilde{\cD}_{a} &=& \cD_{a} - \hf \cM_{a}~, 
\eea
the graded commutation relations become
\begin{subequations}
\bea
\big[ \tilde{\cD}_{a}, \tilde{\cD}_{b} \big] &=& \frac 14 \ve_{a b c} \cM^{c}~, \\
\big[ \tilde{\cD}_{a}, {\cD}_{\a \Io} \big] &=& -\frac 14 (\g_{a})_{\a}^{\phantom \a \b} {\cD}_{\b \Io}~, \\
\big[ \tilde{\cD}_{a}, {\cD}_{\a \Iu} \big] &=& \frac 14 (\g_{a})_{\a}^{\phantom \a \b} {\cD}_{\b \Iu}~, \\
\big\{ {\cD}_{\a \Io}, {\cD}_{\b \Jo} \big\}  &=& -2 \ri \d_{\Io \Jo} \tilde{\cD}_{\a\b} 
- \ri \d_{\Io \Jo} \cM_{\a\b} -  \ri \ve_{\a\b} \cN_{\Io\Jo}~, \label{4.30ee}\\
\big\{ {\cD}_{\a \Iu}, {\cD}_{\b \Ju} \big\}  &=& 2 \ri \d_{\Iu \Ju} \tilde{\cD}_{\a\b} 
- \ri \d_{\Iu \Ju} \cM_{\a\b} -  \ri \ve_{\a\b} \cN_{\Iu\Ju}~.\label{4.30ff}
\eea
\end{subequations}
Note the difference in sign on the vector derivative terms in the spinor derivative anti-commutators. The only way to fix this is by rescaling the spinor derivatives by an imaginary factor, however since we are in three dimensions our spinors must be real and so this is not possible. Hence for this choice of isometry group there is no way to reconcile the algebra of covariant derivatives with those \eqref{algebra-KLT-M}. This is a consequence of this particular choice of  $\mathrm{AdS}^{(3|p,q)}$ supergroup not possessing a Poincar\'e limit \cite{AT2}.

For the previous choice of basis elements we had to redefine our covariant derivatives in order to make the bosonic subspace torsion-free. It would be desirable if this redefinition wasn't necessary. Thus, we would like to choose basis elements such that the bosonic generators form a symmetric pair. Instead defining $\mathfrak{m}_{ab} =  \mathfrak{m}^{\rm L}_{ab} \oplus - \mathfrak{m}^{\rm R}_{ab}$ the (anti-)commutation relations are
\begin{subequations} \label{newcommutators}
\begin{align}
\left[ \mathfrak{m}_{ab}, \mathfrak{m}_{cd} \right] &= \eta_{ad} \mathcal{M}_{bc} - \eta_{ac} \mathcal{M}_{bd} + \eta_{bc} \mathcal{M}_{ad} - \eta_{bd} \mathcal{M}_{ac}~, \\
\left[ \mathcal{M}_{ab}, \mathfrak{m}_{cd} \right] &= \eta_{ad} \mathfrak{m}_{bc} - \eta_{ac} \mathfrak{m}_{bd} + \eta_{bc} \mathfrak{m}_{ad} - \eta_{bd} \mathfrak{m}_{ac}~, \\
\left[ \mathcal{M}_{ab}, \mathcal{M}_{cd} \right] &= \eta_{ad} \mathcal{M}_{bc} - \eta_{ac} \mathcal{M}_{bd} + \eta_{bc} \mathcal{M}_{ad} - \eta_{bd} \mathcal{M}_{ac}~, \\
\left[ \mathfrak{m}_{ab}, Q_{\a \Io} \right] &= - (\S_{ab})_{\a}^{\phantom \a \b} Q_{\b \Io}~, \quad \left[ \mathfrak{m}_{ab}, Q_{\a \Iu} \right] =   (\S_{ab})_{\a}^{\phantom \a \b} Q_{\b \Iu}~, \\
\{ Q_{\a \Io}, Q_{\b \Jo} \} &= 2 \ri \d_{\Io \Jo} (\S^{ab})_{\a\b} ( \hf \mathfrak{m}_{ab} + \hf \mathcal{M}_{ab}) + \ri \ve_{\a\b} \cN_{\Io \Jo}~, \\
\{ Q_{\a \Iu}, Q_{\b \Ju} \} &= -2 \ri \d_{\Iu \Ju} (\S^{ab})_{\a\b} ( -\hf \mathfrak{m}_{ab} + \hf \mathcal{M}_{ab}) - \ri \ve_{\a\b} \cN_{\Iu \Ju}~, \\
\left[ \mathcal{M}_{ab}, Q_{\a \Io} \right] &= - (\S_{ab})_{\a}^{\phantom \a \b} Q_{\b \Io}~, \quad \left[ \mathcal{M}_{ab}, Q_{\a \Iu} \right] = - (\S_{ab})_{\a}^{\phantom \a \b} Q_{\b \Iu}~, \\
\left[ \cN_{\Io \Jo}, Q_{\a \overline{K}} \right] &= - 2 \d_{\overline{K} [\Io} Q_{\a \Jo]}~, \quad \left[ \cN_{\Iu \Ju}, Q_{\a \underline{K}} \right] = - 2 \d_{\underline{K} [\Iu} Q_{\a \Ju]}~, \\
\left[ \cN_{\Io \Jo}, \cN_{\overline{M} \overline{N}} \right] &= \d_{\Io \overline{N}} \cN_{\Jo \overline{M}} - \d_{\Io \overline{M}} \cN_{\Jo \overline{N}} + \d_{\Jo \overline{M}} \cN_{\Io \overline{N}} - \d_{\Jo \overline{N}} \cN_{\Io \overline{M}}~, \\
\left[ \cN_{\Iu \Ju}, \cN_{\underline{M} \underline{N}} \right] &= \d_{\Iu \underline{N}} \cN_{\Ju \underline{M}} - \d_{\Iu \underline{M}} \cN_{\Ju \underline{N}} + \d_{\Ju \underline{M}} \cN_{\Iu \underline{N}} - \d_{\Ju \underline{N}} \cN_{\Iu \underline{M}}~,
\end{align}
\end{subequations}
with all other (anti-)commutators vanishing. We see that the first three commutation relations satisfy the desired property. However, for elements in $\cW$ taking the form \eqref{algcompform} it is not possible to split these basis elements into those generating $\cH$ and those generating $\cW$. We thus consider a different choice of freedom and algebra $\cW$ keeping symmetry between sectors.

\subsection{A more symmetric choice}

We may make a different choice of freedom than \eqref{gauge-fixedY} in order to preserve some symmetry between the left and right sectors. Indeed, let us instead consider the choice
\bea \label{gauge2}
\bm y = x^{-1} \l_{\rm R}~.
\eea
For this choice we define elements $X = (X_{\rm L}, X_{\rm R})$ of $\cW$ to have the form
\bea \label{algcompform2}
X_{\rm L}
=\left(
\begin{array}{c|c}
 A & - \ve^{-1} B_{\rm L}^{\rm T} \\
 \hline 
\ri B_{\rm L} &    0
\end{array}
\right)~, \quad 
X_{\rm R}
=\left(
\begin{array}{c|c}
 -A &  \ve^{-1} B_{\rm R}^{\rm T} \\
 \hline 
\ri B_{\rm R} &    0
\end{array}
\right)~, \quad A \in \mathfrak{sl} (2,\mathbb{R})~,
\eea
whilst elements $Y= (Y_{\rm L}, Y_{\rm R}) \in \cH$ still take the form
\bea \label{algform2}
Y_{\rm L}
=\left(
\begin{array}{c|c}
 n & 0 \\
 \hline 
0 &    r_{\rm L}
\end{array}
\right)~, \quad 
Y_{\rm R}
=\left(
\begin{array}{c|c}
 n & 0 \\
 \hline 
0 &   r_{\rm R}
\end{array}
\right)~, \quad
n  \in \mathfrak{sl} (2,\mathbb{R})~, \quad r_{\rm L} \in  \mathfrak{so} (p)~, \quad
 r_{\rm R} \in  \mathfrak{so} (q)
~.~~
\eea

The Maurer-Cartan one-form is
\begin{subequations}
\begin{align}
\o_{\rm L} 
&= \left(
\begin{array}{c|c}
\l^{2}_{\rm L} x^{-1} {\rm d} x + \l_{\rm L} {\rm d} \l_{\rm L} \mathbbm{1} + \ri \ve^{-1} \te_{\rm L}^{\rm T} {\rm d} \te_{\rm L} & 
 ~ - \ve^{-1} [ {\rm d} \te_{\rm L}^{\rm T}  - {\rm d} { x}^{\rm T} ({ x}^{-1})^{\rm T} \te_{\rm L}^{\rm T} - \l^{-1}_{\rm L} {\rm d} \l_{\rm L} \te_{\rm L}^{\rm T} ] U_{\rm L}   \\
 \hline 
\ri U_{\rm L} \big[ {\rm d} \te_{\rm L} -  \te_{\rm L} {x}^{-1} {\rm d} { x} - \l^{-1}_{\rm L} {\rm d} \l_{\rm L} \te_{\rm L} \big] & \begin{aligned} &\ri \l^{-2}_{\rm L} U_{\rm L} \te_{\rm L} x^{-1} \ve^{-1} {\rm d} (x^{-1})^{\rm T} \te_{\rm L}^{\rm T} U_{\rm L} + U^{-1}_{\rm L} {\rm d} U_{\rm L} \\
&+ \ri \l^{-2}_{\rm L} U_{\rm L} \te_{\rm L} \ve^{-1} {\rm d} \te_{\rm L}^{\rm T} U_{\rm L}
- \ri \l^{-1}_{\rm L} {\rm d} \l_{\rm L} \te_{\rm L} \ve^{-1} \te_{\rm L}^{\rm T} \end{aligned}
\end{array}
\right)~, \\
\o_{\rm R} 
&=  \left(
\begin{array}{c|c}
\l^{2}_{\rm R} x {\rm d} x^{-1} + \l_{\rm R} {\rm d} \l_{\rm R} \mathbbm{1} - \ri \ve^{-1} \te_{\rm R}^{\rm T} {\rm d} \te_{\rm R} & 
 ~ \ve^{-1} [ {\rm d} \te_{\rm R}^{\rm T}  - {\rm d} ( x^{-1})^{\rm T}  x^{\rm T} \te_{\rm R}^{\rm T} - \l^{-1}_{\rm R} {\rm d} \l_{\rm R} \te_{\rm R}^{\rm T} ] U_{\rm R}   \\
 \hline 
\ri U_{\rm R} \big[ {\rm d} \te_{\rm R} -  \te_{\rm R} {x} {\rm d} { x^{-1}} - \l^{-1}_{\rm R} {\rm d} \l_{\rm R} \te_{\rm R} \big] & \begin{aligned} &-\ri \l^{-2}_{\rm R} U_{\rm R} \te_{\rm R} x \ve^{-1} {\rm d} x^{\rm T} \te_{\rm R}^{\rm T} U_{\rm R} + U^{-1}_{\rm R} {\rm d} U_{\rm R}\\
&- \ri \l^{-2}_{\rm R} U_{\rm R} \te_{\rm R} \ve^{-1} {\rm d} \te_{\rm R}^{\rm T} U_{\rm R} + \ri \l^{-1}_{\rm R} {\rm d} \l_{\rm R} \te_{\rm R} \ve^{-1} \te_{\rm R}^{\rm T}  \end{aligned}
\end{array}
\right)~,
\end{align}
\end{subequations}
which we decompose into matrices with the forms \eqref{algcompform2} and \eqref{algform2} to obtain the vielbein
\begin{subequations}\label{vielbein2}
\bea 
E_{\rm L} 
= \left(
\begin{array}{c|c}
\tilde{E} & - \ve^{-1} \cE_{\rm L}^{\rm T} \\
 \hline 
\ri \cE_{\rm L}  & 0
\end{array}
\right)~, \quad 
E_{\rm R} 
&=& \left(
\begin{array}{c|c}
-\tilde{E} &  \ve^{-1} \cE_{\rm R}^{\rm T} \\
 \hline 
\ri \cE_{\rm R}  & 0
\end{array}
\right)~,
\eea
where
\bea \non
\tilde{E} &=& 
\hf \l^{2}_{\rm L} x^{-1} {\rm d} x - \hf \l_{\rm R}^{2} x {\rm d} x^{-1} + \hf \l_{\rm L} {\rm d} \l_{\rm L} \mathbbm{1} - \hf \l_{\rm R} {\rm d} \l_{\rm R} \mathbbm{1} \\
&+& \frac{\ri}{2} \ve^{-1} \te_{\rm L}^{\rm T} {\rm d} \te_{\rm L} + \frac{\ri}{2} \ve^{-1} \te_{\rm R}^{\rm T}{\rm d} \te_{\rm R}~, \\
\cE_{\rm L} &=& 
U_{\rm L} \big[ {\rm d} \te_{\rm L} -  \te_{\rm L} {x}^{-1} {\rm d} { x} - \l^{-1}_{\rm L} {\rm d} \l_{\rm L} \te_{\rm L} \big]
~, \\
\cE_{\rm R} &=& 
U_{\rm R} \big[ {\rm d} \te_{\rm R} -  \te_{\rm R} {x} {\rm d} { x^{-1}} - \l^{-1}_{\rm R} {\rm d} \l_{\rm R} \te_{\rm R} \big]
~,
\eea
\end{subequations}
and the connection
\begin{subequations} \label{connection2}
\bea
\O_{\rm L} 
= \left(
\begin{array}{c|c}
\tilde{\O} & 0 \\
 \hline 
0  & \O_{\sSO(p)}
\end{array}
\right)~, \quad 
\O_{\rm R} 
&=& \left(
\begin{array}{c|c}
\tilde{\O} & 0 \\
 \hline 
0 & \O_{\sSO(q)}
\end{array}
\right)~,
\eea
where
\bea \non
\tilde{\O} &=& \hf \l^{2}_{\rm L} x^{-1} {\rm d} x + \hf \l_{\rm R}^{2} x {\rm d} x^{-1} + \hf \l_{\rm L} {\rm d} \l_{\rm L} \mathbbm{1} + \hf \l_{\rm R} {\rm d} \l_{\rm R} \mathbbm{1} \\ 
&+& \frac{\ri}{2} \ve^{-1} \te_{\rm L}^{\rm T} {\rm d} \te_{\rm L} - \frac{\ri}{2} \ve^{-1} \te_{\rm R}^{\rm T} {\rm d} \te_{\rm R}~, \\ \non
\O_{\sSO(p)} &=& \ri \l^{-2}_{\rm L} U_{\rm L} \te_{\rm L} x^{-1} \ve^{-1} {\rm d} (x^{-1})^{\rm T} \te_{\rm L}^{\rm T} U_{\rm L} 
+ \ri \l^{-2}_{\rm L} U_{\rm L} \te_{\rm L} \ve^{-1} {\rm d} \te_{\rm L}^{\rm T} U_{\rm L} 
+ U^{-1}_{\rm L} {\rm d} U_{\rm L} \\
&-& \ri \l^{-1}_{\rm L} {\rm d} \l_{\rm L} \te_{\rm L} \ve^{-1} \te_{\rm L}^{\rm T}~, \\ \non
\O_{\sSO(q)} &=& 
-\ri \l^{-2}_{\rm R} U_{\rm R} \te_{\rm R} x \ve^{-1} {\rm d} x^{\rm T} \te_{\rm R}^{\rm T} U_{\rm R}
- \ri \l^{-2}_{\rm R} U_{\rm R} \te_{\rm R} \ve^{-1} {\rm d} \te_{\rm R}^{\rm T} U_{\rm R}
+ U^{-1}_{\rm R} {\rm d} U_{\rm R} \\
&+& \ri \l^{-1}_{\rm R} {\rm d} \l_{\rm R} \te_{\rm R} \ve^{-1} \te_{\rm R}^{\rm T}~.
\eea
\end{subequations}

We calculate the torsion and curvature tensors. They are
\begin{subequations}
\bea
\cT_{\rm L} = \left(
\begin{array}{c|c}
\cT_{1} & - \ve^{-1} \cT_{2}^{\rm T} \\
 \hline 
\ri \cT_{2} & 0
\end{array}
\right)~, \quad 
\cT_{\rm R} 
&=& \left(
\begin{array}{c|c}
-\cT_{1} & \ve^{-1} \cT_{3}^{\rm T} \\
 \hline 
\ri \cT_{3} & 0
\end{array}
\right)~,
\eea
where
\bea \non
\cT_{1} &=& -\l_{\rm L} (\l^{2}_{\rm L} - 1) {\rm d} \l_{\rm L} x^{-1} {\rm d} x + \hf \l^{2}_{\rm L} (\l^{2}_{\rm L} - 1) x^{-1} {\rm d} x x^{-1} {\rm d} x - \ri \l_{\rm L} {\rm d} \l_{\rm L} \ve^{-1} \te_{\rm L}^{\rm T} {\rm d} \te_{\rm L} \\ \non
&+& \frac{\ri}{2} \ve^{-1} {\rm d} \te_{\rm L}^{\rm T} U^{2}_{\rm L} {\rm d} \te_{\rm L} - \frac{\ri}{2} \l^{2}_{\rm L} \ve^{-1} {\rm d} x^{\rm T} (x^{-1})^{\rm T} \te_{\rm L}^{\rm T} {\rm d} \te_{\rm L} - \frac{\ri}{2} \l^{2}_{\rm L} \ve^{-1} {\rm d} \te_{\rm L}^{\rm T} \te_{\rm L} x^{-1} {\rm d} x \\ \non
&-& \l_{\rm R} (\l^{2}_{\rm R} - 1) {\rm d} \l_{\rm R}  {\rm d} x x^{-1} - \hf \l^{2}_{\rm R} (\l^{2}_{\rm R} - 1) {\rm d} x x^{-1} {\rm d} x x^{-1}- \ri \l_{\rm R} {\rm d} \l_{\rm R} \ve^{-1} \te_{\rm R}^{\rm T} {\rm d} \te_{\rm R} \\ 
&+& \frac{\ri}{2} \ve^{-1} {\rm d} \te_{\rm R}^{\rm T} U^{2}_{\rm R} {\rm d} \te_{\rm R} + \frac{\ri}{2} \l^{2}_{\rm R} \ve^{-1} (x^{-1})^{\rm T} {\rm d} x^{\rm T} \te_{\rm R}^{\rm T} {\rm d} \te_{\rm R} + \frac{\ri}{2} \l^{2}_{\rm R} \ve^{-1} {\rm d} \te_{\rm R}^{\rm T} \te_{\rm R}  {\rm d} x x^{-1}~, \\ \non
\cT_{2} &=&  \hf \l_{\rm L}^{2} U_{\rm L} \te_{\rm L} x^{-1} {\rm d} x x^{-1} {\rm d} x - \hf \l_{\rm R}^{2} U_{\rm L} \te_{\rm L} x^{-1} {\rm d} x x {\rm d} x^{-1} - \hf U_{\rm L} \te_{\rm L} x^{-1} {\rm d} x \l_{\rm R} {\rm d} \l_{\rm R} \\ \non
&+& \frac \ri 2 U_{\rm L} \te_{\rm L} x^{-1} {\rm d} x \ve^{-1} \te_{\rm L}^{\rm T} {\rm d} \te_{\rm L} + \frac \ri 2 U_{\rm L} \te_{\rm L} x^{-1} {\rm d} x \ve^{-1} \te_{\rm R}^{\rm T} {\rm d} \te_{\rm R} - \hf \l_{\rm L}^{-1} \l_{\rm R}^{2} {\rm d} \l_{\rm L} U_{\rm L} \te_{\rm L} x {\rm d} x^{-1} \\ \non
&-& \hf \l_{\rm L}^{-1} {\rm d} \l_{\rm L} U_{\rm L} \te_{\rm L} \l_{\rm R} {\rm d} \l_{\rm R} + \frac \ri 2 \l_{\rm L}^{-1} {\rm d} \l_{\rm L} U_{\rm L} \te_{\rm L} \ve^{-1} \te_{\rm L}^{\rm T} {\rm d} \te_{\rm L} + \frac \ri 2 \l_{\rm L}^{-1} {\rm d} \l_{\rm L} U_{\rm L} \te_{\rm L} \ve^{-1} \te_{\rm R}^{\rm T} {\rm d} \te_{\rm R} \\ \non
&-& \hf \l_{\rm L}^{2} U_{\rm L} {\rm d} \te_{\rm L} x^{-1} {\rm d} x + \hf \l_{\rm R}^{2} U_{\rm L} {\rm d} \te_{\rm L} x {\rm d} x^{-1} - \hf U_{\rm L} {\rm d} \te_{\rm L} \l_{\rm L} {\rm d} \l_{\rm L} \\ 
&+& \hf U_{\rm L} {\rm d} \te_{\rm L} \l_{\rm R} {\rm d} \l_{\rm R} - \frac \ri 2 U_{\rm L} {\rm d} \te_{\rm L} \ve^{-1} \te_{\rm L}^{\rm T} {\rm d} \te_{\rm L} - \frac \ri 2 U_{\rm L} {\rm d} \te_{\rm L} \ve^{-1} \te_{\rm R}^{\rm T} {\rm d} \te_{\rm R}~, \\ \non
\cT_{3} &=&  -\hf \l_{\rm L}^{2} U_{\rm R} \te_{\rm R} x {\rm d} x^{-1} x^{-1} {\rm d} x + \hf \l_{\rm R}^{2} U_{\rm R} \te_{\rm R} x {\rm d} x^{-1} x {\rm d} x^{-1} - \hf U_{\rm R} \te_{\rm R} x {\rm d} x^{-1} \l_{\rm L} {\rm d} \l_{\rm L} \\ \non
&-& \frac \ri 2 U_{\rm R} \te_{\rm R} x {\rm d} x^{-1} \ve^{-1} \te_{\rm L}^{\rm T} {\rm d} \te_{\rm L} - \frac \ri 2 U_{\rm R} \te_{\rm R} x {\rm d} x^{-1} \ve^{-1} \te_{\rm R}^{\rm T} {\rm d} \te_{\rm R} - \hf \l_{\rm R}^{-1} \l_{\rm L}^{2} {\rm d} \l_{\rm R} U_{\rm R} \te_{\rm R} x^{-1} {\rm d} x \\ \non
&-& \hf \l_{\rm R}^{-1} {\rm d} \l_{\rm R} U_{\rm R} \te_{\rm R} \l_{\rm L} {\rm d} \l_{\rm L} - \frac \ri 2 \l_{\rm R}^{-1} {\rm d} \l_{\rm R} U_{\rm R} \te_{\rm R} \ve^{-1} \te_{\rm L}^{\rm T} {\rm d} \te_{\rm L} - \frac \ri 2 \l_{\rm R}^{-1} {\rm d} \l_{\rm R} U_{\rm R} \te_{\rm R} \ve^{-1} \te_{\rm R}^{\rm T} {\rm d} \te_{\rm R} \\ \non
&+& \hf \l_{\rm L}^{2} U_{\rm R} {\rm d} \te_{\rm R} x^{-1} {\rm d} x - \hf \l_{\rm R}^{2} U_{\rm R} {\rm d} \te_{\rm R} x {\rm d} x^{-1} + \hf U_{\rm R} {\rm d} \te_{\rm R} \l_{\rm L} {\rm d} \l_{\rm L} \\ 
&-& \hf U_{\rm R} {\rm d} \te_{\rm R} \l_{\rm R} {\rm d} \l_{\rm R} + \frac \ri 2 U_{\rm R} {\rm d} \te_{\rm R} \ve^{-1} \te_{\rm L}^{\rm T} {\rm d} \te_{\rm L} + \frac \ri 2 U_{\rm R} {\rm d} \te_{\rm R} \ve^{-1} \te_{\rm R}^{\rm T} {\rm d} \te_{\rm R}~,
\eea
\end{subequations}
whilst the curvature is given by
\begin{subequations}
\bea
\cR_{\rm L} &=& \left(
\begin{array}{c|c}
\cR_{1}
 & 0 \\
 \hline 
0 & \cR_{2} \end{array}
\right)~, \\
\cR_{\rm R} 
&=& \left(
\begin{array}{c|c}
\cR_{1} & 0 \\
 \hline 
0 & \cR_{3}
\end{array}
\right)~.
\eea
where
\bea \non
\cR_{1} &=& \frac 1 4 \l_{\rm L}^{4} x^{-1} {\rm d} x x^{-1} {\rm d} x - \frac 1 4 \l_{\rm L}^{2} \l_{\rm R}^{2} x^{-1} {\rm d } x x {\rm d } x^{-1} + \frac \ri 4 \l_{\rm L}^{2} x^{-1} {\rm d} x \ve^{-1} \te_{\rm L}^{\rm T} {\rm d} \te_{\rm L} \\ \non
&+& \frac \ri 4 \l_{\rm L}^{2} x^{-1} {\rm d} x \ve^{-1} \te_{\rm R}^{\rm T} {\rm d} \te_{\rm R} - \frac 1 4 \l_{\rm R}^{2} \l_{\rm L}^{2} x {\rm d } x^{-1} x^{-1} {\rm d} x + \frac 1 4 \l_{\rm R}^{4} x {\rm d} x^{-1} x {\rm d} x^{-1} \\ \non
&-& \frac \ri 4 \l_{\rm R}^{2} x {\rm d} x^{-1} \ve^{-1} \te_{\rm L}^{\rm T} {\rm d} \te_{\rm L} - \frac \ri 4 \l_{\rm R}^{2} x {\rm d} x^{-1} \ve^{-1} \te_{\rm R}^{\rm T} {\rm d} \te_{\rm R} + \frac \ri 4 \l_{\rm L}^{2} \ve^{-1} \te_{\rm L}^{\rm T} {\rm d} \te_{\rm L} x^{-1} {\rm d} x \\ \non
&-& \frac \ri 4 \l_{\rm R}^{2} \ve^{-1} \te_{\rm L}^{\rm T} {\rm d} \te_{\rm L} x {\rm d} x^{-1} + \frac \ri 4 \l_{\rm L}^{2} \ve^{-1} \te_{\rm R}^{\rm T} {\rm d} \te_{\rm R} x^{-1} {\rm d} x - \frac \ri 4 \l_{\rm R}^{2} \ve^{-1} \te_{\rm R}^{\rm T} {\rm d} \te_{\rm R} x {\rm d} x^{-1} \\ \non
&-& \frac 1 4 \ve^{-1} \te_{\rm L}^{\rm T} {\rm d} \te_{\rm L} \ve^{-1} \te_{\rm L}^{\rm T} {\rm d} \te_{\rm L} - \frac 1 4 \ve^{-1} \te_{\rm L}^{\rm T} {\rm d} \te_{\rm L} \ve^{-1} \te_{\rm R}^{\rm T} {\rm d} \te_{\rm R} - \frac 1 4 \ve^{-1} \te_{\rm R}^{\rm T} {\rm d} \te_{\rm R} \ve^{-1} \te_{\rm L}^{\rm T} {\rm d} \te_{\rm L}\\ 
&-&\frac 1 4 \ve^{-1} \te_{\rm R}^{\rm T} {\rm d} \te_{\rm R} \ve^{-1} \te_{\rm R}^{\rm T} {\rm d} \te_{\rm R}~, \\ \non
\cR_{2} &=& \ri \l^{-1}_{\rm L} {\rm d} \l_{\rm L} U_{\rm L} \te_{\rm L} [ x^{-1} {\rm d} x \ve^{-1} - \ve^{-1} {\rm d} x^{\rm T} (x^{-1})^{\rm T} ] \te_{\rm L}^{\rm T} U_{\rm L} - \ri U_{\rm L} {\rm d} \te_{\rm L} \ve^{-1} {\rm d} \te_{\rm L}^{\rm T} U_{\rm L}  \\ \non
&+& \ri U_{\rm L} [ \te_{\rm L} x^{-1} {\rm d} x \ve^{-1} {\rm d} \te_{\rm L}^{\rm T}  + {\rm d} \te_{\rm L} \ve^{-1} {\rm d} x^{\rm T} (x^{-1})^{\rm T} \te_{\rm L}^{\rm T} ]U_{\rm L} - \ri U_{\rm L} \te_{\rm L} x^{-1} {\rm d} x \ve^{-1} {\rm d} x^{\rm T} (x^{-1})^{\rm T} \te_{\rm L}^{\rm T} U_{\rm L} \\ 
&+& \ri \l^{-1}_{\rm L} {\rm d} \l_{\rm L} U_{\rm L} [ \te_{\rm L} \ve^{-1} {\rm d} \te_{\rm L}^{\rm T} - {\rm d} \te_{\rm L} \ve^{-1} \te_{\rm L}^{\rm T} ]U_{\rm L} ~, \\ \non
\cR_{3} &=& - \ri \l^{-1}_{\rm R} {\rm d} \l_{\rm R} U_{\rm R} \te_{\rm R} [ x {\rm d} x^{-1} \ve^{-1} - \ve^{-1} {\rm d} (x^{-1})^{\rm T} x^{\rm T} ] \te_{\rm R}^{\rm T} U_{\rm R} + \ri U_{\rm R} {\rm d} \te_{\rm R} \ve^{-1} {\rm d} \te_{\rm R}^{\rm T} U_{\rm R}  \\ \non
&-& \ri U_{\rm R} [ \te_{\rm R} x {\rm d} x^{-1} \ve^{-1} {\rm d} \te_{\rm R}^{\rm T}  + {\rm d} \te_{\rm R} \ve^{-1} {\rm d} (x^{-1})^{\rm T} x^{\rm T} \te_{\rm R}^{\rm T} ]U_{\rm R} + \ri U_{\rm R} \te_{\rm R} x {\rm d} x^{-1} \ve^{-1} {\rm d} (x^{-1})^{\rm T} x^{\rm T} \te_{\rm R}^{\rm T} U_{\rm R} \\ 
&-& \ri \l^{-1}_{\rm R} {\rm d} \l_{\rm R} U_{\rm R} [ \te_{\rm R} \ve^{-1} {\rm d} \te_{\rm R}^{\rm T} - {\rm d} \te_{\rm R} \ve^{-1} \te_{\rm R}^{\rm T} ]U_{\rm R} ~.
\eea
\end{subequations}

Again we may express both the torsion and the curvature in terms of the vielbein. They read
\bea
\cT_{\rm L} &=& \left(
\begin{array}{c|c}
 \frac{\ri}{2} \ve^{-1} \cE_{\rm L}^{\rm T} \wedge \cE_{\rm L} + \frac{\ri}{2} \ve^{-1} \cE_{\rm R}^{\rm T} \wedge \cE_{\rm R} &  \tilde{E} \wedge \ve^{-1} \cE_{\rm L}^{\rm T} \\
 \hline 
-\ri \cE_{\rm L} \wedge \tilde{E} & 0
\end{array}
\right)~, \\
\cT_{\rm R} 
&=& \left(
\begin{array}{c|c}
-[ \frac{\ri}{2} \ve^{-1} \cE_{\rm L}^{\rm T} \wedge \cE_{\rm L} + \frac{\ri}{2} \ve^{-1} \cE_{\rm R}^{\rm T} \wedge \cE_{\rm R} ]&  \tilde{E} \wedge \ve^{-1} \cE_{\rm R}^{\rm T} \\
 \hline 
 \ri \cE_{\rm R} \wedge \tilde{E} & 0
\end{array}
\right)~, \\
\cR_{\rm L} &=& \left(
\begin{array}{c|c}
\tilde{E} \wedge \tilde{E}  - \frac{\ri}{2} \ve^{-1} \cE_{\rm L}^{\rm T} \wedge \cE_{\rm L} + \frac{\ri}{2} \ve^{-1} \cE_{\rm R}^{\rm T} \wedge \cE_{\rm R} & 0 \\
 \hline 
0 & - \ri \cE_{\rm L} \wedge \ve^{-1} \cE_{\rm L}^{\rm T}
\end{array}
\right)~,\\
\cR_{\rm R} 
&=&  \left(
\begin{array}{c|c}
\tilde{E} \wedge \tilde{E}  - \frac{\ri}{2} \ve^{-1} \cE_{\rm L}^{\rm T} \wedge \cE_{\rm L} + \frac{\ri}{2} \ve^{-1} \cE_{\rm R}^{\rm T} \wedge \cE_{\rm R} & 0 \\
 \hline 
0 &  \ri \cE_{\rm R} \wedge \ve^{-1} \cE_{\rm R}^{\rm T}
\end{array}
\right)~.
\eea

Once again we must now choose a basis for $\cW$ and $\cH$. For this choice it is convenient to use the generators \eqref{newcommutators}, where we have defined $\mathfrak{m}_{ab} =  \mathfrak{m}^{\rm L}_{ab} \oplus - \mathfrak{m}^{\rm R}_{ab}$. We then take $\mathfrak{m}_{ab}$, $Q_{\a \Io}$ and $Q_{\a \Iu}$ as the basis of $\cW$ whilst the basis for $\cH$ stays the same as $\mathcal{M}_{ab}$, $\cN_{\Io \Jo}$ and $\cN_{\Iu \Ju}$.
The (anti-)commutation relations for this basis are \eqref{newcommutators}, only differing from those \eqref{algcommutators} in the commutators
\begin{subequations}
\begin{align}
\left[ \mathfrak{m}_{ab}, \mathfrak{m}_{cd} \right] &= \eta_{ad} \mathcal{M}_{bc} - \eta_{ac} \mathcal{M}_{bd} + \eta_{bc} \mathcal{M}_{ad} - \eta_{bd} \mathcal{M}_{ac}~, \\
[\mathfrak{m}_{ab}, Q_{\a \Io}] &= - (\S_{ab})_{\a}^{\phantom \a \b} Q_{\b \Io}~, \quad [\mathfrak{m}_{ab}, Q_{\a \Iu}] =  (\S_{ab})_{\a}^{\phantom \a \b} Q_{\b \Iu}~.
\end{align}
\end{subequations}

The non-vanishing (dualised) components of the torsion and curvature are
\begin{subequations}
\bea
\cT_{\a \Io \b \Jo}^{\phantom \a \phantom \Io \phantom \b \phantom \Jo a} = \ri \d_{\Io \Jo} (\g^{a})_{\a\b}~, &\quad& \cT_{\a \Iu \b \Ju}^{\phantom b \phantom \Jo \phantom \a \phantom \Io a} = \ri \d_{\Iu \Ju} (\g^{a})_{\a\b}~, \\
\cT_{a \a \Io}^{\phantom a \phantom \a \phantom \Io \b \Jo} = -\frac{1}{2} \d^{\Jo}_{\Io} (\g_{a})_{\a}^{\phantom \a \b}~, &\quad& \cT_{a \a \Iu}^{\phantom a \phantom \a \phantom \Iu \b \Ju} = \frac{1}{2} \d^{\Ju}_{\Iu} (\g_{a})_{\a}^{\phantom \a \b}~,\\
\cR_{ab}^{\phantom a \phantom b c} &=& - \ve_{ab}^{\phantom a \phantom b c}~, \\
\cR_{\a \Io \b \Jo}^{\phantom \a \phantom \Io \phantom \b \phantom \Jo a} = \ri \d_{\Io \Jo} (\g^{a})_{\a\b}~, &\quad& \cR_{\a \Iu \b \Ju}^{\phantom \b \phantom \Ju \phantom \a \phantom \Iu a} = -\ri \d_{\Iu \Ju} (\g^{a})_{\a\b}~, \\
\cR_{\a \overline{I} \b \overline{J}}^{\phantom \a \phantom {\overline{I}} \phantom \b \phantom {\overline{J}} \overline{M} \overline{N}} = -2 \ri \ve_{\a\b} \d_{[\overline{I}}^{\overline{M}} \d_{\overline{J}]}^{\overline{N}} ~, &\quad& \cR_{\a \underline{I} \b \underline{J}}^{\phantom \a \phantom {\underline{I}} \phantom \b \phantom {\underline{J}} \underline{M} \underline{N}} = 2 \ri \ve_{\a\b} \d_{[\underline{I}}^{\underline{M}} \d_{\underline{J}]}^{\underline{N}}~,
\eea
\end{subequations}
from which we construct the graded commutation relations
\begin{subequations}  \label{derivcomm}
\bea
\big[ \cD_{a}, \cD_{b} \big] &=& \ve_{a b c} \cM^{c}~, \\
\big[ \cD_{a}, {\cD}_{\a \Io} \big] &=& - \frac 12 (\g_{a})_{\a}^{\phantom \a \b} {\cD}_{\b \Io}~, \\
\big[ \cD_{a}, {\cD}_{\a \Iu} \big] &=& \frac 12 (\g_{a})_{\a}^{\phantom \a \b} {\cD}_{\b \Iu}~, \\
\big\{ {\cD}_{\a \Io}, {\cD}_{\b \Ju} \big\} &=& 0~, \\
\big\{ {\cD}_{\a \Io}, {\cD}_{\b \Jo} \big\}  &=&  -\ri \d_{\Io \Jo} \cD_{\a\b} 
- \ri \d_{\Io \Jo} \cM_{\a\b} -  \ri \ve_{\a\b} \cN_{\Io\Jo}~, \\
\big\{ {\cD}_{\a \Iu}, {\cD}_{\b \Ju} \big\}  &=&  -\ri \d_{\Iu \Ju} \cD_{\a\b} 
+ \ri \d_{\Iu \Ju} \cM_{\a\b} +  \ri \ve_{\a\b} \cN_{\Iu\Ju}~.
\eea
\end{subequations}
Thus, we observe that the choice of algebra \eqref{algcompform2} corresponds to covariant derivatives that are already torsion-free in the vector commutator. After a rescaling of the vector derivative these (anti-)commutation relations can be seen to agree with those \eqref{4.30}, corresponding to $S = -\frac 14$, up to an overall negative sign in the anti-commutators of spinor derivatives.


\section{Poincar\'e coordinates for ${\rm AdS}^{(3|p,q)}$} \label{Section6}

Poincar\'e coordinates $(z, x^a)$ for AdS$_d$ 
(with $a=0, 1, \dots, d-2$)
are used in many applications including the AdS/CFT duality.
They are naturally defined in terms of the 
embedding coordinates $X^{\hat a} $,
eq. \eqref{1.1b},
on the open subset of AdS$_d$ where, say,  $z^{-1} := X^{d-1} + X^d >0$,
\bea
X^{\hat a} = (X^a, X^{d-1}, X^d) = \frac{1}{z} \Big( x^a\,, \frac{1-x^2 - (\ell z)^2 }{2}\, ,
 \frac{1+x^2 + (\ell z)^2 }{2}\Big)~, 
\eea
where $x^2 =\eta_{ab} x^a x^b$ and $\eta_{ab} = {\rm diag} (-1, 1, \dots, 1) $ is the  metric on Minkowski space ${\mathbb M}^{d-1}$. In the Poincar\'e  chart, AdS$_d$  is foliated
into a union of Minkowski spaces ${\mathbb M}^{d-1}$.


\subsection{Poincar\'e coordinate patch} 

The freedom to perform equivalence transformations \eqref{ER2} may be used in such a way 
that the two-planes are parametrised in terms of local coordinates. Specifically, we would like to make a choice corresponding to Poincaré coordinates. Motivated by the non-supersymmetric case, we require that our normalised two-planes are upper triangular (lower triangular) in their bosonic part in the left (right) sector. The remaining freedom may be used to equate the bottom right component of the bosonic part of the left two-plane with the top left component of the bosonic part of the right two-plane. With these conditions we obtain the following form for our two-planes in Poincaré coordinates
\begin{subequations}\label{poincare}
\bea 
\cP_{\rm L} &=&  \frac{1}{\sqrt{z}} \left(
\begin{array}{cc}
 z + \frac{\ri}{2} \te_{\Io}{}^{\a} \te_{ \Io \a} ~&~ - u^{\pp} \\
0 ~&~ 1  \\
\hline
 \ri \te_{\Io}{}^{1} & \ri \te_{\Io}{}^{2}
\end{array}
\right)
= \frac{1}{\sqrt{z}} \left(
\begin{array}{cc}
 z + \ri \te_{\Io}^+ \te_{ \Io}^- ~&~ - u^{\pp} \\
0 ~&~ 1  \\
\hline
 \ri \te_{\Io}^- & \ri \te_{\Io}^+
\end{array}
\right)
~, \\
\cP_{\rm R} &=&  \frac{1}{\sqrt{z}} \left(
\begin{array}{cc}
1 ~& 0 ~\\
- u^{=} ~& ~ z - \frac{\ri}{2} \te_{\Iu}{}^{\a} \te_{ \Iu \a} \phantom{\Big|} \\
\hline
 \ri \te_{\Iu}{}^{1} & \ri \te_{\Iu}{}^{2}
 \end{array}
\right)
 = \frac{1}{\sqrt{z}} \left(
\begin{array}{cc}
1 ~& 0 ~\\
- u^{=} ~& ~ \phantom{\Big|}
z -\ri \te_{\Iu}^+ \te_{ \Iu}^-  \\
\hline
 \ri \te_{\Iu}^- & \ri \te_{\Iu}^+
\end{array}
\right)~.
\eea
\end{subequations}
Corresponding to these two-planes are the following coset representatives:
\begin{subequations}
\begin{align}
S_{\rm L} &= \left(
\begin{array}{cc|c}
\l_{\rm L}^{2} \sqrt{z} ~&~ -\frac{1}{\sqrt{z}} u^{\pp} ~& \l_{\rm L} \te^+_{\Jo} + 
(z \l_{\rm L})^{-1}
u^{\pp} \te^-_{ \Jo}   \\
0 ~& \frac{1}{\sqrt{z}} ~&  -
(z \l_{\rm L})^{-1}
\te^-_{ \Jo} \phantom{\Big|}
\\ \hline
\frac{\ri}{\sqrt{z}} \te^{-}_{\Io} ~& \frac{\ri}{\sqrt{z}} \te^{+}_{\Io} ~& (U_{\rm L})_{\Io \Jo}
\end{array}
\right)~, 
\quad  U_{\rm L} = \Big( {\mathbbm 1}_p +\ri \frac{ \q_{\rm L} \ve^{-1} \q_{\rm L}^{\rm T} }{\l_{\rm L}^{2} } \Big)^{-\hf}~, \\
S_{\rm R} &= \left(
\begin{array}{cc|c}
\frac{1}{\sqrt{z}} &~ 0 ~& - 
(z \l_{\rm R})^{-1}
\te^{+}_{ \Ju} \\
-\frac{1}{\sqrt{z}} u^{=} ~& ~\l_{\rm R}^{2} \sqrt{z} ~& 
\l_{\rm R} \te^{-}_{ \Ju} + 
(z \l_{\rm R})^{-1}
u^{=} \te^{+}_{ \Ju} \phantom{\Big|}\\ \hline
 \frac{\ri}{\sqrt{z}} \te^{-}_{\Iu} ~& \frac{\ri}{\sqrt{z}} \te^{+}_{\Iu} ~& (U_{\rm R})_{\Iu \Ju}
\end{array}
\right)~, \quad U_{\rm R} = \Big( {\mathbbm 1}_q -\ri \frac{ \q_{\rm R} \ve^{-1} \q_{\rm R}^{\rm T} }{\l_{\rm R}^{2} } \Big)^{-\hf}~,
\end{align}
\end{subequations}
where we have defined
\bea
\l_{\rm L}^{2} = 1 + \frac{\ri}{2z} \te^{\phantom{\Io} \a}_{\Io} \te_{\Io \a}~, \quad \l_{\rm R}^{2} = 1 - \frac{\ri}{2z} \te^{\phantom{\Iu} \a}_{\Iu} \te_{\Iu \a}~.
\eea


\subsection{Isometry transformations}

The bosonic $(u^{\pp}, u^=)$ and fermionic $(\te_{\Io}^+,\te_{\Iu}^-) $ variables 
may be identified with the coordinates of a two-dimensional $(p,q)$ Minkowski superspace ${\mathbb M}^{(2|p,q)}$. This interpretation is supported by the fact that the transformations from the AdS supergroup \eqref{isometry} act on ${\mathbb M}^{(2|p,q)}$ as two-dimensional  superconformal transformations in the limit 
\bea
z =0~, \quad \te_{\Io}^-=0~, \quad \te_{\Iu}^+=0~.
\label{AdSboundary}
\eea

A Lorentz transformation corresponds to the group element
\bea
g^{\rm (Lor)}_{\rm L} (\L)  = \left(
\begin{array}{cc|c}
\L ~&~ 0 ~& 0   \\
0 ~& ~\L^{-1} ~&  0
\\ \hline
0 ~& 0 ~& {\mathbbm 1}_p
\end{array}
\right)~, \qquad 
g^{\rm (Lor)}_{\rm R} (\L)  = \left(
\begin{array}{cc|c}
\L ~&~ 0 ~& 0   \\
0 ~&~ \L^{-1} ~&  0
\\ \hline
0 ~& 0 ~& {\mathbbm 1}_q
\end{array}
\right)~.
\eea
Its action on \eqref{poincare} is given by 
\begin{subequations}
\bea
(u^{\pp})' &=& \L^2 u^{\pp} ~, \qquad (\te_{\Io}^+)' = \L \te_{\Io}^+~,\\
(u^{=})' &=& \L^{-2} u^{=} ~, \qquad (\te_{\Io}^-)' = \L^{-1} \te_{\Io}^-~,\\
z'&=&z~, \quad  (\te_{\Io}^-)' = \L^{-1} \te_{\Io}^-~, \quad (\te_{\Io}^+)' = \L \te_{\Io}^+~.
\eea
\end{subequations}

A scale/dilatation  transformation corresponds to the group element
\bea
g^{\rm (dil)}_{\rm L} (\z)  = \left(
\begin{array}{cc|c}
\z ~&~ 0 ~& 0   \\
0 ~& ~\z^{-1} ~&  0
\\ \hline
0 ~& 0 ~& {\mathbbm 1}_p
\end{array}
\right)~, \qquad 
g^{\rm (dil)}_{\rm R} (\z)  = \left(
\begin{array}{cc|c}
\z^{-1} ~&~ 0 ~& 0   \\
0 ~& ~\z ~&  0
\\ \hline
0 ~& 0 ~& {\mathbbm 1}_q
\end{array}
\right)~.
\eea
Its action on \eqref{poincare} is given by 
\begin{subequations}
\bea
(u^{\pp})' &=& \z^2 u^{\pp} ~, \qquad (\te_{\Io}^+)' = \z \te_{\Io}^+~,\\
(u^{=})' &=& \z^{2} u^{=} ~, \qquad (\te_{\Io}^-)' = \z \te_{\Io}^-~,\\
z'&=&\z^2 z~, \quad  (\te_{\Io}^-)' = \z \te_{\Io}^-~, \quad (\te_{\Io}^+)' = \z \te_{\Io}^+~.
\eea
\end{subequations}

Spacetime translations correspond to  group elements of the form
\bea
g^{\rm (P)}_{\rm L} (a)  = \left(
\begin{array}{cc|c}
1 ~& -a^{\pp} ~& 0   \\
0 ~& ~1 ~&  0
\\ \hline
0 ~& 0 ~& {\mathbbm 1}_p
\end{array}
\right)~, \qquad 
g^{\rm (P)}_{\rm R} (a)  = \left(
\begin{array}{cc|c}
1 ~&~ 0 ~& 0   \\
-a^= &~ 1 ~&  0
\\ \hline
0 ~& 0 ~& {\mathbbm 1}_q
\end{array}
\right)~.
\eea
They  act on \eqref{poincare} as
\bea
(u^{\pp})' =  u^{\pp} +a^{\pp} ~, \qquad 
(u^{=})' =   u^{=} + a^=~,
\eea
and the other coordinates remain unchanged. 

Let us turn to special conformal transformations. 
We consider the special conformal transformation generated by a parameter $b^=$
\bea
g^{\rm (SC)}_{\rm L} (b^=)  = \left(
\begin{array}{cc|c}
1 ~&  ~0~& 0   \\
-b^= & ~1 ~&  0
\\ \hline
0 ~& 0 ~& {\mathbbm 1}_p
\end{array}
\right)~, \qquad 
g^{\rm (P)}_{\rm R} (b^=)  =  {\mathbbm 1}_{2+q}~.
\eea
It acts as follows
\begin{subequations}
\bea
(u^{\pp})' &=& \frac{ u^{\pp} }{1+b^= u^{\pp}} ~, \qquad 
(\te_{\Io}^+)' = \frac{ \te_{\Io}^+}{1+b^= u^{\pp}}  ~,\\
(u^{=})' &=& u^{=} - \frac{b^= (z \l_{\rm L}\l_{\rm R} )^2}{1+b^= u^{\pp}} 
~, \qquad (\te_{\Iu}^-)' =  \te_{\Iu}^- + 
\frac{\te_{\Iu}^+   b^= z\l_{\rm L}^2}{1+b^= u^{\pp}} 
~,\\
z'&=& \frac{z}{1+b^= u^{\pp}} ~, \quad  
(\te_{\Io}^-)' =  \te_{\Io}^- +  \frac{ \te_{\Io}^+ b^= z \l_{\rm L}^2} {1+b^= u^{\pp}} 
~, \quad (\te_{\Iu}^+)' = \frac{ \te_{\Iu}^+}{1+b^= u^{\pp}} ~.
\eea
\end{subequations}
The special conformal transformation generated by a parameter $b^{\pp}$ is
\bea
g^{\rm (P)}_{\rm L} (b^{\pp})  =  {\mathbbm 1}_{2+p}~, \qquad 
g^{\rm (SC)}_{\rm R} (b^{\pp})  = \left(
\begin{array}{cc|c}
1 ~&  -b^{\pp}~& 0   \\
0 ~& ~1 ~&  0
\\ \hline
0 ~& 0 ~& {\mathbbm 1}_p
\end{array}
\right)~.
\eea
It acts as follows
\begin{subequations}
\bea
(u^{\pp})' &=& u^{\pp} - \frac{b^{\pp} (z \l_{\rm L}\l_{\rm R} )^2}{1+b^{\pp} u^{=}} 
~, \qquad (\te_{\Io}^+)' =  \te_{\Io}^+ + 
\frac{\te_{\Io}^-  b^{\pp} z\l_{\rm R}^2}{1+b^{\pp} u^{=}} 
~,\\
(u^{=})' &=& \frac{ u^{=} }{1+b^{\pp} u^{=}} ~, \qquad 
(\te_{\Iu}^-)' = \frac{ \te_{\Iu}^-}{1+b^{\pp} u^{=}}  ~,\\
z'&=& \frac{z}{1+b^{\pp} u^{=}} ~, \quad  
(\te_{\Io}^-)' =  \frac{\te_{\Io}^-}  {1+b^{\pp} u^{=}} 
~, \quad (\te_{\Iu}^+)' =  \te_{\Iu}^+ +
\frac{ \te_{\Iu}^- b^{\pp} z\l_{\rm R}^2} {1+b^{\pp} u^{=}} ~.
\eea
\end{subequations}

It remains to consider $Q$ and $S$-supersymmetry transformations. 
A $Q$-supersymmetry transformation is described by group elements of the form
\bea
g_{\rm L}^{(Q)} (\e) = \left(
\begin{array}{cc|c}
1 ~&~ 0 ~& \e^+_{\Jo}    \\
0 ~& ~1 ~&  0
\\ \hline
0 ~& {\ri} \e^{+}_{\Io} ~& \d_{\Io \Jo}
\end{array}
\right)~, \qquad
g_{\rm R}^{(Q)} (\e) = \left(
\begin{array}{cc|c}
1 ~&~ 0 ~&  0  \\
0 ~& ~1 ~&  \e^-_{\Ju} 
\\ \hline
{\ri} \e^{-}_{\Iu} ~& 0 ~& \d_{\Iu \Ju}
\end{array}
\right)~, 
\eea
It acts as follows
\begin{subequations}
\bea
(u^{\pp})' &=& u^{\pp} - \ri \e_{\Io}^+ \te_{\Io}^+
~, \qquad 
(\te_{\Io}^+)' =  \te_{\Io}^+ + \e_{\Io}^+
~,\\
(u^{=})' &=&  u^{=} -\ri  \e_{\Iu}^- \te_{\Iu}^- ~, \qquad 
(\te_{\Iu}^-)' =  \te_{\Iu}^- + \e_{\Iu}^-  ~,
\\
z'&=& {z} ~, \quad  
(\te_{\Io}^-)' =  \te_{\Io}^-
~, \quad (\te_{\Iu}^+)' =  \te_{\Iu}^+   ~.
\eea
\end{subequations}
These imply the two-dimensional spinor covariant derivatives
\bea
D_{+ \Io} = \frac{\pa}{\pa \q^+_\Io} + \ri \q^+_{\Io} \frac{\pa}{\pa u^{\pp}}~,
\qquad 
D_{- \Iu } = \frac{\pa}{\pa \q^-_{\Iu} } + \ri \q^-_{\Iu} \frac{\pa }{\pa u^= }~.
\eea
They obey the anti-commutation relations
\bea
\{ D_{+ \Io} , D_{+ \Jo} \} = 2\ri \d_{\Io \Jo} \pa_{\pp} ~,\qquad 
\{ D_{- \Iu} , D_{- \Ju} \} = 2\ri \d_{\Iu \Ju} \pa_{=} ~,
\eea
which correspond to the $(p,q)$ Poincar\'e supersymmetry in two dimensions. 

Finally, the $S$-supersymmetry transformation 
corresponding to a  parameter $\eta_{\Io}^-$ is 
\bea
g_{\rm L}^{(S)} (\eta_{\Io}^-) = \left(
\begin{array}{cc|c}
1 ~&~ 0 ~&  0  \\
0 ~& ~1 ~&-  \eta^-_{\Jo} 
\\ \hline
{\ri} \eta^{-}_{\Io}  ~& 0~& \d_{\Io \Jo}
\end{array}
\right)~, \qquad
g_{\rm R}^{(S)} (\eta_{\Io}^-) ={\mathbbm 1}_{2+q}~.
\eea
It acts as follows
\begin{subequations}
\bea
(u^{\pp})' &=& \frac{ u^{\pp} }{1 - \ri \eta^- \cdot \q^+} ~, \qquad 
(\te_{\Io}^+)' = \frac{ \te_{\Io}^+ - \eta^-_{\Io} u^{\pp}}{1 - \ri \eta^- \cdot \q^+}  ~,\\
(u^{=})' &=& u^{=} - \ri  \frac{\eta^-_{\Io} \q^-_{\Io} z \l_{\rm R} ^2}
{1 - \ri \eta^- \cdot \q^+}
~, \qquad 
(\te_{\Iu}^-)' =  \te_{\Iu}^- 
+ \ri \frac{\te_{\Iu}^+ \eta^-_{\Jo} \q^-_{\Jo} }
{1 - \ri \eta^- \cdot \q^+}
~,\\
z'&=& \frac{z}{1 - \ri \eta^- \cdot \q^+} ~, \quad  
(\te_{\Io}^-)' =  \te_{\Io}^- +  \eta^-_{\Io} z \l_{\rm L}^2 
+ \ri \frac{ (\q^+_{\Io} - \eta^-_{\Io} u^{\pp} ) \eta_{\Jo}^- \q^-_{\Jo} } 
{1 - \ri \eta^- \cdot \q^+}
~, \\
 (\te_{\Iu}^+)' &=& \frac{ \te_{\Iu}^+}{1 - \ri \eta^- \cdot \q^+} ~,
\eea
\end{subequations}
where we have denoted $ \eta^- \cdot \q^+ = \eta^-_{\Io} \q^+_{\Io}$.
The $S$-supersymmetry transformation 
corresponding to a  parameter $\eta_{\Iu}^+$ is 
\bea
g_{\rm L}^{(S)} (\eta_{\Iu}^+) ={\mathbbm 1}_{2+p}~, \qquad
g_{\rm R}^{(S)} (\eta_{\Iu}^+) = \left(
\begin{array}{cc|c}
1 ~&~ 0 ~&  -  \eta^+_{\Ju}  \\
0 ~& ~1 ~&  0
\\ \hline
0  ~&~ {\ri} \eta^{+}_{\Iu}~& \d_{\Iu \Ju}
\end{array}
\right)
~.
\eea
It acts as follows
\begin{subequations}
\bea
(u^{\pp})' &=&  u^{\pp} 
- \ri \frac{ \eta^+_{\Iu} \q^+_{\Iu} z \l_{\rm L}^2 }
{1 - \ri \eta^+ \cdot \q^-}  ~, \qquad 
(\te_{\Io}^+)' =  \te_{\Io}^+ 
+\ri \frac{ \q^-_{\Io} \eta^+_{\Ju} \q^+_{\Ju} }{1 - \ri \eta^+ \cdot \q^-}  ~,\\
(u^{=})' &=& \frac{u^{=} }{1 - \ri \eta^+ \cdot \q^-}
~, \qquad 
(\te_{\Iu}^-)' = \frac{ \te_{\Iu}^-  - \eta^+_{\Iu} u^=}{1 - \ri \eta^+ \cdot \q^-} 
~,\\
z'&=& \frac{z}{1 - \ri \eta^+ \cdot \q^-}
~, \quad  
(\te_{\Io}^-)' = \frac{ \te_{\Io}^- }{1 - \ri \eta^+ \cdot \q^-} 
 ~, \\
 (\te_{\Iu}^+)' &=&  \te_{\Iu}^+ +\eta^+_{\Iu} z \l_{\rm R}^2 
 +\ri \frac{ (\q^-_{\Iu} - \eta^+_{\Iu} u^{=}) \eta^+_{\Ju} \q^+_{\Ju}  }{1 - \ri \eta^+ \cdot \q^-}
  ~,
\eea
\end{subequations}
where we have denoted $\eta^+ \cdot  \q^- = \eta^+_{\Iu} \q^-_{\Iu}$

The AdS isometry transformations, which we have described above, have a well defined limit to the boundary of $(p,q)$ AdS superspace, eq. \eqref{AdSboundary}. 


\subsection{Superspace geometry}

The Maurer-Cartan one-form $\o = (\o_{\rm L}, \o_{\rm R})$ is then
\begin{subequations}
\begin{align}
\o_{\rm L} &=  \left(
\begin{array}{c:c|c}
\frac{1}{2z} {\rm d} z - \frac{\ri}{z} \te_{\Io}^{-} {\rm d} \te_{\Io}^{+} ~& -\frac{1}{z} {\rm d} u^{\pp} - \frac{\ri}{z} \te_{\Io}^{+} {\rm d} \te_{\Io}^{+}  ~& \frac{1}{z^{3/2}} \l_{\rm L}^{-1}  {\rm d} u^{\pp} \te^{-}_{\Io} + \frac{1}{\sqrt{z}} (U_{\rm L})_{\Io \Jo} {\rm d} \te_{\Jo}^{+}  \\ \hdashline
 \frac{\ri}{z} \te_{\Io}^{-} {\rm d} \te_{\Io}^{-} ~& -\frac{1}{2z} {\rm d} z + \frac{\ri}{z} \te_{\Io}^{-} {\rm d} \te_{\Io}^{+}  ~& \begin{aligned} &\tfrac{1}{z^{3/2}} \l_{\rm L} {\rm d} z \te^{-}_{\Io} + \tfrac{1}{\sqrt{z}} \l_{\rm L}^{-1} {\rm d} \l_{\rm L}^{2} \te^{-}_{\Io} \\ &\qquad \quad- \tfrac{1}{\sqrt{z}} (U_{\rm L})_{\Io \Jo} {\rm d} \te_{\Jo}^{-} \end{aligned} \\ \hline
\begin{aligned} &- \tfrac{\ri}{z^{3/2}} \l_{\rm L} {\rm d} z \te_{\Io}^{-} \\ &-\tfrac{\ri}{\sqrt{z}} \l_{\rm L}^{-1} {\rm d} \l_{\rm L}^{2} \te_{\Io}^{-} \\ &+ \tfrac{\ri}{\sqrt{z}} (U_{\rm L})_{\Io \Jo} {\rm d} \te_{\Jo}^{-} \end{aligned} ~&  \begin{aligned}&\tfrac{\ri}{z^{3/2}} \l_{\rm L}^{-1} {\rm d} u^{\pp} \te_{\Io}^{-} \\ &+\tfrac{\ri}{\sqrt{z}} (U_{\rm L})_{\Io \Jo} {\rm d} \te_{\Jo}^{+} \end{aligned} ~& \begin{aligned}&\quad- \tfrac{\ri}{z}\l_{\rm L}^{-1} {\rm d} \l_{\rm L} [ \te_{\Io}^{-} \te^{+}_{\Jo} + \te_{\Io}^{+} \te^{-}_{\Jo} ] \\ &+ \tfrac{\ri}{z} \te_{\Io}{}^{\a} {\rm d} \te_{\Jo \a} - \tfrac{\ri}{z^{2}} \l_{\rm L}^{-2} {\rm d} u^{\pp} \te_{\Io}^{-} \te^{-}_{\Jo} \\ &- \tfrac{\ri}{z^{2}} {\rm d} z \te_{\Io}^{+} \te^{-}_{\Jo} + (U_{\rm L})_{\Io \overline{K}} ({\rm d} U_{\rm L})_{\overline{K} \Jo} \end{aligned}
\end{array}
\right)~,\\
\o_{\rm R} &= \left(
\begin{array}{c:c|c}
-\frac{1}{2z} {\rm d} z + \frac{\ri}{z} \te_{\Iu}^{+} {\rm d} \te_{\Iu}^{-} ~&  \frac{\ri}{z} \te_{\Iu}^{+} {\rm d} \te_{\Iu}^{+} ~& \begin{aligned} &\tfrac{1}{z^{3/2}} \l_{\rm R} {\rm d} z \te^{+}_{\Iu} + \tfrac{1}{\sqrt{z}} \l_{\rm R}^{-1} {\rm d} \l_{\rm R}^{2} \te^{+}_{\Iu} \\ &\qquad \quad- \tfrac{1}{\sqrt{z}} (U_{\rm R})_{\Iu \Ju} {\rm d} \te_{\Ju}^{+} \end{aligned} \\ \hdashline
-\frac{1}{z} {\rm d} u^{=} - \frac{\ri}{z} \te_{\Iu}^{-} {\rm d} \te_{\Iu}^{-} ~& \frac{1}{2z} {\rm d} z - \frac{\ri}{z} \te_{\Iu}^{+} {\rm d} \te_{\Iu}^{-}  ~&  \frac{1}{z^{3/2}} \l_{\rm R}^{-1} {\rm d} u^{=} \te^{+}_{\Iu} + \frac{1}{\sqrt{z}} (U_{\rm R})_{\Iu \Ju} {\rm d} \te_{\Ju}^{-} \\ \hline
\begin{aligned}&\tfrac{\ri}{z^{3/2}} \l_{\rm R}^{-1} {\rm d} u^{=} \te_{\Iu}^{+} \\&+ \tfrac{\ri}{\sqrt{z}} (U_{\rm R})_{\Iu \Ju} {\rm d} \te_{\Ju}^{-} \end{aligned} ~& \begin{aligned} &-\tfrac{\ri}{z^{3/2}} \l_{\rm R} {\rm d} z \te_{\Iu}^{+} \\&- \tfrac{\ri}{\sqrt{z}} \l_{\rm R}^{-1} {\rm d} \l_{\rm R}^{2}  \te_{\Iu}^{+} \\&+ \tfrac{\ri}{\sqrt{z}} (U_{\rm R})_{\Iu \Ju} {\rm d} \te_{\Ju}^{+} \end{aligned} ~&  \begin{aligned} &\quad -\tfrac{\ri}{z} \l_{\rm R}^{-1} {\rm d} \l_{\rm R} [ \te_{\Iu}^{-} \te^{+}_{\Ju} + \te_{\Iu}^{+} \te^{-}_{\Ju} ]  \\&-  \tfrac{\ri}{z} \te_{\Iu}{}^{\a} {\rm d} \te_{\Ju \a} - \tfrac{\ri}{z^{2}} \l_{\rm R}^{-2} {\rm d} u^{=} \te_{\Iu}^{+} \te^{+}_{\Ju} \\&- \tfrac{\ri}{z^{2}} {\rm d} z \te_{\Iu}^{-} \te^{+}_{\Ju} + (U_{\rm R})_{\Iu \underline{K}} ({\rm d} U_{\rm R})_{\underline{K} \Ju} \end{aligned}
\end{array}
\right)~,
\end{align}
\end{subequations}
We choose to decompose $\o$ into the vielbein and connection having the previously discussed forms \eqref{vielbein2} and \eqref{connection2} respectively. For the vielbein $E = (E_{\rm L}, E_{\rm R})$ we obtain
\begin{subequations}
\bea 
E_{\rm L} 
= \left(
\begin{array}{c|c}
\tilde{E} & - \ve^{-1} \cE_{\rm L}^{\rm T} \\
 \hline 
\ri \cE_{\rm L}  & 0
\end{array}
\right)~, \quad 
E_{\rm R} 
&=& \left(
\begin{array}{c|c}
-\tilde{E} &  \ve^{-1} \cE_{\rm R}^{\rm T} \\
 \hline 
\ri \cE_{\rm R}  & 0
\end{array}
\right)~,
\eea
where
\begin{align}
\tilde{E} &= \left(
\begin{array}{cc}
\frac{1}{2z} \big( 
{\rm d} z - 
\ri \te_{\Iu}^{+} {\rm d} \te_{\Iu}^{-} - 
 \ri \te_{\Io}^{-} {\rm d} \te_{\Io}^{+} \big)~& 
 -\frac{1}{2z} \big( {\rm d} u^{\pp} 
 +\ri 
 \te_{\Io}^{+} {\rm d} \te_{\Io}^{+} 
 +\ri 
  \te_{\Iu}^{+} {\rm d} \te_{\Iu}^{+}\big) \\
\frac{1}{2z} \big(
{\rm d} u^{=} + 
\ri \te_{\Io}^{-} {\rm d} \te_{\Io}^{-} + 
\ri \te_{\Iu}^{-} {\rm d} \te_{\Iu}^{-} \big)~& 
-\frac{1}{2z} \big( {\rm d} z 
-\ri
\te_{\Iu}^{+} {\rm d} \te_{\Iu}^{-} 
-\ri 
\te_{\Io}^{-} {\rm d} \te_{\Io}^{+}\big)
\end{array}
\right)~, \\
\cE_{\rm L} &= \left(
\begin{array}{cc}
\begin{aligned} &- \tfrac{1}{z^{3/2}} \l_{\rm L} {\rm d} z \te_{\Io}^{-} - \tfrac{1}{\sqrt{z}} \l_{\rm L}^{-1} {\rm d} \l_{\rm L}^{2}  \te_{\Io}^{-} \\ &\qquad \quad + \tfrac{1}{\sqrt{z}} (U_{\rm L})_{\Io \Jo} {\rm d} \te_{\Jo}^{-} \end{aligned} ~& \frac{1}{z^{3/2}} \l_{\rm L}^{-1} {\rm d} u^{\pp} \te_{\Io}^{-} + \frac{1}{\sqrt{z}} (U_{\rm L})_{\Io \Jo} {\rm d} \te_{\Jo}^{+}
\end{array}
\right)~,\\
\cE_{\rm R} &= \left(
\begin{array}{cc}
 \frac{1}{z^{3/2}} \l_{\rm R}^{-1} {\rm d} u^{=} \te_{\Iu}^{+} + \frac{1}{\sqrt{z}} (U_{\rm R})_{\Iu \Ju} {\rm d} \te_{\Ju}^{-} ~& \begin{aligned} &- \tfrac{1}{z^{3/2}} \l_{\rm R}  {\rm d} z \te_{\Iu}^{+} - \tfrac{1}{\sqrt{z}} \l_{\rm R}^{-1} {\rm d} \l_{\rm R}^{2} \te_{\Iu}^{+} \\ &\qquad \quad + \tfrac{1}{\sqrt{z}} (U_{\rm R})_{\Iu \Ju} {\rm d} \te_{\Ju}^{+} \end{aligned} 
\end{array}
\right)~,
\end{align}
\end{subequations}
whilst for the connection $\O = (\O_{\rm L}, \O_{\rm R})$ we obtain
\begin{subequations}
\bea
\O_{\rm L} 
= \left(
\begin{array}{c|c}
\tilde{\O} & 0 \\
 \hline 
0  & \O_{\sSO(p)}
\end{array}
\right)~, \quad 
\O_{\rm R} 
&=& \left(
\begin{array}{c|c}
\tilde{\O} & 0 \\
 \hline 
0 & \O_{\sSO(q)}
\end{array}
\right)~,
\eea
where $\tilde{\O}$ is given by
\begin{align}
\tilde{\O} &= \left(
\begin{array}{cc}
\frac{\ri}{2z} \big( \te_{\Iu}^{+} {\rm d} \te_{\Iu}^{-} - \te_{\Io}^{-} {\rm d} \te_{\Io}^{+} \big)~& -\frac{1}{2z} \big( {\rm d} u^{\pp} +\ri \te_{\Io}^{+} {\rm d} \te_{\Io}^{+} -\ri \te_{\Iu}^{+} {\rm d} \te_{\Iu}^{+} \big) \\
-\frac{1}{2z} \big( {\rm d} u^{=} -\ri \te_{\Io}^{-} {\rm d} \te_{\Io}^{-} +\ri \te_{\Iu}^{-} {\rm d} \te_{\Iu}^{-} \big) ~&  -\frac{\ri}{2z} \big( \te_{\Iu}^{+} {\rm d} \te_{\Iu}^{-} - \te_{\Io}^{-} {\rm d} \te_{\Io}^{+} \big)
\end{array}
\right)~,
\end{align}
and $\O_{\sSO(p)}$, $\O_{\sSO(q)}$ are
\bea \non
\O_{\sSO(p)} &=& -\frac{\ri}{z}\l_{\rm L}^{-1} {\rm d} \l_{\rm L} [ \te_{\Io}^{-} \te^{+}_{\Jo} + \te_{\Io}^{+} \te^{-}_{\Jo} ] + \frac{\ri}{z} \te_{\Io}{}^{\a} {\rm d} \te_{\Jo \a} - \frac{\ri}{z^{2}} \l_{\rm L}^{-2} {\rm d} u^{\pp} \te_{\Io}^{-} \te^{-}_{\Jo} \\
&-& \frac{\ri}{z^{2}} {\rm d} z \te_{\Io}^{+} \te^{-}_{\Jo} + (U_{\rm L})_{\Io \overline{K}} ({\rm d} U_{\rm L})_{\overline{K} \Jo} ~, \\ \non
\O_{\sSO(q)} &=& -\frac{\ri}{z} \l_{\rm R}^{-1} {\rm d} \l_{\rm R} [ \te_{\Iu}^{-} \te^{+}_{\Ju} + \te_{\Iu}^{+} \te^{-}_{\Ju} ] -  \frac{\ri}{z} \te_{\Iu}{}^{\a} {\rm d} \te_{\Ju \a} - \frac{\ri}{z^{2}} \l_{\rm R}^{-2} {\rm d} u^{=} \te_{\Iu}^{+} \te^{+}_{\Ju} \\ 
&-& \frac{\ri}{z^{2}} {\rm d} z \te_{\Iu}^{-} \te^{+}_{\Ju} + (U_{\rm R})_{\Iu \underline{K}} ({\rm d} U_{\rm R})_{\underline{K} \Ju} ~.
\eea
\end{subequations}

From here we can calculate the torsion and curvature. The torsion $\cT = (\cT_{\rm L}, \cT_{\rm R})$ is calculated as
\begin{subequations}
\bea
\cT_{\rm L} = \left(
\begin{array}{c|c}
\cT_{1} & - \ve^{-1} \cT_{2}^{\rm T} \\
 \hline 
\ri \cT_{2} & 0
\end{array}
\right)~, \quad 
\cT_{\rm R} 
&=& \left(
\begin{array}{c|c}
-\cT_{1} & \ve^{-1} \cT_{3}^{\rm T} \\
 \hline 
\ri \cT_{3} & 0
\end{array}
\right)~,
\eea
where
\begin{align}
\cT_{1} &= -\left(
\begin{array}{cc}
\begin{aligned} &\tfrac{1}{2z^{2}} \big( \ri \te_{\Iu}^{+} {\rm d} \te_{\Iu}^{-} {\rm d} z + \ri \te_{\Io}^{-} {\rm d} \te_{\Io}^{+} {\rm d} z \\ &- \ri \te_{\Iu}^{+} {\rm d} \te_{\Iu}^{+} {\rm d} u^{=} + \ri z {\rm d} \te_{\Iu}^{+} {\rm d} \te_{\Iu}^{-}  \\ &- \ri \te_{\Io}^{-} {\rm d} \te_{\Io}^{-} {\rm d} u^{\pp} + \ri z {\rm d} \te_{\Io}^{+} {\rm d} \te_{\Io}^{-} \\ &+ \te_{\Iu}^{+} {\rm d} \te_{\Iu}^{+} \te_{\Ju}^{-} {\rm d} \te_{\Ju}^{-} + \te_{\Io}^{-} {\rm d} \te_{\Io}^{-} \te_{\Jo}^{+} {\rm d} \te_{\Jo}^{+}\big) \end{aligned} ~& \begin{aligned} &- \tfrac{1}{z^{2}} \big( \ri {\rm d} z \te_{\Iu}^{+} {\rm d} \te_{\Iu}^{+} -\ri {\rm d} u^{\pp} \te_{\Io}^{-} {\rm d} \te_{\Io}^{+} \\ &- \tfrac{\ri z}{2} {\rm d} \te_{\Iu}^{+} {\rm d} \te_{\Iu}^{+} - \tfrac{\ri z}{2} {\rm d} \te_{\Io}^{+} {\rm d} \te_{\Io}^{+} \\ &+ \te_{\Iu}^{+} {\rm d} \te_{\Iu}^{-} \te_{\Ju}^{+} {\rm d} \te_{\Ju}^{+} + \te_{\Io}^{+} {\rm d} \te_{\Io}^{+} \te_{\Jo}^{-} {\rm d} \te_{\Jo}^{+} \big) \end{aligned} \\
\begin{aligned} &\tfrac{1}{z^{2}} \big( \ri {\rm d} z \te_{\Io}^{-} {\rm d} \te_{\Io}^{-} - \ri {\rm d} u^{=} \te_{\Iu}^{+} {\rm d} \te_{\Iu}^{-} \\ &- \tfrac{\ri z}{2} {\rm d} \te_{\Iu}^{-} {\rm d} \te_{\Iu}^{-} - \tfrac{\ri z}{2} {\rm d} \te_{\Io}^{-} {\rm d} \te_{\Io}^{-} \\ &+ \te_{\Io}^{-} {\rm d} \te_{\Io}^{+} \te_{\Jo}^{-} {\rm d} \te_{\Jo}^{-} + \te_{\Iu}^{-} {\rm d} \te_{\Iu}^{-} \te_{\Ju}^{+} {\rm d} \te_{\Ju}^{-} \big) \end{aligned} ~& \begin{aligned} &- \tfrac{1}{2z^{2}} \big( \ri \te_{\Iu}^{+} {\rm d} \te_{\Iu}^{-} {\rm d} z + \ri \te_{\Io}^{-} {\rm d} \te_{\Io}^{+} {\rm d} z   \\ &-\ri \te_{\Iu}^{+} {\rm d} \te_{\Iu}^{+} {\rm d} u^{=} +\ri z {\rm d} \te_{\Iu}^{+} {\rm d} \te_{\Iu}^{-} \\&-\ri \te_{\Io}^{-} {\rm d} \te_{\Io}^{-} {\rm d} u^{\pp} +\ri z {\rm d} \te_{\Io}^{+} {\rm d} \te_{\Io}^{-}\\ &+\ri \te_{\Iu}^{+} {\rm d} \te_{\Iu}^{+} \te_{\Ju}^{-} {\rm d} \te_{\Ju}^{-} +\ri \te_{\Io}^{-} {\rm d} \te_{\Io}^{-} \te_{\Jo}^{+} {\rm d} \te_{\Jo}^{+} \big) \end{aligned}
\end{array}
\right)~, \\
\cT_{2} &= -\left(
\begin{array}{cc}
\begin{aligned} &(- \tfrac{1}{z^{3/2}} \l_{\rm L} {\rm d} z \te_{\Io}^{-} - \tfrac{1}{\sqrt{z}} \l_{\rm L}^{-1} {\rm d} \l_{\rm L}^{2}  \te_{\Io}^{-} \\ &+ \tfrac{1}{\sqrt{z}} (U_{\rm L})_{\Io \Jo} {\rm d} \te_{\Jo}^{-} ) \\ &\times ( \tfrac{1}{2z} {\rm d} z - \tfrac{\ri}{2z} \te_{\Iu}^{+} {\rm d} \te_{\Iu}^{-} - \tfrac{\ri}{2z} \te_{\Io}^{-} {\rm d} \te_{\Io}^{+} ) \\ &+ (\tfrac{1}{z^{3/2}} \l_{\rm L}^{-1} {\rm d} u^{\pp} \te_{\Io}^{-} + \tfrac{1}{\sqrt{z}} (U_{\rm L})_{\Io \Jo} {\rm d} \te_{\Jo}^{+}  ) \\ &\times ( \tfrac{1}{2z} {\rm d} u^{=} + \tfrac{\ri}{2z} \te_{\Io}^{-} {\rm d} \te_{\Io}^{-} + \tfrac{\ri}{2z} \te_{\Iu}^{-} {\rm d} \te_{\Iu}^{-} ) \end{aligned} ~& \begin{aligned} &(- \tfrac{1}{z^{3/2}} \l_{\rm L} {\rm d} z \te_{\Io}^{-} - \tfrac{1}{\sqrt{z}} \l_{\rm L}^{-1} {\rm d} \l_{\rm L}^{2}  \te_{\Io}^{-} \\
&+ \tfrac{1}{\sqrt{z}} (U_{\rm L})_{\Io \Jo} {\rm d} \te_{\Jo}^{-} ) \\ &\times ( -\tfrac{1}{2z} {\rm d} u^{\pp} - \tfrac{\ri}{2z} \te_{\Io}^{+} {\rm d} \te_{\Io}^{+} -\tfrac{\ri}{2z} \te_{\Iu}^{+} {\rm d} \te_{\Iu}^{+} ) \\ &+ (\tfrac{1}{z^{3/2}} \l_{\rm L}^{-1} {\rm d} u^{\pp} \te_{\Io}^{-} + \tfrac{1}{\sqrt{z}} (U_{\rm L})_{\Io \Jo} {\rm d} \te_{\Jo}^{+}  ) \\ &\times ( -\tfrac{1}{2z} {\rm d} z + \tfrac{\ri}{2z} \te_{\Iu}^{+} {\rm d} \te_{\Iu}^{-} + \tfrac{\ri}{2z} \te_{\Io}^{-} {\rm d} \te_{\Io}^{+} ) \end{aligned}\end{array}
\right)~,\\
\cT_{3} &= \left(
\begin{array}{cc}
\begin{aligned} &( \tfrac{1}{z^{3/2}} \l_{\rm R}^{-1} {\rm d} u^{=} \te_{\Iu}^{+} + \tfrac{1}{\sqrt{z}} (U_{\rm R})_{\Iu \Ju} {\rm d} \te_{\Ju}^{-} ) \\ &\times ( \tfrac{1}{2z} {\rm d} z - \tfrac{\ri}{2z} \te_{\Iu}^{+} {\rm d} \te_{\Iu}^{-} - \tfrac{\ri}{2z} \te_{\Io}^{-} {\rm d} \te_{\Io}^{+} ) \\
&+ (- \tfrac{1}{z^{3/2}} \l_{\rm R}  {\rm d} z \te_{\Iu}^{+} - \tfrac{1}{\sqrt{z}} \l_{\rm R}^{-1} {\rm d} \l_{\rm R}^{2} \te_{\Iu}^{+} \\
&+ \tfrac{1}{\sqrt{z}} (U_{\rm R})_{\Iu \Ju} {\rm d} \te_{\Ju}^{+} ) \\ &\times ( \tfrac{1}{2z} {\rm d} u^{=} + \tfrac{\ri}{2z} \te_{\Io}^{-} {\rm d} \te_{\Io}^{-} + \tfrac{\ri}{2z} \te_{\Iu}^{-} {\rm d} \te_{\Iu}^{-} ) \end{aligned} ~& \begin{aligned} &( \tfrac{1}{z^{3/2}} \l_{\rm R}^{-1} {\rm d} u^{=} \te_{\Iu}^{+} + \tfrac{1}{\sqrt{z}} (U_{\rm R})_{\Iu \Ju} {\rm d} \te_{\Ju}^{-} ) \\ &\times (  -\tfrac{1}{2z} {\rm d} u^{\pp} - \tfrac{\ri}{2z} \te_{\Io}^{+} {\rm d} \te_{\Io}^{+} -\tfrac{\ri}{2z} \te_{\Iu}^{+} {\rm d} \te_{\Iu}^{+} ) \\
&+ (- \tfrac{1}{z^{3/2}} \l_{\rm R}  {\rm d} z \te_{\Iu}^{+} - \tfrac{1}{\sqrt{z}} \l_{\rm R}^{-1} {\rm d} \l_{\rm R}^{2} \te_{\Iu}^{+} \\
&+ \tfrac{1}{\sqrt{z}} (U_{\rm R})_{\Iu \Ju} {\rm d} \te_{\Ju}^{+} ) \\ &\times ( -\tfrac{1}{2z} {\rm d} z + \tfrac{\ri}{2z} \te_{\Iu}^{+} {\rm d} \te_{\Iu}^{-} + \tfrac{\ri}{2z} \te_{\Io}^{-} {\rm d} \te_{\Io}^{+} ) \end{aligned}\end{array}
\right)~.
\end{align}
\end{subequations}
The curvature $\cR = (\cR_{\rm L}, \cR_{\rm R})$ takes the form
\begin{subequations}
\bea
\cR_{\rm L} = \left(
\begin{array}{c|c}
\cR_{1}
 & 0 \\
 \hline 
0 & \cR_{2} \end{array}
\right)~, \quad
\cR_{\rm R} 
= \left(
\begin{array}{c|c}
\cR_{1} & 0 \\
 \hline 
0 & \cR_{3}
\end{array}
\right)~,
\eea
where $\cR_{1}$ is a bosonic $2\times 2$ matrix with elements
\bea
\cR_{1} &=& \left(
\begin{array}{cc}
r_{1} ~& r_{2} \\
r_{3} ~& -r_{1}
\end{array}
\right)~,\\ \non
r_{1} &=& (-\frac{1}{2z} {\rm d} u^{\pp} - \frac{\ri}{2z} \te_{\Io}^{+} {\rm d} \te_{\Io}^{+} -\frac{\ri}{2z} \te_{\Iu}^{+} {\rm d} \te_{\Iu}^{+}) ( \frac{1}{2z} {\rm d} u^{=} + \frac{\ri}{2z} \te_{\Io}^{-} {\rm d} \te_{\Io}^{-} + \frac{\ri}{2z} \te_{\Iu}^{-} {\rm d} \te_{\Iu}^{-} ) \\ \non 
&-& \frac{\ri}{2z^{2}} \te_{\Iu}^{+} {\rm d} \te_{\Iu}^{-} {\rm d} z + \frac{\ri}{2z^{2}} \te_{\Io}^{-} {\rm d} \te_{\Io}^{+} {\rm d} z - \frac{\ri}{2z} {\rm d} \te_{\Iu}^{+} {\rm d} \te_{\Iu}^{-} + \frac{\ri}{2z} {\rm d} \te_{\Io}^{+} {\rm d} \te_{\Io}^{-} +  \frac{\ri}{2z^{2}} \te_{\Iu}^{+} {\rm d} \te_{\Iu}^{+} {\rm d} u^{=} \\ 
 &-& \frac{\ri}{2z^{2}} \te_{\Io}^{-} {\rm d} \te_{\Io}^{-} {\rm d} u^{\pp} - \frac{1}{2z^{2}} \te_{\Iu}^{+} {\rm d} \te_{\Iu}^{+} \te_{\Ju}^{-} {\rm d} \te_{\Ju}^{-} + \frac{1}{2z^{2}} \te_{\Io}^{-} {\rm d} \te_{\Io}^{-} \te_{\Jo}^{+} {\rm d} \te_{\Jo}^{+}~, \\ \non
r_{2} &=& 2( \frac{1}{2z} {\rm d} z - \frac{\ri}{2z} \te_{\Iu}^{+} {\rm d} \te_{\Iu}^{-} - \frac{\ri}{2z} \te_{\Io}^{-} {\rm d} \te_{\Io}^{+} ) (-\frac{1}{2z} {\rm d} u^{\pp} - \frac{\ri}{2z} \te_{\Io}^{+} {\rm d} \te_{\Io}^{+} -\frac{\ri}{2z} \te_{\Iu}^{+} {\rm d} \te_{\Iu}^{+}) \\ \non
&+& \frac{\ri}{z^{2}} {\rm d} z \te_{\Iu}^{+} {\rm d} \te_{\Iu}^{+} + \frac{\ri}{z^{2}} {\rm d} u^{\pp} \te_{\Io}^{-} {\rm d} \te_{\Io}^{+} - \frac{\ri}{2z} {\rm d} \te_{\Iu}^{+} {\rm d} \te_{\Iu}^{+} + \frac{\ri}{2z} {\rm d} \te_{\Io}^{+} {\rm d} \te_{\Io}^{+} + \frac{1}{z^{2}} \te_{\Iu}^{+} {\rm d} \te_{\Iu}^{-} \te_{\Ju}^{+} {\rm d} \te_{\Ju}^{+} \\ 
&-& \frac{1}{z^{2}} \te_{\Io}^{+} {\rm d} \te_{\Io}^{+} \te_{\Jo}^{-} {\rm d} \te_{\Jo}^{+}~, \\ \non
r_{3} &=& 2( \frac{1}{2z} {\rm d} u^{=} + \frac{\ri}{2z} \te_{\Io}^{-} {\rm d} \te_{\Io}^{-} + \frac{\ri}{2z} \te_{\Iu}^{-} {\rm d} \te_{\Iu}^{-}  ) ( \frac{1}{2z} {\rm d} z - \frac{\ri}{2z} \te_{\Iu}^{+} {\rm d} \te_{\Iu}^{-} - \frac{\ri}{2z} \te_{\Io}^{-} {\rm d} \te_{\Io}^{+}  ) \\ \non
&+& \frac{\ri}{z^{2}} {\rm d} z \te_{\Io}^{-} {\rm d} \te_{\Io}^{-} + \frac{\ri}{z^{2}} {\rm d} u^{=} \te_{\Iu}^{+} {\rm d} \te_{\Iu}^{-} + \frac{\ri}{2z} {\rm d} \te_{\Iu}^{-} {\rm d} \te_{\Iu}^{-} - \frac{\ri}{2z} {\rm d} \te_{\Io}^{-} {\rm d} \te_{\Io}^{-} + \frac{1}{z^{2}} \te_{\Io}^{-} {\rm d} \te_{\Io}^{+} \te_{\Jo}^{-} {\rm d} \te_{\Jo}^{-} \\ 
&-& \frac{1}{z^{2}} \te_{\Io}^{-} {\rm d} \te_{\Io}^{-} \te_{\Jo}^{+} {\rm d} \te_{\Jo}^{-}~, 
\eea
whilst $\cR_{2}$ and $\cR_{3}$ are given by
\bea \non
\cR_{2} &=& 2 \ri (- \frac{1}{z^{3/2}} \l_{\rm L} {\rm d} z \te_{\Io}^{-} - \frac{1}{\sqrt{z}} \l_{\rm L}^{-1} {\rm d} \l_{\rm L}^{2} \te_{\Io}^{-} + \frac{1}{\sqrt{z}} (U_{\rm L})_{\Io \overline{K}} {\rm d} \te_{\overline{K}}^{-}) \\ 
&\times& (\frac{1}{z^{3/2}} \l_{\rm L}^{-1} {\rm d} u^{\pp} \te_{\Jo}^{-} + \frac{1}{\sqrt{z}} (U_{\rm L})_{\Jo \overline{L}} {\rm d} \te_{\overline{L}}^{+})~, \\ \non
\cR_{3} &=& 2 \ri (- \frac{1}{z^{3/2}} \l_{\rm R} {\rm d} z \te_{\Iu}^{+} - \frac{1}{\sqrt{z}} \l_{\rm R}^{-1} {\rm d} \l_{\rm R}^{2} \te_{\Iu}^{+} + \frac{1}{\sqrt{z}} (U_{\rm R})_{\Iu \underline{K}} {\rm d} \te_{\underline{K}}^{+}  ) \\ 
&\times&  (\frac{1}{z^{3/2}} \l_{\rm R}^{-1} {\rm d} u^{=} \te_{\Ju}^{+} + \frac{1}{\sqrt{z}} (U_{\rm R})_{\Ju \underline{L}} {\rm d} \te_{\underline{L}}^{-} )~.
\eea
\end{subequations}

Since the vielbein was chosen to have the form \eqref{vielbein2} we know from the previous section that the (anti-)commutation relations of the covariant derivatives will take the form \eqref{derivcomm}. Indeed, for an explicit choice of local coordinates such as Poincaré coordinates it is possible to calculate explicit expressions for the covariant derivatives. We obtain
\begin{subequations} \label{derivatives}
\begin{align}
\cD_{0} &= z \l_{\rm L}^{2} \pa_{\pp} + z \l_{\rm R}^{2} \pa_{=} - \q^{-}_{\Io} \frac{\pa}{\pa \q^{+}_{\Io}} -  \q^{+}_{\Iu} \frac{\pa}{\pa \q^{-}_{\Iu}} - \cM_{2}~, \\
\cD_{1} &= z \pa_{z} + \te^{-}_{\Io} \frac{\pa}{\pa \te^{-}_{\Io}} + \te^{+}_{\Iu} \frac{\pa}{\pa \te^{+}_{\Iu}}~, \\
\cD_{2} &= z \l_{\rm L}^{2} \pa_{\pp} - z \l_{\rm R}^{2} \pa_{=} - \q^{-}_{\Io} \frac{\pa}{\pa \q^{+}_{\Io}} + \q^{+}_{\Iu} \frac{\pa}{\pa \q^{-}_{\Iu}} -  \cM_{0}~, \\ \non
\cD_{- \Io} &= \ri \sqrt{z} \l_{\rm L}^{-1} \l_{\rm R}^{2} \te^{-}_{\Io} \pa_{=} + \sqrt{z} [ (U_{\rm L})_{\Io \Jo} - \frac{\ri}{z} \l_{\rm L}^{-1} \te^{-}_{\Io} \te^{+}_{\Jo} ] \frac{\pa}{\pa \te^{-}_{\Jo}} - \frac{\ri}{\sqrt{z}} \l_{\rm L}^{-1} \te^{-}_{\Io} \te^{+}_{\Ju} \frac{\pa}{\pa \te^{-}_{\Ju}} \\ \non
&+ \frac{\ri}{\sqrt{z}} \l_{\rm L}^{-1} \te^{-}_{\Io} ( \cM_{0} - \cM_{2} ) + \frac{1}{2} [ \frac{1}{(\l_{\rm L} + 1)} \frac{\ri}{\sqrt{z}} ( \delta_{\Io \overline{M}} \te^{+}_{\overline{N}} - \delta_{\Io \overline{N}} \te^{+}_{\overline{M}} ) \\
&+ \frac{1}{\l_{\rm L} (\l_{\rm L} + 1)^{2}} \frac{1}{z^{3/2}} \te^{-}_{\Io} \te^{+}_{\overline{M}} \te^{+}_{\overline{N}} - \frac{1}{2 \l_{\rm L} (\l_{\rm L} + 1)^{2}} \frac{1}{z^{3/2}} \te^{+}_{\Io} ( \te^{-}_{\overline{M}} \te^{+}_{\overline{N}}  +\te^{+}_{\overline{M}} \te^{-}_{\overline{N}}  ) ] \cN_{\overline{M} \overline{N}}~, \\ \non
\cD_{+ \Io} &= \ri \sqrt{z} \l_{\rm L} \te^{+}_{\Io} \pa_{\pp} - \ri \sqrt{z} \l_{\rm L}^{-1} \te^{-}_{\Io} \pa_{z} + \sqrt{z} [ (U_{\rm L})_{\Io \Jo} - \frac{\ri}{z} \l_{\rm L}^{-1} \te^{-}_{\Io} \te^{+}_{\Jo} ] \frac{\pa}{\pa \te^{+}_{\Jo}} \\ \non
&- \frac{\ri}{\sqrt{z}} \l_{\rm L}^{-1} \te^{-}_{\Io} \te^{+}_{\Ju} \frac{\pa}{\pa \te^{+}_{\Ju}} - \frac{\ri}{\sqrt{z}} \l_{\rm L}^{-1} \te^{-}_{\Io} \cM_{1} + \frac{1}{2} [ -\frac{1}{(\l_{\rm L} + 1)} \frac{\ri}{\sqrt{z}} ( \delta_{\Io \overline{M}} \te^{-}_{\overline{N}} - \delta_{\Io \overline{N}} \te^{-}_{\overline{M}} ) \\ 
&+ \frac{1}{\l_{\rm L} (\l_{\rm L} + 1)^{2}} \frac{1}{z^{3/2}} \te^{+}_{\Io} \te^{-}_{\overline{M}} \te^{-}_{\overline{N}} - \frac{1}{2 \l_{\rm L} (\l_{\rm L} + 1)^{2}} \frac{1}{z^{3/2}} \te^{-}_{\Io} ( \te^{-}_{\overline{M}} \te^{+}_{\overline{N}}  +\te^{+}_{\overline{M}} \te^{-}_{\overline{N}}  ) ] \cN_{\overline{M} \overline{N}}~, \\ \non
\cD_{- \Iu} &= - \ri \sqrt{z} \l_{\rm R}^{-1} \te^{+}_{\Iu} \pa_{z} + \ri \sqrt{z} \l_{\rm R} \te^{-}_{\Iu} \pa_{=} + \sqrt{z} [ (U_{\rm R})_{\Iu \Ju} - \frac{\ri}{z} \l_{\rm R}^{-1} \te^{+}_{\Iu} \te^{-}_{\Ju} ] \frac{\pa}{\pa \te^{-}_{\Ju}} \\ \non
& - \frac{\ri}{\sqrt{z}} \l_{\rm R}^{-1} \te^{+}_{\Iu} \te^{-}_{\Jo} \frac{\pa}{\pa \te^{-}_{\Jo}} + \frac{\ri}{\sqrt{z}} \l_{\rm R}^{-1} \te^{+}_{\Iu} \cM_{1} + \frac{1}{2} [ - \frac{1}{(\l_{\rm R} + 1)} \frac{\ri}{\sqrt{z}} (\delta_{\Iu \underline{M}} \te^{+}_{\underline{N}} - \delta_{\Iu \underline{N}} \te^{+}_{\underline{M}} ) \\ 
&+ \frac{1}{\l_{\rm R} (\l_{\rm R} + 1)^{2}} \frac{1}{z^{3/2}} \te^{-}_{\Iu} \te^{+}_{\underline{M}} \te^{+}_{\underline{N}} - \frac{1}{2 \l_{\rm R} (\l_{\rm R} + 1)^{2} } \frac{1}{z^{3/2}} \te^{+}_{\Iu} ( \te^{-}_{\underline{M}} \te^{+}_{\underline{N}} + \te^{+}_{\underline{M}} \te^{-}_{\underline{N}} ) ] \cN_{\underline{M} \underline{N}}~, \\ \non
\cD_{+ \Iu} &= \ri \sqrt{z} \l_{\rm R}^{-1} \l_{\rm L}^{2} \te^{+}_{\Iu} \pa_{\pp} - \frac{\ri}{\sqrt{z}} \l_{\rm R}^{-1} \te^{+}_{\Iu} \te^{-}_{\Jo} \frac{\pa}{\pa \te^{+}_{\Jo}} +  \sqrt{z} [ (U_{\rm R})_{\Iu \Ju} - \frac{\ri}{z} \l_{\rm R}^{-1} \te^{+}_{\Iu} \te^{-}_{\Ju} ] \frac{\pa}{\pa \te^{+}_{\Ju}} \\ \non
&- \frac{\ri}{\sqrt{z}} \l_{\rm R}^{-1} \te^{+}_{\Iu} ( \cM_{0} + \cM_{2} ) + \frac{1}{2} [  \frac{1}{(\l_{\rm R} + 1)} \frac{\ri}{\sqrt{z}} (\delta_{\Iu \underline{M}} \te^{-}_{\underline{N}} - \delta_{\Iu \underline{N}} \te^{-}_{\underline{M}} ) \\ 
&+ \frac{1}{\l_{\rm R} (\l_{\rm R} + 1)^{2}} \frac{1}{z^{3/2}} \te^{+}_{\Iu} \te^{-}_{\underline{M}} \te^{-}_{\underline{N}} - \frac{1}{2 \l_{\rm R} (\l_{\rm R} + 1)^{2} } \frac{1}{z^{3/2}} \te^{-}_{\Iu} ( \te^{-}_{\underline{M}} \te^{+}_{\underline{N}} + \te^{+}_{\underline{M}} \te^{-}_{\underline{N}} ) ] \cN_{\underline{M} \underline{N}}~.
\end{align}
\end{subequations}
We may further define the derivatives $\cD_{\pp} = \frac{1}{2} ( \cD_{0} + \cD_{2})$ and $\cD_{=} = \frac{1}{2} ( \cD_{0} - \cD_{2})$. Explicitly
\begin{subequations}
\begin{align}
\cD_{\pp} &= z \l_{\rm L}^{2} \pa_{\pp} - \q^{-}_{\Io} \frac{\pa}{\pa \q^{+}_{\Io}} - \frac{1}{2} ( \cM_{0} + \cM_{2} )~, \\
\cD_{=} &= z \l_{\rm R}^{2} \pa_{=} - \q^{+}_{\Iu} \frac{\pa}{\pa \q^{-}_{\Iu}} + \frac{1}{2} ( \cM_{0} - \cM_{2} ) .
\end{align}
\end{subequations}
Whilst the vector covariant derivatives have a simple structure, the spinor covariant derivatives have complicated forms. We will elaborate on these derivatives in section 7.


\subsection{Bi-supertwistor construction} \label{section6.4}

The freedom  \eqref{supertwistorgauge} may be fixed in order to obtain the bi-supertwistors, and hence two-point functions, in a specific coordinate system. We make the choice corresponding to Poincar\'e  coordinates \eqref{poincare}. 

For the bi-supertwistors  \eqref{bisupertwistors} we obtain:
\bea
\mathbb{Z}_{\Ao \Bu} 
&=& \frac{1}{z} \left(
\begin{array}{cc|c}
-u^{\pp}~ & u^{\pp} u^{=} - z^{2} \l_{\rm L}^{2} \l_{\rm R}^{2}~ & - \ri u^{\pp} \te_{\Ju}^{-} - \ri z \l_{\rm L}^{2} \te_{\Ju}^{+}  \\
1~ & - u^{=}~ & \ri \te_{\Ju}^{-} \\ [1pt]
 \hline 
\ri \te_{\Io}^{+}~ & -\ri \te_{\Io}^{+} u^{=} - \ri z \l_{\rm R}^{2} \te_{\Io}^{-}~ & - \te_{\Io}^{+} \te_{\Ju}^{-} + \te_{\Io}^{-} \te_{\Ju}^{+}
\end{array}
\right) ~,\\
\mathbb{X}_{\Ao \Bo}
&=& \frac{1}{z} \left(
\begin{array}{cc|c}
0 & -z - \ri \te_{\Io}^{+} \te_{\Io}^{-}~ & - \ri u^{\pp} \te_{\Jo}^{-} - \ri z \l_{\rm L}^{2} \te_{\Jo}^{+}  \\
z + \ri \te_{\Io}^{+} \te_{\Io}^{-}~ & 0~ & \ri \te_{\Jo}^{-} \\ [1pt]
 \hline 
\quad\ri u^{\pp} \te_{\Io}^{-}  + \ri z \l_{\rm L}^{2} \te^{+}_{\Io} & - \ri \te_{\Io}^{-}~ & - \te_{\Io}^{+} \te_{\Jo}^{-} + \te_{\Io}^{-} \te_{\Jo}^{+}
\end{array}
\right)~, \\
\mathbb{Y}_{\Au \Bu} 
&=& \frac{1}{z} \left(
\begin{array}{cc|c}
0 & -z + \ri \te_{\Iu}^{+} \te_{\Iu}^{-}~ & - \ri \te_{\Ju}^{+}  \\
z - \ri \te_{\Iu}^{+} \te_{\Iu}^{-} & 0~ & ~\ri u^{=} \te_{\Ju}^{+} + \ri z \l_{\rm R}^{2} \te_{\Ju}^{-} \\ [1pt]
 \hline 
\ri \te_{\Iu}^{+} & - \ri u^{=} \te_{\Iu}^{+} - \ri z \l_{\rm R}^{2} \te_{\Iu}^{-}~ & - \te_{\Iu}^{+} \te_{\Ju}^{-} + \te_{\Iu}^{-} \te_{\Ju}^{+}
\end{array}
\right)~.
\eea

The two-point functions in Poincar\'e coordinates are then
\begin{align}
\mathrm {str} (\tilde{\mathbb{Z}} \mathbb{Z}) =& \frac{1}{ z \tilde{z}} \big[ - (u^{\pp} - \tilde{u}^{\pp}) (u^{=} - \tilde{u}^{=})
- \ri (u^{\pp} - \tilde{u}^{\pp}) \tilde{\te}_{\underline{I}}^{-} \te_{\underline{I}}^{-}
- \ri (u^{=} - \tilde{u}^{=} ) \tilde{\te}_{\overline{I}}^{+} \te_{\overline{I}}^{+} \non \\ 
&+ \tilde{\te}_{\Io}^{+} \tilde{\te}_{\Iu}^{-} \te_{\Iu}^{-} \te_{\Io}^{+} + z^{2} \l_{\rm L}^{2} \l_{\rm R}^{2} + \tilde{z}^{2} \tilde{\l}_{\rm L}^{2} \tilde{\l}_{\rm R}^{2} -\ri  z \l_{\rm L}^{2} \tilde{\te}_{\underline{I}}^{-} \te_{\underline{I}}^{+} + \ri \tilde{z} \tilde{\l}_{\rm L}^{2} \tilde{\te}_{\underline{I}}^{+} \te_{\underline{I}}^{-}  \non\\ 
&- \ri  z \l_{\rm R}^{2} \tilde{\te}_{\overline{I}}^{+} \te_{\overline{I}}^{-} +\ri  \tilde{z} \tilde{\l}_{\rm R}^{2} \tilde{\te}_{\overline{I}}^{-} \te_{\overline{I}}^{+} + \tilde{\te}_{\Io}^{-} \tilde{\te}_{\Iu}^{+} \te_{\Iu}^{+} \te_{\Io}^{-} - \tilde{\te}_{\Io}^{-} \tilde{\te}_{\Iu}^{+} \te_{\Iu}^{-} \te_{\Io}^{+} - \tilde{\te}_{\Io}^{+} \tilde{\te}_{\Iu}^{-} \te_{\Iu}^{+} \te_{\Io}^{-} \big]~, \\
\mathrm{str} (\tilde{\mathbb{X}} \mathbb{X}) =& \frac{2}{z \tilde{z}} \big[ z \tilde{z} \l_{\rm L}^{2} \tilde{\l}_{\rm L}^{2} + \ri (u^{\pp} - \tilde{u}^{\pp} ) \tilde{\te}_{\Io}^{-} \te_{\Io}^{-} + \ri  z \l_{\rm L}^{2} \tilde{\te}_{\Io}^{-} \te_{\Io}^{+} - \ri  \tilde{z} \tilde{\l}_{\rm L}^{2} \tilde{\te}_{\Io}^{+} \te_{\Io}^{-}  \non\\ 
 &+  \tilde{\te}_{\overline{I}}^{+}  \tilde{\te}_{\Jo}^{-} \te_{\Jo}^{+} \te_{\overline{I}}^{-} -  \tilde{\te}_{\overline{I}}^{+}  \tilde{\te}_{\Jo}^{-} \te_{\Jo}^{-} \te_{\overline{I}}^{+} \big]~, \\
\mathrm{str} (\tilde{\mathbb{Y}} \mathbb{Y}) =& \frac{2}{z \tilde{z}} \big[ z \tilde{z} \l_{\rm R}^{2} \tilde{\l}_{\rm R}^{2} + \ri (u^{=} - \tilde{u}^{=} ) \tilde{\te}_{\Iu}^{+} \te_{\Iu}^{+} + \ri  z \l_{\rm R}^{2} \tilde{\te}_{\Iu}^{+} \te_{\Iu}^{-} - \ri  \tilde{z} \tilde{\l}_{\rm R}^{2} \tilde{\te}_{\Iu}^{-} \te_{\Iu}^{+} \non \\ 
 &+  \tilde{\te}_{\underline{I}}^{-}  \tilde{\te}_{\Ju}^{+} \te_{\Ju}^{-} \te_{\underline{I}}^{+} -  \tilde{\te}_{\underline{I}}^{-}  \tilde{\te}_{\Ju}^{+} \te_{\Ju}^{+} \te_{\underline{I}}^{-} \big]~.
\end{align}
Separately these two-point functions do not admit simpler forms, however they may be combined to obtain a single two-point function $s^{2}$ with suggestive structure. We have
\begin{align}
s^2 \equiv & \mathrm {str} (\tilde{\mathbb{Z}} \mathbb{Z}) - \frac 12 \mathrm{str} (\tilde{\mathbb{X}} \mathbb{X}) - \frac 12 \mathrm{str} (\tilde{\mathbb{Y}} \mathbb{Y}) \non \\ \non
=& \frac{1}{ z \tilde{z}} \Big[ \big( z - \tilde{z} + \ri ( \te^{+}_{\Io} - \tilde{\te}^{+}_{\Io} ) \tilde{\te}^{-}_{\Io} + \ri ( \te^{-}_{\Iu} - \tilde{\te}^{-}_{\Iu} ) \te^{+}_{\Iu} \big) 
\big( z - \tilde{z} + \ri ( \te^{+}_{\Io} - \tilde{\te}^{+}_{\Io} ) \te^{-}_{\Io} + \ri ( \te^{-}_{\Iu} - \tilde{\te}^{-}_{\Iu} ) \tilde{\te}^{+}_{\Iu} \big) \\ 
&- \big( u^{\pp} - \tilde{u}^{\pp} + \ri \tilde{\te}^{+}_{\Io} \te^{+}_{\Io} + \ri \tilde{\te}^{+}_{\Iu} \te^{+}_{\Iu} \big) \big( u^{=} - \tilde{u}^{=} + \ri \tilde{\te}^{-}_{\Io} \te^{-}_{\Io} + \ri \tilde{\te}^{-}_{\Iu} \te^{-}_{\Iu} \big) \Big]~,
\end{align}
which, for infinitesimally separated points, reads
\begin{align} \non
\rd s^2 =& \frac{1}{z^{2}} \big[ ( \mathrm{d} z + \ri \mathrm{d} \te^{+}_{\Io} \te^{-}_{\Io} + \ri \mathrm{d} \te^{-}_{\Iu} \te^{+}_{\Iu} )^{2} \\ 
& - ( \mathrm{d} u^{\pp} - \ri \mathrm{d} \te^{+}_{\Io} \te^{+}_{\Io} - \ri \mathrm{d} \te^{+}_{\Iu} \te^{+}_{\Iu} ) ( \mathrm{d} u^{=} - \ri \mathrm{d} \te^{-}_{\Io} \te^{-}_{\Io} - \ri \mathrm{d} \te^{-}_{\Iu} \te^{-}_{\Iu} ) \big]~,
\end{align}
in analogy with the non-supersymmetric case.


\section{Conclusion}
In this paper the bi-supertwistor formulation of $\textrm{AdS}^{(3|p,q)}$ was presented, providing the supersymmetric analogue of the embedding \eqref{1.1b}. The use of bi-supertwistors faciliates the simple construction of two-point functions \eqref{twopoint}. The supercoset construction of $\textrm{AdS}^{(3|p,q)}$ was then given and from it the superspace geometry of $\textrm{AdS}^{(3|p,q)}$ obtained, before being used to explore the superspace geometry of $\textrm{AdS}^{(3|p,q)}$ for a particular local coordinate system. Explicit realisations of the covariant derivatives were obtained for this coordinate system.

As stated at the beginning of this paper, the first step in generalising the Ba\~nados metric to a $(p,q)$ supersymmetric analogue should be to derive a Poincar\'e coordinate patch in which the covariant derivatives $\cD_{A} = \{\cD_{a}, \cD^{I}_{\a}\}$ are conformally related to the covariant derivatives $D_{A} = \{ \partial_{a}, D_{\a}^{I}\}$ of Minkowski superspace $\mathbb{M}^{3|p+q}$. Whilst it is true that the local coordinates introduced in section 6 can be identified as Poincaré coordinates, the obtained covariant derivatives are not directly conformally related to the Minkowski superspace derivatives. Indeed, the finite forms of the super-Weyl transformations in 3D $\cN$-extended conformal supergravity are given in \cite{KLT-M12} and imply the following relations between conformally flat $\textrm{AdS}^{(3|p,q)}$ and Minkowski superspace
\begin{subequations} \label{conflata-d}
\begin{align} \label{conflata}
    \mathcal{D}^{I}_{\alpha} &= \mathrm{e}^{\frac{1}{2} \sigma} \biggr( D^{I}_{\alpha} + (D^{\beta I}\sigma) \mathcal{M}_{\alpha \beta} + (D_{\alpha J} \sigma ) \mathcal{N}^{IJ} \biggr)~, \\ \nonumber \label{conflatb}
    \mathcal{D}_{a} &= \mathrm{e}^{\sigma} \biggr( D_{a} + \frac{\ri}{2} (\gamma_{a})^{\alpha \beta} ( D^{K}_{(\alpha} \sigma )  D_{\beta) K} + \varepsilon_{abc} (D^{b} \sigma ) \mathcal{M}^{c} + \frac{\ri}{16} (\gamma_{a})^{\alpha \beta} ( [D^{[K}_{(\alpha}, D^{L]}_{\beta )} ] \sigma ) \mathcal{N}_{KL} \\
    &- \frac{\ri}{8} (\gamma_{a})^{\alpha\beta} ( D^{\rho}_{K} \sigma )( D^{K}_{\rho} \sigma ) \mathcal{M}_{\alpha \beta} + \frac{3\ri}{8} (\gamma_{a})^{\alpha \beta} ( D^{[K}_{(\alpha} \sigma ) ( D^{L]}_{\beta)} \sigma ) \mathcal{N}_{KL} \biggr)~, \\ \label{weyltorsion}
    S^{IJ} &= - \frac{\ri}{4} (D^{\rho(I} D^{J)}_{\rho}) \mathrm{e}^{\sigma} +\frac{\ri}{2} \mathrm{e}^{-\sigma} ( \delta^{I}_{K} \delta^{J}_{L} - \frac{1}{4} \delta^{IJ} \delta_{KL} ) ( D^{\rho(K} \mathrm{e}^{\sigma} ) ( D^{L)}_{\rho} \mathrm{e}^{\sigma} )~, \\ \label{weylconstraint}
    0 &= D^{[I}_{(\alpha} D^{J]}_{\beta)}  \mathrm{e}^{\sigma}~, 
\end{align}
\end{subequations}
where $I, J \in \{1, ..., \cN \}$ and $S^{IJ}$ is a dimension-1 torsion parameter satisfying
\bea \label{weyltorsionconstraint}
\cD_{A} S^{IJ} = 0~, \quad S^{IK} S_{KJ} = S^{2} \d^{I}_{J}~, \quad S^{2} = \frac{1}{\cN} S^{KL} S_{KL}~.
\eea

Our construction will be similar to the conformally flat realisation 
for the $\cN=2$ $\rm AdS$ superspace in four dimensions that makes use of Poincar\'e coordinates 
for $\rm AdS_4$ \cite{BKLT-M}.
Beginning with the three-dimensional (3D) gamma matrices $\g_{{a}}$, where ${a} \in \{0,1,2\}$, we may perform a $2+1$ splitting of 3D vectors by first deleting $\g_{1}$ 
\begin{align}
\g_a := \big( (\g_{{a}})_{{\a} {\b}}\big) = ( \mathbbm{1}, \sigma_{1}, \sigma_{3}) 
~\longrightarrow ~ \g_{\hat a} := \big( (\g_{\hat a})_{\hat \a \hat \b} \big) 
= ( \mathbbm{1}, \sigma_{3} )~, \quad \hat a \in \{0, 1\}~,
\end{align}
in order to obtain 2D gamma matrices, $(\g_{\hat a})_{\hat \a \hat \b} $. A 3D vector $V^{{a}}$ may then be decomposed into a 2D vector $V^{\hat a}$ and a scalar $\mathfrak U$ according to 
\begin{align}
V_{{\a} {\b}} = V^{{a}}(\g_{{a}})_{{\a} {\b}} ~ \longrightarrow ~
V_{\hat \a \hat \b} + {\mathfrak U} \,{\mathfrak C}_{\hat \a\hat \b}
~, 
\quad 
V_{\hat \a\hat \b} = V^{\hat a} (\g_{\hat a})_{\hat \a\hat \b} 
= \left(
\begin{array}{cc}
V^\pp &~0\\
0 & ~V^=
\end{array}
\right)
~,
\end{align}
where ${\mathfrak U} = V^1$, ${\mathfrak C} := \big( {\mathfrak C}_{\hat \a\hat \b} \big) = \s_1$, and 
$V^\pp := V^0 + V^2$, $V^= := V^0 - V^2$.
Choosing $V^a = \pa^a = (-\pa_0 , \pa_1 , \pa_2)$ gives 
\bea
\pa_{\hat \a\hat \b} =  (\g_{\hat a})_{\hat \a\hat \b} \pa^{\hat a}
= -\left(
\begin{array}{cc}
\pa_= &~0\\
0 & ~\pa_\pp
\end{array}
\right)
~,\quad \pa_= = \pa_0 - \pa_2~, \quad \pa_\pp = \pa_0 + \pa_2~.
\eea

Upon the  $2+1$ splitting, the  spinor derivatives $D_{\a }^I$ of 3D $\cN$-extended Minkowski superspace ${\mathbb M}^{3|\cN}$ turn into those corresponding to 2D $\cN$-extended Minkowski superspace with a central charge
\begin{subequations} 
\bea
D_{\hat \a}^{I} = \frac{\partial}{\partial \te^{\hat \a}_{I}} + \ri \te^{\hat \b I} \partial_{\hat \a \hat \b} + \ri  {\mathfrak C}_{\hat \a\hat \b}  \te^{\hat \b I} \partial_{z}~, 
\eea
since
the operators $D_{\hat \a}^{I} $ satisfy the anti-commutation relations
\bea
\{ D_{\hat \a}^{I}, D_{\hat \b}^{J} \} = 2\ri \d^{IJ} \partial_{\hat \a \hat \b} + 2\ri \d^{IJ} 
{\mathfrak C}_{\hat \a \hat \b} \partial_{z}~.
\eea
\end{subequations}
The central charge variable 
$z$ denotes the 3D coordinate $x^1$.

To solve the equations \eqref{weyltorsion}, \eqref{weylconstraint} and \eqref{weyltorsionconstraint}, we make 
an $\sISO(1,1)$ invariant ansatz for the Weyl parameter
\bea
\re^{\s} = A(z) + \te_{IJ} B^{IJ}(z) + \te^{IJ} \te_{IJ} C(z)~,
\eea
where $A(z)$ and $C(z)$ are real, $B^{IJ}(z)$ is symmetric and imaginary, and $\te_{IJ} = \te^{\hat \a}_{I} \te_{\hat \a J}$. Applying the constraint \eqref{weylconstraint} we obtain
\bea
C(z) = \partial_{z} B^{IJ} (z) = \partial^{2}_{z} A(z) = 0~,
\eea
and thus our Weyl parameter takes the form
\bea
\re^{\s} = a + bz - \ri s^{IJ} \te_{IJ}~,
\eea
where $a, b \in \mathbb{R}$ and $s^{IJ} = s^{JI} \in \mathbb{R}$. If we now employ \eqref{weyltorsion} in tandem with \eqref{weyltorsionconstraint} we acquire the further constraints
\bea
 s^{IK} s_{KJ} = s^{2} \d^{I}_{J}~, \quad b = 2s~.
 \eea
 The constant $a$ is then chosen as $a=0$. Thus, the desired covariant derivatives $\cD_{A} = \{\cD_{a}, \cD^{I}_{\a}\}$ should be related to the Minkowski superspace derivatives $D_{A} = \{ \partial_{a}, D_{\a}^{I}\}$ through a Weyl parameter taking the form
\bea
\re^{\s} = 2sz - \ri s^{IJ} \te_{IJ}~.
\eea
Clearly, the derivatives \eqref{derivatives} are not conformally related to the Minkowski superspace derivatives $D_{A} = \{ \partial_{a}, D_{\a}^{I}\}$. The explanation for this becomes apparent by considering the torsion tensor $S^{IJ}$. Through this analysis we obtain an explicit expression for the torsion tensor\footnote{The appearance of $\mathfrak C$ is related to the explicit 2+1 splitting performed.}
\bea
S^{IJ} = s^{IJ} + 2\ri \frac{s^{2} \te^{IJ} - \te_{NM} s^{IN} s^{JM} - 2s \te^{\hat \a (I } \te^{\hat \b}_{M} 
{\mathfrak C}_{\hat \a \hat \b} s^{J)M}} { 2sz - \ri s^{PQ} \te_{PQ}}~,
\eea
satisfying 
\begin{align}
S^{2} = s^{2}~, &\quad \mathcal{S} \equiv \frac{1}{\mathcal{N}} \d_{IJ} S^{IJ} = \frac{1}{\mathcal{N}} \d_{IJ} s^{IJ} 
\quad \implies \quad \mathcal{D}_{A} S^{2} =\mathcal{D}_{A} \mathcal{S} = 0~,
\end{align} 
which are sufficient conditions to ensure that $S^{IJ}$ is covariantly constant $\mathcal{D}_{A} S^{IJ} = 0$ \cite{KLT-M12}. However, $S^{IJ}$ is clearly not in the diagonal form \eqref{diagonal}. Indeed, in order to reconcile this result with the coset construction a local $\sSO(\mathcal{N})$ transformation must be applied to diagonalise $S^{IJ}$ and obtain the $\sSO(p) \times \sSO(q)$ local group. Said transformation will also act on the spinor derivatives, resulting in the complicated forms \eqref{derivatives} which are no longer conformally related to the covariant derivatives of Minkowski superspace. Thus, $\textrm{AdS}^{(3|p,q)}$ in Poincar\'e coordinates is only conformally flat with the $\sSO(\mathcal{N})$ local group left intact.

Associated with the conformally flat derivatives 
\eqref{conflata-d} 
are a set of vielbeins $E^{A}$. Using the obtained Weyl parameter we may calculate these one-forms, which in lightcone coordinates take the form
\begin{subequations}
\begin{align}
E^{\pp} &= \mathrm{e}^{- \sigma} ( \mathrm{d} u^{\pp} + \ri \mathrm{d} \te^{+}_{I} \te^{+}_{I} )~, \\
E^{z} &= \mathrm{e}^{- \sigma} ( \mathrm{d} z + \ri \mathrm{d} \te^{+}_{I} \te^{-}_{I} + \ri \mathrm{d} \te^{-}_{I} \te^{+}_{I} )~, \\
E^{=} &= \mathrm{e}^{- \sigma} (\mathrm{d} u^{=} + \ri \mathrm{d} \te^{-}_{I} \te^{-}_{I} )~, \\
E^{-}_{I} &= \mathrm{e}^{-\frac12 \sigma} ( \mathrm{d} \te^{-}_{I} - \te^{-J} ( s \d_{IJ} - s_{IJ}) E^{z} + 2 \te^{+J} ( s \d_{IJ} + s_{IJ}) E^{=} )~, \\
E^{+}_{I} &= \mathrm{e}^{-\frac12 \sigma} ( \mathrm{d} \te^{+}_{I} - \te^{+J} ( s \d_{IJ} + s_{IJ}) E^{z} + 2 \te^{-J} ( s \d_{IJ} - s_{IJ}) E^{\pp} )~.
\end{align}
\end{subequations}
It is further possible to diagonalise $s^{IJ}$ to simplify these expressions, but this is not necessary.

The Ba\~nados metric \eqref{1.4} is a deformation of the AdS$_3$ metric by a two-dimensional conformal 
energy-momentum tensor, with its components $\cT_{\pp \pp}$ and 
$\cT_{==}$ satisfying the conservation equations
\bea
\pa_{=} \cT_{\pp \pp} =0~, \qquad \pa_{\pp} \cT_{==}=0~.
\eea
It is natural to expect that a supersymmetric extension of the Ba\~nados metric should be a deformation of the $(p,q)$ AdS superspace geometry \eqref{alg-AdS} by a two-dimensional conformal $(p,q)$ supercurrent multiplet. Such a supercurrent is determined by two conformal primary superfields $\hat{\cJ}_{+(4-p)}$ 
and $\check{\cJ}_{-(4-q)}$ 
defined on $(p,q)$ Minkowski superspace ${\mathbb M}^{(2|p,q)}$, which satisfy the equations
\begin{subequations}
\bea
 q&>&0: \quad D_-^{\underline I} \hat{\cJ}_{+(4-p)} =0~,
\qquad  q=0:\quad \pa_= \hat{\cJ}_{+(4-p)} =0~; \\
 p&>&0: \quad D_+^{\overline I} \check{\cJ}_{-(4-q)} =0~,
\qquad  p=0:\quad \pa_\pp \check{\cJ}_{-(4-q)} =0~.
\eea
\end{subequations}
These equations are superconformal  \cite{Kuzenko:2022qnb}, and the
dimensions of $\hat{\cJ}_{+(4-p)}$ and $\check{\cJ}_{-(4-q)}$  are $\hf (4- p)$ and $\hf (4-q)$, respectively. The functional structure of the conformal supercurrents 
is dictated by their top components
\begin{subequations}
\bea
\hat{\cJ}_{+(4-p)} (x^\pp, \q^+) &\propto &
\dots + \frac{\ri^{\hf p(p-1)}}{p!} \ve^{\Io_1 \dots \Io_p} \q^+_{\Io_1} \dots \q^+_{\Io_p} 
\cT_{\pp \pp} (x^\pp)~,\\
\check{\cJ}_{-(4-q)} (x^=, \q^-) &\propto &
\dots + \frac{\ri^{\hf q(q-1)}}{q!} \ve^{\Iu_1 \dots \Iu_q} \q^-_{\Iu_1} \dots \q^-_{\Iu_q} 
\cT_{= =} (x^=)~.
 \eea
 \end{subequations}
 The structure of conformal supercurrents implies that conformal $(p,q)$ supergravity 
 \cite{Kuzenko:2022qnb} is characterised by two unconstrained prepotentials, 
 $ \hat{H}^{+(4-q)}$ and $\check{H}^{-(4-p)}$, which couple to the supercurrents as follows 
 \bea 
I= \int \rd^{(2|p,q)}_{(p-q)} \Big\{ \hat{\cJ}_{+(4-p)} \hat{H}^{+(4-q)} 
 +\check{\cJ}_{-(4-q)} \check{H}^{-(4-p)}  \Big\}~,
 \eea 
 where $\rd^{(2|p,q)}_{(p-q)} $ denotes the full superspace integration measure 
 for ${\mathbb M}^{(2|p,q)}$. In the $p,q >0$ case, the prepotentials are defined modulo gauge transformations
 \bea
 \d  \hat{H}^{+(4-q)} = \ri^q D_-^{\Iu} \hat{\L}^{+(3-q)}_{\Iu}~,\qquad 
 \d  \check{H}^{-(4-p)} = \ri^p D_+^{\Io} \check{\L}^{-(3-p)}_{\Io}~,
 \eea
 with unconstrained real gauge parameters. Explicit construction of supersymmetric extensions of the Ba\~nados metric will be described elsewhere.
  
Our conclusions about the structure of conformal supercurrents and associated prepotentials are in agreement with the well-known prepotential descriptions 
of $(1,0)$ supergravity \cite{GMT,GGMT}, $(1,1)$ (or $\cN=1$) supergravity \cite{RvNZ, Gates:1985vk},
$(p,0)$ supergravity \cite{EO}, $(2,2)$ (or $\cN=2$) supergravity \cite{GW}, and $(4,4)$ (or $\cN=4$) supergravity \cite{KUM,BI}.
It is instructive to compare the $d=2$ conformal $(p,q)$ supercurrents with those corresponding to $\cN$-extended conformal supersymmetry in the $d=3 $
\cite{BKNT-M,KNT-M} (see also \cite{BKS}) 
and $d=4$ cases (see \cite{HST} and references therein).
\\

\noindent
{\bf Acknowledgements:}\\
We are grateful to Stefan Theisen for bringing Ref. \cite{Banados} to our attention, 
and to Gabriele Tartaglino-Mazzucchelli for email correspondence.
The work of SMK is supported in part by the Australian 
Research Council, projects DP200101944 and DP230101629.
The work of KT is supported by the Australian Government Research Training Program Scholarship.


\appendix

\section{Conventions and notation} \label{Appendix}

Our 3D notation and conventions follow \cite{KLT-M11}. 
In particular, the real gamma matrices 
satisfy the relations
\begin{align} \label{gammacommutator}
\{ \g_{a}, \g_{b} \} = 2 \eta_{ab} \mathbbm{1}~, \qquad a,b = 0,1,2~, 
\end{align}
where the 3D Minkowski metric is $\eta_{ab}  = \mathrm{diag}(-1, +1, +1$). 
The following realisation of the $\g$-matrices is used
\begin{subequations}
\begin{align}
(\g_{a})_{\a}{}^{ \b} 
= (-\ri \sigma_{2}, \sigma_{3}, -\sigma_{1})~,
\label{A.2a}
\end{align}
and therefore 
\begin{align} 
\g_{a} \g_{b} = \eta_{a b} \mathbbm{1} + \ve_{abc} \g^{c} ~,
\label{A.2b}
\end{align}
\end{subequations}
where the Levi-Civita tensor is normalised as $\ve_{012} = - \ve^{012} = -1$. 
In three dimensions, there are two inequivalent irreducible representations of the Clifford algebra \eqref{gammacommutator}, which may be chosen to be $ \g_a$ and 
$\tilde \g_a =  -\g_a$. In the latter case, the sign of the second term in the right-hand side of \eqref{A.2b} is opposite.

Spinor indices are raised and lowered using the $\sSL(2, \mathbb{R})$ invariant tensors
\begin{align}
\ve^{\a\b} = \left(
\begin{array}{cc}
0 & 1 \\
-1 & 0
\end{array}
\right)~, \quad \ve_{\a\b} = \left(
\begin{array}{cc}
0 & -1 \\
1 & 0
\end{array}
\right) \quad \implies \quad 
\ve_{\a\b} \ve^{\b \g} = \d_{\a}{}^{\g}~,
\end{align}
according to the convention
\begin{align} 
\psi_{\a} = \ve_{\a\b} \psi^{\b}~, \quad \psi^{\a} = \ve^{\a\b} \psi_{\b}~.
\end{align}
In particular, lowering the second spinor index of $(\g_{a})_{\a}{}^{ \b} $ leads to the matrices
\begin{align}
(\g_{a})_{\a\b} = (\g_{a})_{\b\a} = ( \mathbbm{1}, \sigma_{1}, \sigma_{3})~,
\end{align}
which may be used to prove the well-known isomorphism 
$\sSO_0(2,1)\cong \sSL(2,{\mathbb R} )/{\mathbb Z}_2 $.

The gamma matrices  satisfy some useful identities, including the following:
\begin{subequations}
\begin{align}
(\g^{a})_{\a\b} (\g_{a})_{\rho \delta} =& 2 \ve_{\a (\rho} \ve_{\delta) \b}~, \\
\ve_{abc} (\g^{b})_{\a\b} (\g^{c})_{\rho \delta} =& \ve_{\rho(\a} (\g_{a})_{\b ) \d} + \ve_{\d ( \a} (\g_{a})_{\b)\rho}~, \\
\mathrm{tr}[ \g_{a} \g_{b} \g_{c} \g_{d} ] =& 2 \eta_{ab} \eta_{cd} - 2 \eta_{ac} \eta_{db} + 2 \eta_{ad} \eta_{bc}~.
\end{align}
\end{subequations}
The gamma matrices may be used to express any three-vector $V_{a}$ as a symmetric rank-two spinor $V_{\a\b} = V_{\b\a}$. This correspondence is given by
\begin{align}
V_{\a\b} = (\g^{a})_{\a\b} V_{a}~, \quad V_{a} = -\frac{1}{2} (\g_{a})^{\a\b}  V_{\a\b}~.
\end{align}

In three dimensions, an anti-symmetric tensor $F_{ab} = -F_{ba}$ is Hodge-dual to a three-vector $F_{a}$ through the correspondence 
\begin{align} \label{Hodge}
F_{a} = \frac 12 \ve_{abc} F^{bc}~, \quad F_{ab} = -\ve_{abc} F^{c}~.
\end{align}
Then, the symmetric spinor $F_{\a\b}$ associated with $F_{a}$ can alternatively be expressed in terms of $F_{ab}$
\begin{align} \label{SpinorHodge}
F_{\a\b} = (\g^{a})_{\a\b} F_{a} = \frac 12 (\g^{a})_{\a\b} \ve_{abc} F^{bc}~.
\end{align}
The three objects $F_{a}$, $F_{ab}$ and $F_{\a\b}$ are in one-to-one correspondence with each other. The corresponding inner products are related as
\begin{align}
- F^{a} G_{a} = \frac 12 F^{ab} G_{ab} = \frac 12 F^{\a\b} G_{\a\b}~.
\end{align}

Let $\{ \mathcal{M}_{ab} = - \mathcal{M}_{ba}\}$ be 
the Lorentz generators or, equivalently, the generators of  $\mathfrak{sl}(2, \mathbb{R})$.
They satisfy the commutation relations
\begin{align}
\big[ \cM_{ab}, \cM_{cd} \big] = \eta_{ad} \cM_{bc} - \eta_{ac} \cM_{bd} 
+ \eta_{bc} \cM_{ad} - \eta_{bd} \cM_{ac}~.
\end{align}
The generator $\cM_{ab}$ acts on a vector $V_{c}$ as
\begin{align}
\mathcal{M}_{ab} V_{c} = 2\eta_{c[a}V_{b]}~,
\end{align}
and on a spinor $\psi_{\g}$ as
\begin{align}
\mathcal{M}_{ab} \psi_{\g} = (\S_{ab})_{\g}{}^{ \d} \psi_{\d}~, 
\quad 
(\S_{ab})_{\g}{}^{\d} = \frac{1}{4} [ \g_{a}, \g_{b} ]_{\g}{}^{\d}~.
\end{align}
In accordance with \eqref{Hodge} and \eqref{SpinorHodge} the Lorentz generator $\mathcal{M}_{ab}$ may equivalently be expressed as a vector $\mathcal{M}_{a}$ or a symmetric spinor $\mathcal{M}_{\a\b}$ such that
\begin{align}
\mathcal{M}_{a} \psi_{\g} = - \frac 12 (\g^{a})_{\g}{}^{\d} \psi_{\d}~, \quad \mathcal{M}_{\a\b} \psi_{\g} = \ve_{\g (\a} \psi_{\b)}~.
\end{align}

Let $\{\mathcal{N}_{IJ} = -\mathcal{N}_{JI} \}$ be the generators of $\sSO(n)$. 
They act on an $n$-vector $V_{K}$ as
\begin{align}
\mathcal{N}_{IJ} V_{K} = 2 \d_{K [ I} V_{J]}~,
\end{align}
and obey the commutation relations 
\bea
\left[ \cN_{I J}, \cN_{{M} {N}} \right] = 
 \d_{I {N}} \cN_{J {M}}
- \d_{I {M}} \cN_{J {N}} 
+\d_{J {M}} \cN_{I {N}} - \d_{J {N}} \cN_{I {M}} ~.
\eea

Our super $p$-form conventions are as follows. With respect to a set of basis super one-forms $E^{A}$ a super $p$-form $\o$ can be decomposed as
\begin{align}
\o = \frac{1}{p!} E^{A_{1}} \wedge ... \wedge E^{A_{p}} \o_{A_{p} ... A_{1}}~.
\end{align}
Given a super $p$-form $A$ and a super $q$-form $B$ we have
\begin{align}
\mathrm{d} (A \wedge B) = A \wedge \mathrm{d} B + (-1)^{q} \mathrm{d} A \wedge B.
\end{align}


\begin{footnotesize}

\end{footnotesize}

\end{document}